\documentclass{aa}  

\usepackage{txfonts}

\usepackage[utf8]{inputenc} %
\usepackage[english]{babel} %
\usepackage{amssymb}        %
\usepackage{amsmath}        %
\usepackage{bm} %
\usepackage{graphicx}  %
\usepackage[dvipsnames,table,xcdraw]{xcolor}
\newcommand{\Prot}{P_{\mathrm{rot}}}
\newcommand{\DeltaTsp}{\Delta T_\mathrm{sp}}

\usepackage{placeins} %

\usepackage{tikz} %
\usetikzlibrary{decorations.pathreplacing,calc,arrows.meta,positioning}

\usepackage{subcaption} %

\usepackage{xurl}

\usepackage{natbib} %
\setcitestyle{notesep={}} %
\usepackage{aas_macros} %

\usepackage[breaklinks,colorlinks,urlcolor=blue,linkcolor=blue,citecolor=blue]{hyperref} %
\bibpunct{(}{)}{;}{a}{}{,}             %
\hypersetup{
  colorlinks,
  citecolor=blue,%
  linkcolor=[rgb]{0.8, 0.2, 1.0},
  urlcolor=blue,%
}

\usepackage{arydshln} %

\usepackage{siunitx} %

\usepackage{array}%
\newcolumntype{L}{>{$}l<{$}} %
\newcolumntype{C}{>{$}c<{$}} %
\usepackage[labelfont=bf]{caption} %

\usepackage{dsfont} %

\newcommand{\UNSPOTTER}{\texttt{UnSPOTTER}}
\newcommand{\UNSPOTTERexp}{Uncontaminated Spectra from Photospheric Observables using Transformers for TLS Effect Removal}%

\begin{document}

    \title{Correcting stellar contamination in transmission spectroscopy with contemporaneous monitoring: Application to GJ\,1214\,b\thanks{Based on observations collected at Centro Astron\'omico Hispano en Andaluc\'ia, Calar Alto, jointly operated by the Instituto de Astrof\'isica de Andaluc\'ia (IAA-CSIC) and Junta de Andaluc\'ia.}}

   \author{Ò.~Porqueras-León\inst{1}\fnmsep\inst{2}
          \and
          M. Perger\inst{1}\fnmsep\inst{2}
          \and
          J. C. Morales\inst{1}\fnmsep\inst{2}
          \and
          I. Ribas\inst{1}\fnmsep\inst{2}
          \and
          J. Blanco-Pozo\inst{1}\fnmsep\inst{2}\fnmsep\inst{3}
          \and
          H. Tabernero\inst{1}\fnmsep\inst{2}
          \and
          E. Herrero\inst{1}
          \and%
          E.~Pall\'e\inst{4}\fnmsep\inst{5}
          \and
          H. Parviainen\inst{4}\fnmsep\inst{5}
          \and
          F. Murgas\inst{4}\fnmsep\inst{5}
          \and
          C. Rodríguez-López\inst{6}
          \and
          P. H. Hauschildt\inst{7}
          \and
          A. Schweitzer\inst{7}
          \and
          N.-E. Nèmec\inst{1}\fnmsep\inst{2}
          \and%
          D. Baroch\inst{1}
          \and
          N. R. C. Gomes\inst{1}\fnmsep\inst{2}
          \and
          G. Anglada-Escudé\inst{1}\fnmsep\inst{2}
          \and
          J. Aceituno\inst{8}
          \and
          N. Narita\inst{4}\fnmsep\inst{9}\fnmsep\inst{10}
          \and
          A. Fukui\inst{4}\fnmsep\inst{9}
          \and
          M. Mori\inst{10}\fnmsep\inst{11}
          \and%
          Y.~Smolyarova\inst{12}
          }
    
    \titlerunning{Correcting stellar contamination in transmission spectroscopy with contemporaneous monitoring}

   \institute{Institut d'Estudis Espacials de Catalunya (IEEC), Edifici RDIT, Campus PMT-UPC, 08860 Castelldefels (Barcelona), Spain\\ %
              \email{porqueras@ieec.cat}
         \and
             Institut de Ciències de l’Espai (ICE, CSIC), Campus UAB, c/Can Magrans s/n, 08193 Bellaterra (Barcelona), Spain %
        \and
            Facultat de Física, Universitat de Barcelona (UB), Martí Franquès 1, 08028, Barcelona, Spain
        \and
            Instituto de Astrofísica de Canarias (IAC), 38200 La Laguna, Tenerife, Spain
        \and
            Departamento de Astrofísica, Universidad de La Laguna (ULL), 38206 La Laguna, Tenerife, Spain
        \and
            Instituto de Astrofísica de Andalucía (IAA-CSIC), Glorieta de la Astronomía s/n, 18008 Granada, Spain
        \and
            Hamburger Sternwarte, Gojenbergsweg 112, 21029 Hamburg, Germany
        \and
            Centro Astronómico Hispano en Andalucía, Observatorio de Calar Alto, Sierra de los Filabres, 04550, Gérgal, Almería, Spain
        \and
            Komaba Institute for Science, The University of Tokyo, 3-8-1 Komaba, Meguro, Tokyo 153-8902, Japan
        \and
            Astrobiology Center, 2-21-1 Osawa, Mitaka, Tokyo 181-8588, Japan
        \and
            National Astronomical Observatory of Japan, 2-21-1 Osawa, Mitaka, Tokyo 181-8588, Japan
        \and
            Barcelona School of Economics (BSE), 08005 Barcelona, Spain
             }

   \date{Received 15 May 2026 / Accepted 26 October 2026}

  \abstract
   {Transmission spectroscopy has emerged as an essential tool for characterising the atmospheres of transiting exoplanets. However, stellar surface inhomogeneities contaminate transmission spectra, posing a major challenge, particularly to high-precision observations like those from the \textit{James Webb Space Telescope} (JWST).}
   {We develop and apply practical methodologies to correct for stellar contamination in transmission spectroscopy data, thereby improving the accuracy of exoplanet atmospheric characterisation.} %
   {We obtained coordinated multi-band photometry and high-resolution spectroscopy around JWST observations of GJ\,1214\,b and applied two correction frameworks: (i) an inversion with the \texttt{StarSim} code, in which we infer stellar activity parameters from the monitoring data and propagate them to corrections; and (ii) \UNSPOTTER{}, a data-driven approach in which neural networks (NNs) are trained on a large suite of \texttt{StarSim} simulations to learn the mapping from the monitoring data to the corrections.}
   {Both frameworks agree within uncertainties and indicate that the JWST observations of GJ\,1214 occurred at a favourable rotation phase with low stellar contamination. For \UNSPOTTER{}, the inferred corrections can be constrained at the $\sim$10--30 ppm level across the JWST wavelength range, implying post-correction residuals of the same order.
   If the same system were observed near activity maximum, the predicted contamination would be strongly chromatic and substantially affect atmospheric retrievals, highlighting the importance of coordinated ground-based monitoring and robust activity correction techniques to accurately interpret transmission spectra.}%
   {The two approaches are complementary but serve different roles. The inversion yields physically interpretable parameters and surface maps, whereas \UNSPOTTER{} marginalises more efficiently over surface-configuration degeneracies and provides tighter predictive corrections once trained and validated within the relevant simulation domain. The NNs carry an upfront training cost, but then enable fast predictions within their training distribution. Our pipeline can potentially be generalised to other active stars: with multi-band and multi-technique monitoring around JWST visits, it can provide activity-level forecasts for scheduling and activity corrections for atmospheric analyses.}


   \maketitle

\section{Introduction}

Over the past three decades, exoplanetary research has witnessed ground-breaking advancements, marked by the discovery of exoplanets through methods like the Doppler technique and photometric transits \citep[e.g.][]{Mayor:1995aa,Charbonneau_2000,Anglada-Escude:2016aa,Gillon:2017aa}. Expanding beyond detection, the scientific community now also endeavours to characterise the atmospheres of known transiting planets. Transmission spectroscopy has emerged as a pivotal method, where the chromatic dependence of the planetary transit depth reveals crucial information about the planet's atmosphere, enabling the study of its chemical composition, abundances, and physical properties \citep[e.g.][]{Seager_2000,Charbonneau_2002,Tinetti:2007aa,Fortney_2010,Tsiaras:2019aa,Madhusudhan2019}. %

In practice, the transit depth is measured as the square of the planet-to-star radius ratio, $p^2 (\lambda) \coloneqq (R_\mathrm{p}/R_\star)^2$ for a homogeneous stellar disc, with its wavelength ($\lambda$) dependence tracing atmospheric opacity. However, the presence of magnetically-driven evolving active regions and granulation patterns on stellar surfaces introduces challenges to the accurate measurement of transit depths by inducing a variety of signals \citep{sing2011,McCullough_2014,Rackham_2017}. The inhomogeneous distribution of bright and dark magnetic features (e.g. faculae and spots) means that the conventional approach of using the entire visible stellar disc as the reference light source during a planet's transit may not accurately represent the flux occulted by the planet along its transit chord. This phenomenon is known as the transit light source (TLS) effect \citep{Rackham_2018,Rackham_2019,Rackham_2023}. Failure to account for the TLS effect results in the contamination of the transmission spectrum with stellar features, biasing the retrieval of planetary atmospheric properties. For instance, the \textit{James Webb Space Telescope} (JWST) NIRISS transmission spectra of TRAPPIST-1\,b exhibit robust evidence of stellar contamination \citep{Lim_2023}, and \citet{Edwards_2021} encountered challenges using \textit{Hubble Space Telescope} (HST) observations of LHS\,1140b. Changes in visit-to-visit contamination and offsets between instruments have also been reported in JWST spectra of GJ\,3090\,b \citep{Ahrer_2025}. Despite significant research efforts, characterising stellar contamination in transmission spectroscopy remains a challenge that highlights the critical need for advanced tools to address and mitigate stellar effects, ensuring more robust and conclusive assessments of exo-atmospheres. %

Several strategies have been used to mitigate these effects. One approach is forward-modelling of active photospheres, in which chromatic biases are predicted by simulating spots/faculae, e.g. SOAP \citep{2012A&A...545A.109B,2025A&A...702A..84C}, \texttt{StarSim} \citetext{\citealp{herrero_starsim,rosich_starsim}; Gomes et al., in prep.}, \texttt{SAGE} \citep{Chakraborty2024}, \texttt{actress} \citep{2021MNRAS.504.4751J} or \texttt{StARPA} \citep{Cracchiolo2021a,Cracchiolo2021b,Thompson2024}. These tools can be used to interpret monitoring data or compute correction factors under an assumed surface configuration. Another approach is joint planet--star retrieval, which embeds stellar heterogeneity directly in the fit so that activity parameters are inferred alongside atmospheric properties, e.g. \textsc{AURA} \citep{2018MNRAS.480.5314P} or \texttt{ASteRA} \citep{Thompson2024}. In practice, many of these frameworks describe the stellar contamination through simplified disc-averaged prescriptions. While such treatments can be adequate in some favourable regimes, recent work has shown that they can significantly bias the inferred stellar-contamination correction, and therefore the corrected transmission spectrum, when limb darkening, the spatial distribution of active regions, and the transit geometry become important \citep{2026A&A...706A.281S}. Surface-based approaches such as \texttt{StarSim}, by contrast, retain these geometric effects explicitly. %

In parallel, other developments are expanding the practical toolkit. Data-driven and machine learning (ML) approaches have been explored mainly for related tasks such as detecting and correcting spot-crossing signatures in transit light curves \citep{2023RASTI...2..695N} or accelerating retrievals \citep{2022A&A...662A.108A}, with early demonstrations on simulated and limited observed datasets. Meanwhile, theory is refining the spectra and centre-to-limb behaviour of magnetic components, with advances in 3D radiative magnetohydrodynamics (MHD) models informing facular and spot contrasts, improving the priors and templates that forward models and joint retrievals rely on, and quantifying when 1D atmosphere models are insufficient to reproduce the behaviour of magnetic regions \citep{Witzke_2022,2025A&A...693A.264B,2025ApJ...978L..13S}.

Despite this progress, there remains a practical gap between activity monitoring and correction: given contemporaneous time series around a transit, there is no standardised route to derive epoch-specific chromatic corrections with uncertainties that can be propagated to the observed transmission spectra.

In this work, we introduce a pipeline to mitigate stellar contamination in transmission spectroscopy, and apply it to JWST/NIRSpec G395H observations of GJ\,1214\,b. The pipeline combines coordinated monitoring around the transits (multi-band photometry and high-resolution spectroscopy in the case of GJ\,1214\,b) with two complementary correction frameworks: (i) an inversion approach with \texttt{StarSim} that fits the monitoring data to infer physically interpretable stellar activity parameters and propagates them into wavelength-dependent transit-depth corrections; and (ii) \UNSPOTTER{},\footnote{\UNSPOTTERexp.} a data-driven approach in which a transformer-encoder neural network (NN) ensemble is trained on a suite of \texttt{StarSim} simulations to learn the mapping from the same monitoring observables to the chromatic contamination, marginalising over surface-configuration degeneracies. Section~\ref{sec:methodology} details the methodology and defines the two correction frameworks. Section~\ref{sec:a_model_for_gj_1214} describes the GJ\,1214 system, the monitoring and JWST datasets, and the data reduction. Section~\ref{sec:results} presents the inferred stellar-contamination corrections at the observed JWST epochs and near activity maximum. Section~\ref{sec:discussion} discusses the implications, agreement between frameworks, and key limitations. Section~\ref{sec:conclusions} summarises our conclusions.

\section{Methodology}\label{sec:methodology}

\begin{figure*}[t]
\centering
\resizebox{\textwidth}{!}{%
\begin{tikzpicture}[
  font=\small,
  node distance=0.85cm,
  block/.style args={#1}{
    draw, rounded corners=2pt, thick,
    align=center, inner ysep=8pt, inner xsep=8pt,
    minimum height=1.85cm, text width=#1
  },
  input/.style={fill=blue!5},
  simulation/.style={fill=orange!10},
  inference/.style={fill=green!7},
  output/.style={fill=violet!7},
  edge/.style={-Latex, very thick},
  dataedge/.style={-Latex, very thick, dashed, draw=blue!60!black},
  swatch/.style={draw, rounded corners=1pt, thick, minimum width=0.50cm,
                 minimum height=0.34cm, inner sep=0pt}
]

\def\WInput{3.90cm}
\def\WFit{3.25cm}
\def\WMaps{4.15cm}
\def\WForward{3.75cm}
\def\WResult{3.55cm}
\def\WFinal{5.10cm}

\node (invobs) [block=\WInput, input]
  {\textbf{Observed monitoring}\\[1pt]
   \footnotesize Time series + target\\[-1pt]
   \footnotesize constraints};

\node (invfit) [block=\WFit, inference, right=of invobs]
  {\textbf{\texttt{StarSim} inversion}\\[1pt]
   \footnotesize Fit the monitoring data};

\node (invmaps) [block=\WMaps, inference, right=of invfit]
  {\textbf{Compatible surface models}\\[1pt]
   \footnotesize Ensemble of active-region maps};

\node (invforward) [block=\WForward, simulation, right=of invmaps]
  {\textbf{Forward transit\\simulations}\\[1pt]
   \footnotesize Propagate every\\[-1pt]
   \footnotesize surface model};

\node (invresult) [block=\WResult, output, right=of invforward]
  {\textbf{Contamination spectra}\\[1pt]
   \footnotesize Ensemble of $\Delta p^2(\lambda)$};

\node[font=\large\bfseries, anchor=south]
  at ([yshift=0.32cm]invmaps.north)
  {Inversion framework};

\draw[edge] (invobs) -- (invfit);
\draw[edge] (invfit) -- (invmaps);
\draw[edge] (invmaps) -- (invforward);
\draw[edge] (invforward) -- (invresult);

\node (simdesign) [block=\WInput, input, below=2.20cm of invobs]
  {\textbf{Target-informed design}\\[1pt]
   \footnotesize Priors, cadence, uncertainties,\\[-1pt]
   \footnotesize and observed variability level};

\node (simforward) [block=\WFit, simulation, right=of simdesign]
  {\textbf{\texttt{StarSim} simulations}\\[1pt]
   \footnotesize Sample the \\[-1pt]
   \footnotesize training domain};

\node (simpairs) [block=\WMaps, simulation, right=of simforward]
  {\textbf{Synthetic labelled dataset}\\[1pt]
   \footnotesize Monitoring time series + true\\[-1pt]
   \footnotesize $\Delta p^2(\lambda)$ for each simulation};

\node (train) [block=\WForward, inference, right=of simpairs]
  {\textbf{Train \UNSPOTTER{}}\\[1pt]
   \footnotesize Neural-network ensemble};

\node (apply) [block=\WResult, output, right=of train] %
  {\textbf{Apply to observations}\\[1pt]
   \footnotesize Predicted $\Delta p^2(\lambda)$};

\node[font=\large\bfseries, anchor=south]
  at ([yshift=0.32cm]simpairs.north)
  {\UNSPOTTER{} framework};

\draw[edge] (simdesign) -- (simforward);
\draw[edge] (simforward) -- (simpairs);
\draw[edge] (simpairs) -- (train);
\draw[edge] (train) -- (apply);

\coordinate (databus) at ($(invobs.south)!0.47!(simdesign.north)$);
\coordinate (datajunction) at (invobs.south |- databus);
\draw[very thick, dashed, draw=blue!60!black]
  (invobs.south) -- (datajunction);
\draw[dataedge]
  (datajunction) -- (simdesign.north);
\draw[dataedge]
  (datajunction)
  -- node[pos=0.78, above, font=\footnotesize, text=blue!60!black]
     {Observed monitoring}
  (apply.north |- databus) -- (apply.north);

\coordinate (resultmid) at ($(invresult)!0.5!(apply)$);
\coordinate (finalwest) at ($(resultmid)+(2.925cm,0)$);
\node (final) [block=\WFinal, output, minimum height=2.15cm,
               inner xsep=14pt, anchor=west] at (finalwest)
  {\textbf{Correct the observed}\\[-1pt]
   \textbf{transmission spectrum}\\[2pt]
   \footnotesize Epoch-specific stellar-contamination\\[-1pt]
   \footnotesize correction with uncertainties};

\draw[edge] (invresult.east) -- ++(0.45cm,0)
  |- ([yshift=0.43cm]final.west);
\draw[edge] (apply.east) -- ++(0.45cm,0)
  |- ([yshift=-0.43cm]final.west);

\coordinate (legendbase) at ([yshift=-0.95cm]simpairs.south);
\node (keyinput) [swatch, input, anchor=east]
  at ($(legendbase)+(-5.25cm,0)$) {}; %
\node[anchor=west, font=\footnotesize] at ([xshift=0.10cm]keyinput.east)
  {Observations / Target inputs};

\node (keysim) [swatch, simulation, anchor=east]
  at ($(legendbase)+(-0.75cm,0)$) {};
\node[anchor=west, font=\footnotesize] at ([xshift=0.10cm]keysim.east)
  {Forward simulations};

\node (keyinf) [swatch, inference, anchor=east]
  at ($(legendbase)+(2.85cm,0)$) {}; %
\node[anchor=west, font=\footnotesize] at ([xshift=0.10cm]keyinf.east)
  {Fitting / Learning};

\node (keyout) [swatch, output, anchor=east]
  at ($(legendbase)+(6.10cm,0)$) {}; %
\node[anchor=west, font=\footnotesize] at ([xshift=0.10cm]keyout.east)
  {Correction products};

\end{tikzpicture}%
}
\caption{Schematic overview of the two stellar-contamination correction
frameworks. Both approaches use the same underlying \texttt{StarSim} physical
model but differ in how they use the monitoring observations (see text).} %
\label{fig:method_overview}
\end{figure*}
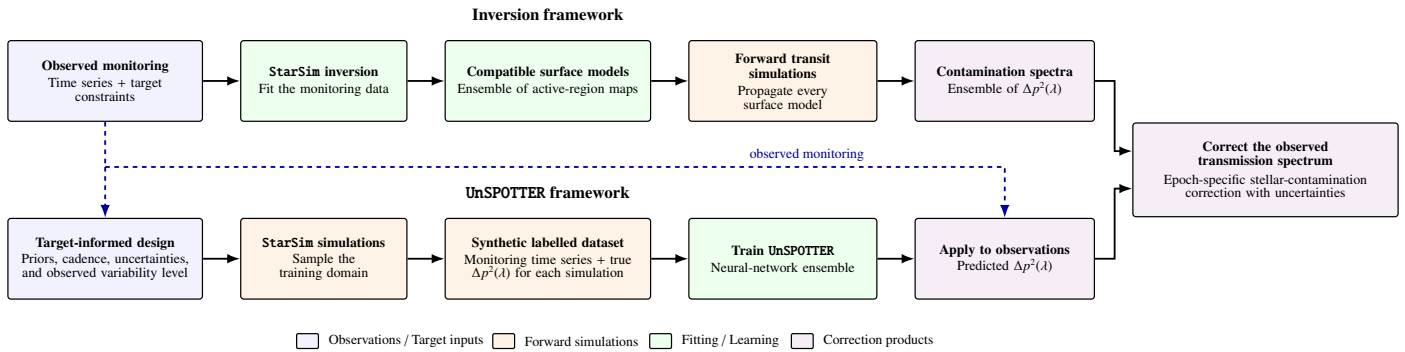

We use two complementary frameworks to assess the consistency of the inferred stellar-contamination corrections and compare their uncertainties. Both use \texttt{StarSim},\footnote{\url{https://github.com/dbarochlopez/starsim}} \citetext{\citealp{herrero_starsim,rosich_starsim}; Gomes et al., in prep.} a modelling code that simulates the effects of stellar activity on photometric and spectroscopic time series. Both frameworks aim to infer the wavelength-dependent stellar contamination
correction to $p^2(\lambda)$, which we denote by
$\Delta p^2(\lambda)$. In the inversion framework,  \texttt{StarSim} is used to directly fit the observed monitoring time series to infer an ensemble of stellar surface configurations, which are then propagated into contamination corrections. In the \UNSPOTTER{} framework,  \texttt{StarSim} is instead used in forward mode to generate a synthetic training set spanning the target-relevant parameter space and observing characteristics, from which a neural-network ensemble learns the mapping from monitoring observables to the associated contamination spectrum. Once trained, \UNSPOTTER{} is applied directly to the observed monitoring data. A schematic overview is shown in Fig.~\ref{fig:method_overview}. %

\subsection{The \texttt{StarSim} modelling code}\label{subsec:starsim_overview}

In \texttt{StarSim}, active elements are represented as circular regions with effective temperatures lower/higher than that of the quiet photosphere; we denote these as spots/faculae, respectively. The code divides the stellar photosphere into a grid of elements (spots, faculae or quiet photosphere) and assigns synthetic stellar spectra Doppler-shifted according to their projected velocity, with corresponding effective temperature. The centre-to-limb variation can either be specified through an analytic limb-darkening law or obtained directly from atmosphere-model specific intensities; in this work, we use the latter approach for both spots and the quiet photosphere. Using the atmosphere-model intensities directly provides a more accurate representation of their centre-to-limb variation than an approximate analytic limb-darkening law. \texttt{StarSim} can additionally model faculae using a limb-brightening prescription \citep[see][]{herrero_starsim}. Observables are computed by integrating the specific intensities over the visible hemisphere, thus working with $\mu$-resolved spectra $I_\lambda(\mu)$, with $\mu$ being the cosine of the angle between the surface normal and the line of sight.

Recent 3D MHD simulations suggest that the solar spot/facula dichotomy does not straightforwardly carry over to low-mass stars: magnetic regions that would be called ``faculae'' in the solar context may only be weakly brighter than the quiet photosphere or even have negative contrast in the optical/near-IR \citep{2015A&A...581A..42B,2021MNRAS.504.4751J,2023MNRAS.524.1139N,Kostogryz2026submitted,2026ApJ..1008...24S}. Since the target analysed in this work is an M4 dwarf (Section~\ref{sec:a_model_for_gj_1214}), for which such simulations do not predict solar-like bright faculae to dominate, we adopt a simplified two-component photosphere in which all magnetically active surface inhomogeneities are represented by cooler dark components. For simplicity and consistency with \texttt{StarSim}, we refer to these dark components as spots throughout.

A set of seven parameters are used to characterise an active region: the time of appearance, lifetime, location on the surface (colatitude $\theta_\mathrm{sp}$ and longitude $\phi_\mathrm{sp}$), angular radius ($r_\mathrm{sp}$), temperature contrast ($\Delta T_\mathrm{sp}\coloneqq T_\star - T_\mathrm{sp}$), and time-evolution law. The last two variables are assumed to be identical for all active elements.

The \texttt{StarSim} modelling code also takes into account all the relevant geometric, physical and wavelength-dependent features related to the presence of active regions, such as limb darkening/brightening, convective blueshift \citep{baroch_carm_starsim}, and the projection of each surface element. \texttt{StarSim} can also simulate the effects of a planet on the time series. In this work, the planet is treated as an opaque occulting disc and its contribution to the system flux is neglected. At each timestamp, the projected position of the planet is computed and the contribution from surface elements occulted by the planetary disc is removed from the disc-integrated flux. Consequently, spot or facular crossings are naturally included when an active region intersects the transit chord. A number of stellar and planet input parameters can be adjusted, such as the effective temperature of the star, its rotation period, its radius, and the surface gravity.

After integrating the whole surface, the code is able to deliver light curves and cross-correlation functions (CCFs) for a variety of instruments. From the CCF, the radial velocity (RV) and common activity indicators such as the bisector inverse slope (BIS), full width at half maximum (FWHM) and contrast, as well as the chromatic index \citep[CRX,][]{serval_paper}, can be accessed.

\subsection{Transmission spectroscopy with \texttt{StarSim}}\label{sec:transmission_spectroscopy_with_starsim}

To simulate the transmission spectrum of an exoplanet, we first adapt the resolution and sampling of the synthetic spectra used by \texttt{StarSim} \citep[e.g. PHOENIX,][see Appendix\,\ref{app:preparation_synthetic_spectra} for further details]{husser_phoenix}. Then, we compute 
a time series of disc-integrated stellar spectra before, during, and after transit, and construct synthetic spectral light curves by integrating the spectra over user-defined wavelength bins and/or weighting by instrumental throughputs (e.g. JWST/NIRSpec, or \textit{Ariel}/AIRS; \citealt{Tinetti:2018aa}). We estimate $p^2(\lambda)=(R_\mathrm{p}/R_\star)^2$ from these light curves by fitting analytic transit models computed with \texttt{batman}\footnote{\url{https://lkreidberg.github.io/batman}} \citep{batman_paper}. As a validation step, we also performed Bayesian transit fits with \texttt{juliet}\footnote{\url{https://juliet.readthedocs.io/en/latest/}} \citep{juliet_paper} and verified that $p^2(\lambda)$ values were consistent for representative channels.\footnote{These routines were implemented within \texttt{StarSim} and are planned for inclusion in a forthcoming public release (Gomes et al., in prep.).}

The transit model is parametrised by the planet-to-star radius ratio $p$, semi-major axis $a/R_\star$, orbital inclination $i$, time of mid-transit $T_c$, orbital period $P_\mathrm{p}$, eccentricity $e$, argument of periastron $\omega$, and limb-darkening (LD) coefficient vector $\bm{u}$. For each transit, we first fit the white light curve to obtain the global white-light transit solution after removing a linear out-of-transit baseline to account for rotational modulation of the spotted stellar surface. For spectroscopic channels, we renormalise each by a channel-specific linear baseline, fix the global white-light solution, and fit per-channel $\{p(\lambda),\,\bm{u}(\lambda)\}$. Limb darkening is fitted in compact, physically-bounded parameter spaces following the parametrisations in
\citet{2013MNRAS.435.2152K,2016MNRAS.455.1680K}.
We adopted the quadratic LD law for this work. As an additional check, we tested the three-parameter law, which has been shown to mitigate some wavelength-dependent biases \citep{2024ApJ...977L...7K}, but we did not find significant differences for the GJ\,1214 system analysed in this paper (see Appendix\,\ref{appendix:limb_darkening_laws} for details).

The fitted compact parameters are transformed to the $\bm{u}$ coefficients used by \texttt{batman} for model evaluation. Optimisation uses Levenberg–Marquardt \citep[LM, ][]{1978LNM...630..105M} when no bounds are required and Trust-Region-Reflective \citep[TRF, ][]{10.1137/S1064827595289108} when enforcing physical bounds (e.g. $p\in[0,1]$, compact LD parameters $\in[0,1]$), as implemented in \texttt{scipy.optimize.curve\_fit} \citep{Virtanen:2020aa}. Because the simulated light curves have no per-point uncertainties we perform unweighted least-squares. The stellar contamination correction is then defined as the differential signal relative to an inactive photosphere,
\begin{equation}
\Delta p^2(\lambda) \coloneqq p^2_{\rm active}(\lambda) - p^2_{\rm inactive}(\lambda).
\end{equation}

\subsection{Inverse modelling correction framework}\label{sec:inverse_approach}

 \texttt{StarSim} can also be run in inverse mode to infer a stellar activity model that reproduces the observed time series. This is solved with a two-level scheme: an outer Monte Carlo sampling draws $N_{\mathrm{steps}}$ realisations of the global model parameters (e.g. $P_{\rm rot}$ or $\Delta T_{\rm sp}$) and, at each step, a simulated-annealing optimiser performs $N_{\mathrm{iters,SA}}$ iterations to search for a spot configuration that maximises the joint log-likelihood $\ell=\ln\mathcal{L}$ of the monitoring data. Fitting multiple bands and spectroscopic activity indicators simultaneously helps reduce degeneracies such as between spot area coverage and temperature contrast (see \citealt{rosich_starsim}).

 \subsubsection{Stellar surface inference from the monitoring data}\label{subsubsec:spot_map_inference}

   We fit the monitoring time series with the inverse mode, using priors informed by literature values (when available) and fixing parameters that are weakly constrained by the monitoring dataset.

The number of spots is fixed in \texttt{StarSim}'s inverse mode. We therefore fit a sequence of models with increasing spot count and select $N_\mathrm{sp}$ by balancing goodness of fit against model complexity using the Akaike information criterion (AIC; \citealt{Akaike1974}). For each candidate number of spots $N$, we evaluate the maximised log-likelihood $\hat{\ell}_N$ and compute AIC using the corresponding number of free parameters $k_N$. In our setup each additional spot adds three parameters $(\phi_\mathrm{sp},\theta_\mathrm{sp},r_\mathrm{sp})$, so $k_N=k_0+3N$, with $k_0$ common to all models. We also compute the Bayesian information criterion (BIC; \citealt{Schwarz1978}) and adjacent-model log-likelihood gains $\Delta\hat{\ell}_{N-1\rightarrow N}=\hat{\ell}_{N}-\hat{\ell}_{N-1}$ as diagnostic cross-checks. Because the data are time series with correlated residuals and spot models are non-regular, these metrics are used heuristically rather than as formal hypothesis tests. The full diagnostics for the case of GJ 1214 are given in Appendix~\ref{app:spot_number_ic}.

   \subsubsection{Ensemble summaries for $\Delta p^2(\lambda)$}
    \label{subsubsec:inv_ensemble_inference}

   Once $N_\mathrm{sp}$ is selected, we first discard inversion runs that did not return a valid likelihood or physically admissible surface configuration. Among the remaining inversion runs, we retain all solutions with $\Delta \ell_{N_\mathrm{sp}} = \hat{\ell}_{N_\mathrm{sp}}-\ell_{N_\mathrm{sp}} \leq 15$ relative to the maximum-likelihood solution. This leaves $N_{\rm sol}$ valid solutions. For each remaining surface map, we simulate spectral light curves at the transmission-spectroscopy epochs with and without active regions, and compute the corresponding correction spectrum $\Delta p^2(\lambda)$ using the procedure in Section~\ref{sec:transmission_spectroscopy_with_starsim}.

   Because many distinct surface configurations fit the monitoring data comparably well, we propagate this map degeneracy into the correction by
    treating all retained solutions equally. At each wavelength bin, we adopt the ensemble mean of 
    $\Delta p^2(\lambda)$
    as our point estimate. Uncertainties are quantified by the central 68\% interval, defined by the 16$^{\rm th}$ and 84$^{\rm th}$ percentiles of the ensemble of $\Delta p^2(\lambda)$ values at $\lambda$. This approach propagates the surface-configuration degeneracy into the final correction.

\subsection{Deep learning correction framework}\label{sec:deep_learning_correction_framework}

   We train \UNSPOTTER{}, a NN ensemble to predict stellar contamination corrections directly from the monitoring time series. %

\subsubsection{Synthetic dataset generation}\label{subsubsec:synthetic_dataset_generation}

   We generate a synthetic dataset of $N_\mathrm{sims}$ \texttt{StarSim} simulations. For each simulation, we generate the corresponding time series (e.g. multi-band photometry, RVs, activity indices) and the associated corrections $\Delta p^2(\lambda)$, computed using the procedure detailed in Section \ref{sec:transmission_spectroscopy_with_starsim}. We sample model parameters from priors designed to span the range consistent with the target system: parameters that have literature constraints are sampled from priors centred on the literature values, while epoch-dependent parameters are drawn independently for each simulation from distributions tuned so that the simulated monitoring time series reproduce the empirical variability level of the observed monitoring dataset. We also removed a small number of simulated realisations that produced pathological correction spectra, such as unphysical values. %

   To make the simulated data more representative of real observations and improve noise robustness, we augment each independent base simulation by generating \(N_\mathrm{noisy}\) noisy realisations. We add Gaussian noise to all observables, with a standard deviation equal to the quadrature sum of (i) the measurement uncertainties of the real data and (ii) the data--model residuals from the best-fit inversion. This accounts for reported uncertainties and residual variability and regularises the network against spurious fluctuations \citep[see also][]{manuel_paper}.

    The train, validation, and test sets are defined at the level of the independent \texttt{StarSim} base simulations, before noise augmentation. We first reserve 20\% of the base simulations as an independent test set and then assign 20\% of the remaining simulations to validation, yielding an approximate 64/16/20 train/validation/test split. All \(N_\mathrm{noisy}\) realisations derived from a given base simulation remain in the same partition, preventing noisy versions of the same stellar-surface configuration from appearing in both training and evaluation data. The validation set is used for hyperparameter selection, training diagnostics, and checkpoint selection, whereas the test set is held out from model fitting and is used only for final performance assessment and calibration.

     This split allows a substantial number of simulations to be reserved for validation and testing while still allocating the majority to training. In particular, for the GJ~1214 application presented in Sect.~\ref{subsec:results_unspotter}, the learning-curve analysis in Appendix~\ref{app:unspotter_learning_curve} shows that the adopted training-set size lies in a regime of diminishing returns.%

   \subsubsection{Preprocessing}
\label{sec:preproc}

For preprocessing, we treat each observational modality as a separate \emph{branch} \(b\) (e.g. I-band photometry, V-band photometry, RV residuals, BIS\dots). It is important to note that since branches correspond to different observables, they in general have different timestamps and number of observations, that is, branch \(b\) has \(L_b\) observations at timestamps $\{t^{(b)}_\ell\}_{\ell=1}^{L_b}$. Each branch is supplied to the model as an array of shape $(N_\mathrm{sims}\cdot N_\mathrm{noisy},\, L_b,\, 1)$. %

We apply per-branch normalisation. For the photometric branches and output corrections, we subtract the training-set mean and divide by the training-set standard deviation. For the spectroscopic branches (RV residuals, FWHM,
BIS), we subtract the per-simulation mean and apply training-set standardisation, scaling by the empirical standard deviation measured in the real monitoring series. This addresses a known simulation--observation amplitude mismatch: \texttt{StarSim} models the underlying stellar variability signal, while the observed variability amplitude also reflects instrumental effects, residual systematics, and non-activity Doppler contributions in the RVs. Without rescaling, the NN could misinterpret this amplitude difference as informative structure.

These reference scalings are computed once from the observed monitoring data and subsequently kept fixed for all synthetic inputs. They are therefore determined solely from the input observations as part of the target-specific preprocessing, and do not make use of the simulated $\Delta p^2$ values.

Each observation (branch \(b\), observation index \(\ell\), timestamp \(t^{(b)}_\ell\)) is encoded into a \emph{token}:
\begin{equation}
    \bm{\tau}_{\ell}^{(b)} = \big(v_{\ell}^{(b)},\; \mathrm{PE}\big(t^{(b)}_\ell\big),\; \bm{e}_b\big)^\top,
\end{equation}
where:
\begin{itemize}
  \item \(v_{\ell}^{(b)}\in\mathbb{R}\) is the normalised measurement (flux, RV residual, or activity index) after per-branch normalisation, representing the measured stellar variability with this observation. %
  \item $\mathrm{PE}(t)\in\mathbb{R}^{d_{\rm temp}}$ is a fixed sinusoidal time encoding in the spirit of the original Transformer positional encoding \citep{Vaswani2017}, defined component-wise as
  \begin{equation}
  \mathrm{PE}(t)_{2i} = \sin\left(\frac{\tilde{t}}{n^{2i/d_{\rm temp}}}\right),\quad
  \mathrm{PE}(t)_{2i+1} = \cos\left(\frac{\tilde{t}}{n^{2i/d_{\rm temp}}}\right),
  \end{equation}
  with $n=10{,}000$, $i=0,\dots,d_{\rm temp}/2-1$, and $ \tilde{t} \coloneqq (t-t_{\rm ref})/t_{\rm scale}$ is a normalised timestamp, defined using a fixed reference epoch $t_{\rm ref}$ and fixed timescale $t_{\rm scale}$ to keep the encoding arguments on a numerically well-behaved scale over typical monitoring baselines. This encoding injects information about the observation temporal order and spacing, enabling the NN to reason about quasi-periodic and long-range temporal correlations.

  \item \(\bm{e}_b\in\mathbb{R}^{d_{\rm emb}}\) is a learned embedding vector that encodes the data modality. This allows the model to distinguish and combine different types of observations. Each branch $b$ is assigned an integer index that is mapped, via a trainable embedding layer, to a dense vector $\bm{e}_b$.
  These embeddings are optimised jointly with the rest of the NN.
\end{itemize}

Let $\mathcal{I}=\{(b,\ell): b=1,\dots,B,\ \ell=1,\dots,L_b\}$. We order $\mathcal{I}$ so $t_{\ell_k}^{(b_k)}$ are non-decreasing timestamps ($t_{\ell_k}^{(b_k)}\leq t_{\ell_{k'}}^{(b_{k'})}$ if $k<k'$). Then, the input sequence for a given simulation is the time-ordered concatenation of tokens,
\begin{equation}
X=\big(\bm{\tau}_{\ell_1}^{(b_1)},\ \bm{\tau}_{\ell_2}^{(b_2)},\ \dots,\ \bm{\tau}_{\ell_L}^{(b_L)}\big)^\top \in \mathbb{R}^{L\times d_{\mathrm{tok}}},
\end{equation}
where $d_{\mathrm{tok}}=1+d_{\mathrm{temp}}+d_{\mathrm{emb}}$ and $L=|\mathcal{I}|=\sum_{b=1}^B L_b$.

\subsubsection{Architecture, hyperparameter tuning and training}%

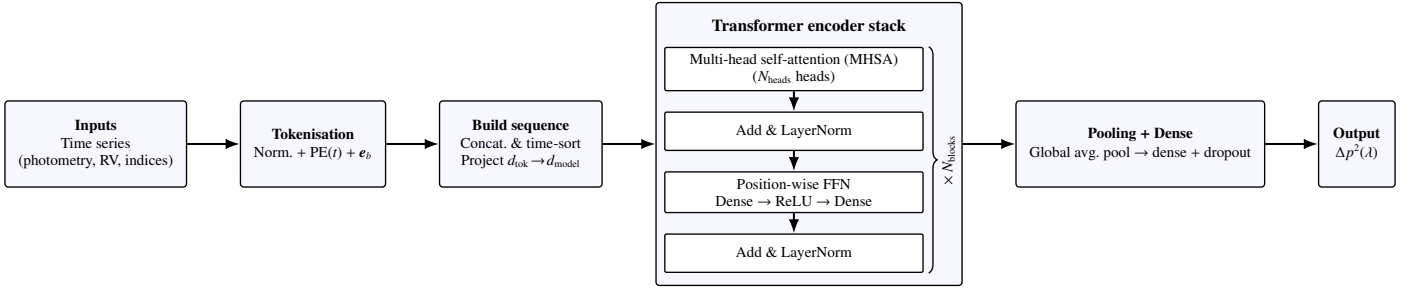
\begin{figure*}[t]
\centering
\resizebox{\textwidth}{!}{%
\begin{tikzpicture}[
  font=\small,
  node distance=1.1cm,
  block/.style args={#1}{
    draw, rounded corners=2pt, thick, fill=blue!3,
    align=center, inner sep=8pt,
    minimum height=1.8cm, minimum width=#1
  },
  block/.default=3.6cm,
  edge/.style={-Latex, very thick},
  inner/.style={
    draw, rounded corners=1pt, thick, fill=white,
    align=center, inner sep=4pt,
    minimum height=\NNHinner, minimum width=\InnerW
  }
]

\def\WIn{3.2cm}
\def\WTok{2.8cm}
\def\WSeq{3.4cm}
\def\TRW{6.4cm}
\def\InnerW{5.4cm} %
\def\WPD{4.8cm}
\def\WOut{1.6cm}

\def\StackXShift{-9pt}
\def\BraceXShift{-3pt}  %
\def\BraceTextXShift{-12pt}

\node (in)  [block=\WIn] {\textbf{Inputs}\\[1 pt]\footnotesize Time series \\[0 pt] (photometry, RV, indices)};
\node (tok) [block=\WTok, right=of in] {\textbf{Tokenisation}\\[1 pt]\footnotesize Norm. + PE$(t)$ + $\bm{e}_b$};
\node (seq) [block=\WSeq, right=of tok] {\textbf{Build sequence}\\[1 pt]\footnotesize Concat. \& time-sort \\[0 pt] Project $d_{\rm tok}\!\to\! d_{\rm model}$};

\ifdefined\NNHinner\else\newlength{\NNHinner}\fi
\ifdefined\NNGap\else\newlength{\NNGap}\fi
\ifdefined\NNHstack\else\newlength{\NNHstack}\fi
\ifdefined\NNHstackHalf\else\newlength{\NNHstackHalf}\fi
\ifdefined\NNHtitle\else\newlength{\NNHtitle}\fi
\ifdefined\NNHpad\else\newlength{\NNHpad}\fi
\ifdefined\NNTRH\else\newlength{\NNTRH}\fi

\setlength{\NNHinner}{0.80cm}
\setlength{\NNGap}{12pt}

\setlength{\NNHstack}{4\NNHinner}
\addtolength{\NNHstack}{3\NNGap}
\setlength{\NNHstackHalf}{0.5\NNHstack}

\setlength{\NNHtitle}{0.75cm}
\setlength{\NNHpad}{0.35cm}

\setlength{\NNTRH}{\NNHstack}
\addtolength{\NNTRH}{\NNHtitle}
\addtolength{\NNTRH}{2\NNHpad}

\node (tr) [block=\TRW, minimum height=\NNTRH, right=of seq] {};

\node (trtitle) [font=\bfseries] at ([yshift=-14pt]tr.north) {Transformer encoder stack};

\coordinate (areaTop)    at ([yshift=-\NNHtitle] tr.north);
\coordinate (areaBottom) at ([yshift= \NNHpad]   tr.south);
\coordinate (areaC)      at ($(areaTop)!0.5!(areaBottom)$);

\node (mhsa) [inner, anchor=north] at ($(areaC)+(\StackXShift,\NNHstackHalf)$)
  {Multi-head self-attention (MHSA)\\ \footnotesize ($N_{\rm heads}$ heads)};
\node (ln1)  [inner, anchor=north] at ($(mhsa.south)+(0,-\NNGap)$)
  {Add \& LayerNorm};
\node (ffn)  [inner, anchor=north] at ($(ln1.south)+(0,-\NNGap)$)
  {Position-wise FFN\\ \footnotesize Dense $\rightarrow$ ReLU $\rightarrow$ Dense};
\node (ln2)  [inner, anchor=north] at ($(ffn.south)+(0,-\NNGap)$)
  {Add \& LayerNorm};

\draw[edge] (mhsa.south) -- (ln1.north);
\draw[edge] (ln1.south)  -- (ffn.north);
\draw[edge] (ffn.south)  -- (ln2.north);

\draw[decorate, decoration={brace, amplitude=5pt}, thick]
  ([xshift=-\BraceXShift]mhsa.north east) -- ([xshift=-\BraceXShift]ln2.south east)
  node[midway, xshift=-\BraceTextXShift, rotate=90, font=\footnotesize] {$\times\,N_{\rm blocks}$};

\node (pd)  [block=\WPD, right=of tr] {\textbf{Pooling + Dense}\\[1 pt]\footnotesize Global avg. pool $\rightarrow$ dense + dropout};
\node (out) [block=\WOut, right=of pd] {\textbf{Output}\\[1 pt]\footnotesize $\Delta p^2(\lambda)$};

\draw[edge] (in) -- (tok);
\draw[edge] (tok) -- (seq);
\draw[edge] (seq) -- (tr);
\draw[edge] (tr) -- (pd);
\draw[edge] (pd) -- (out);

\end{tikzpicture}%
}
\caption{Neural-network architecture used for the \UNSPOTTER{} correction framework.}
\label{fig:nn_architecture}
\end{figure*}

   The sequence of tokens is processed by a transformer encoder-inspired architecture depicted in Fig.~\ref{fig:nn_architecture}, %
   which is implemented in TensorFlow/Keras \citep{2016arXiv160304467A,2018ascl.soft06022C}. The sequence is first projected with a dense layer to the model dimension $d_{\rm model}$. The projected tokens are passed through a stack of $N_{\rm blocks}$ transformer encoder layers. Each layer applies multi-head
   self-attention (MHSA) with $N_\mathrm{heads}$ heads, residual connections, layer normalisation, and a two-layer position-wise (i.e. applied independently to each token) feed-forward network of dimension $d_\mathrm{FFN}$. Attention is bidirectional over the full time-sorted sequence and across branches, allowing tokens from different branches and epochs to attend to one another. We use post-norm blocks, in which layer normalisation is applied after the residual addition, and dropout (randomly zeroing a fraction of activations during training) is applied after the input projection and after the MHSA and feed-forward outputs, with dropout rate (probability that any given neuron will be deactivated) $\mathrm{dropout}_{\mathrm{tr}}$.

    After the final encoder block, the token sequence is aggregated with global average pooling across tokens to produce a fixed-length summary vector. This is processed by \(N_{\rm dense}\) fully connected layers with dimension $d_\mathrm{dense}$ and dropout rate $\mathrm{dropout}_\mathrm{dense}$ and finally a linear output layer returning the predicted contamination vector $\Delta p^2(\lambda)\in \mathbb{R}^{N_\lambda}$ defined on the target wavelength grid. All hidden dense layers use rectified linear unit (ReLU) activations, i.e. zero for negative inputs and linear for non-negative inputs. %

    We optimise the NN architecture and training configuration using \texttt{Optuna} \citep{Akiba2019}, a Bayesian optimisation framework that adaptively explores the hyperparameter space and prunes (early stops) unpromising trials. The hyperparameters included in the search space are 
    $d_{\mathrm{temp}}$, 
    $d_{\mathrm{emb}}$, 
    $d_\mathrm{model}$, 
    $N_\mathrm{blocks}$, 
    $N_\mathrm{heads}$, 
    $d_\mathrm{FFN}$, 
    $\mathrm{dropout}_\mathrm{tr}$, 
    $N_\mathrm{dense}$, 
    $d_\mathrm{dense}$, 
    $\mathrm{dropout}_\mathrm{dense}$, 
    initial learning rate $\eta_0$ (a hyperparameter that controls how much the weights are adjusted in response to estimated errors during training), and batch size (number of training samples processed in one forward/backward pass before updating model weights). We adopt \texttt{Optuna}’s default Tree-structured Parzen Estimator (TPE) sampler, which adaptively explores promising regions of parameter space based on prior trials. To accelerate the search, we use the \texttt{MedianPruner} in combination with the \texttt{TFKerasPruningCallback} to prune trials whose intermediate validation losses are worse than the running median at the same epoch. Architecturally inconsistent configurations are explicitly pruned. Each trial is trained for up to 15 epochs (full passes of the training set through the learning algorithm) with early stopping, sufficient to evaluate relative model quality without expending full training resources. %

    The hyperparameter search is performed on the fixed train/validation partition. The same validation set is used for learning-rate scheduling and checkpoint selection during final training, while the independent test set is not used at either stage and is reserved for final performance assessment and uncertainty calibration.

   The network is trained with the Adam optimiser \citep{Adam_paper}, a training algorithm that computes an adaptive effective step size for each parameter using running estimates of the gradient and its variance, while still being controlled by a global learning rate. We therefore use a \texttt{ReduceLROnPlateau} scheduler to reduce this global learning rate when the validation loss stops improving. We train the final model for 30 epochs, retaining the weights saved from the epoch with lowest validation loss. We adopt mean absolute error (MAE) as the loss function as it reduces sensitivity to outliers and produces less biased predictors in repeated trainings, and monitor mean squared error (MSE) as a complementary metric.

\subsubsection{Posterior-predictive inference and uncertainties}\label{subsubsec:nn_posteriors}

    To predict $\Delta p^2(\lambda)$ for the transit of interest, we use the actual observed monitoring data as inputs to the network. Hence, the sequence of measurements $\tilde v_\ell^{(b)}$ at observation times $t_\ell^{(b)}$ also has some associated measurement uncertainties $\widetilde \sigma_\ell^{(b)}$.
    
    To propagate the uncertainties in the observed monitoring data, we generate $R$ Monte Carlo realisations of the input sequence by sampling each token value from a normal distribution,
    \begin{equation}
    \hat v_\ell^{(b,r)} \sim \mathcal{N}\big(\tilde v_\ell^{(b)}, (\widetilde \sigma_\ell^{(b)})^2\big), 
    \qquad r = 1,\dots,R,
    \end{equation}
    while keeping the temporal encoding and branch embeddings fixed. Hence, each token takes the form
    \begin{equation}
    \bm{\tau}_\ell^{(b,r)} = \big(\hat v_\ell^{(b,r)},\, \mathrm{PE}(t_\ell^{(b)}),\,\bm{e}_b\big)^\top.
    \end{equation}

    Each perturbed input sequence represents a plausible instance of the monitoring data given the measurement errors,
    \begin{equation}
    {\hat{X}}^{(r)} = \big(\bm{\tau}_{\ell_1}^{(b_1,r)},\; \bm{\tau}_{\ell_2}^{(b_2,r)},\; \dots,\; \bm{\tau}_{\ell_L}^{(b_L,r)}\big)^\top, %
    \end{equation}
    that is, the chronologically ordered sequence of tokens for the $r$-th Monte Carlo realisation.

    Let $g(\,\mathord{\cdot}\,;\bm{\omega}):\mathbb{R}^{L\times d_\mathrm{tok}}\rightarrow\mathbb{R}^{N_\lambda}$ denote the neural map from inputs to the wavelength-dependent contamination spectrum $\Delta p^2(\lambda)$, with trained NN weights $\bm{\omega}$. To account for model uncertainty, we perform $D$ stochastic forward passes for each realisation with dropout active, producing predictions %
    \begin{equation}
    \bm{\hat{y}}^{(r,d)} = g({\hat{X}}^{(r)}; \bm{\omega}^{(d)}), \qquad d = 1,\dots,D,
    \end{equation}
    where $\bm{\omega}^{(d)}$ denotes the effective weight configuration obtained by applying a random dropout mask to $\bm{\omega}$ during the $d$-th forward pass. This procedure, known as Monte Carlo dropout \citep{Gal2016}, corresponds to sampling from an approximate posterior distribution over the network weights.

    To reduce biases from individual network initialisations and stochastic training realisations, we train an ensemble of $M$ independently-initialised networks, each sharing the same fixed hyperparameters. Different random seeds produce diverse local optima of the optimisation procedure; pooling across these replicas improves both calibration and coverage \citep{Lakshminarayanan2017}.
    We hence obtain draws $\bm{y}^{(m,r,d)}$ from the approximate posterior predictive distribution (the prediction from NN model $m$, input realisation $r$, and dropout pass $d$).
    The posterior predictive distribution for the contamination spectrum is then approximated by %
    \begin{equation}
    \label{eq:nn_pooled_ppd}
    p(\bm{y}_{\bm{\ast}} \mid \hat{X}, \mathcal{D}_\mathrm{train}) 
    \;\approx\; 
    \frac{1}{S} \sum_{m=1}^{M} \sum_{r=1}^{R} \sum_{d=1}^{D} \delta\big(\bm{y}_{\bm{\ast}} - \bm{y}^{(m,r,d)}\big),
    \end{equation}
    where $S = M R D$ is the total number of pooled draws, and $\mathcal{D}_\mathrm{train}$ is the training dataset. %

    Posterior summaries are computed from the pooled draws, $\bm{y}^{(m,r,d)}$. Let the set of indices for all draws be
    \begin{equation}
    \mathcal{S} = \left\{(m,r,d) \,\middle|\, 1\leq m \leq M,\ 1\leq r \leq R,\ 1\leq d \leq D \right\}.%
    \end{equation}
    For each wavelength bin $\lambda_j$, let
    \begin{equation}
    y_{s,j} \coloneqq y^{(m,r,d)}_j, \quad \text{for } s=(m,r,d) \in \mathcal{S},
    \end{equation}
    denote the pooled draws from the combined input Monte Carlo, dropout, and ensemble procedure. %
    The empirical marginal cumulative distribution function (CDF) at wavelength $\lambda_j$ is then
    \begin{equation}
    \mathrm{CDF}_j(y) \;=\; \frac{1}{S} \sum_{s\in \mathcal{S}} \mathds{1}\big(y_{s,j} \le y\big),
    \end{equation}
    where $\mathds{1}(\cdot)$ is the indicator function, equal to 1 when its argument is true and to 0 otherwise. The $q$-th percentile of the approximate posterior predictive is defined as the generalised inverse of the CDF: %
    \begin{equation}
    P_j(q) \;=\; \inf \Big\{\, y \;:\; \mathrm{CDF}_j(y) \,\ge\, \tfrac{q}{100} \,\Big\}, 
    \qquad q \in (0,100).
    \end{equation}
    Our point estimate for the correction at each wavelength is the posterior mean,
    \begin{equation}
    \overline{\Delta p^2}(\lambda_j) \;\coloneqq\; \frac{1}{S}\sum_{s\in\mathcal{S}} y_{s,j}.
    \end{equation}
    The corresponding raw uncertainty interval is defined by the 16th and 84th percentiles of this distribution, $\big[P_j\,(16),\, P_j\,(84)\big]$.%

    The uncertainty represented by the ensemble, input perturbations, and Monte Carlo dropout is not guaranteed to yield predictive intervals (PIs) with the desired 68\% empirical coverage. Thus, we additionally calibrate the width of the raw intervals using the independent test set. Since the true contamination spectrum is known for every synthetic test simulation, the empirical coverage of the PIs can be measured directly.

    Importantly, this calibration does not determine the wavelength-to-wavelength or case-to-case variation of the uncertainty. These variations are already encoded in the posterior-predictive distributions. Instead, we preserve this structure and scale the deviations of the lower and upper interval bounds from the central prediction by a single common factor. The calibration therefore adjusts only the absolute scale of the uncertainty intervals.

    Since the final prediction is made for a particular set of monitoring observations, we estimate this scale factor from test simulations whose synthetic monitoring signals are compatible with the observed ones.
    For branch \(b\) and observation \(\ell\), we define the standardised residual
    \begin{equation}
        z_{i,\ell}^{(b)}
        =
        \frac{
            x_{i,\ell}^{(b),\mathrm{sim}}
            -
            x_{\ell}^{(b),\mathrm{obs}}
        }{
            \sigma_{\ell}^{(b),\mathrm{aug}}
        },
    \end{equation}
    where \(x_{i,\ell}^{(b),\mathrm{sim}}\) is the value of the corresponding monitoring observable for the test simulation \(i\), \(x_{\ell}^{(b),\mathrm{obs}}\) is its observed value, and \(\sigma_{\ell}^{(b),\mathrm{aug}}\) is the scatter adopted for that measurement in the noise augmentation described in Sect.~\ref{subsubsec:synthetic_dataset_generation}. We then quantify the agreement between simulation \(i\) and the observed monitoring dataset using the Manhattan distance
    \begin{equation}
        d_i
        =
        \frac{1}{L}
        \sum_{b=1}^{B}
        \sum_{\ell=1}^{L_b}
        \left|z_{i,\ell}^{(b)}\right|.%
    \end{equation}

    We define the calibration sample as the test simulations whose monitoring data are compatible with the observations at the level of the adopted augmentation noise. If a simulation reproduces the same underlying monitoring signal as the observations, the standardised residuals defined above follow \(z\sim\mathcal{N}(0,1)\). Thus, we derive the expected distribution of \(d\) under this noise model and use the simulations below the percentile corresponding to a \(3\sigma\) Gaussian probability (99.73\%) to determine the uncertainty scale calibration. %

    For each simulation \(i\) in this calibration sample and wavelength bin \(\lambda_j\), let \(\hat{y}_{i,j}\) denote the central prediction, \(y_{i,j}\) the known simulated correction, and \(L_{i,j}^{\rm raw}\) and \(U_{i,j}^{\rm raw}\) the raw 16th- and 84th-percentile bounds. We rescale the lower and upper bounds using a common multiplicative factor \(s\),
    \begin{equation}
        L_{i,j}^{\rm cal}
        =
        \hat{y}_{i,j}
        -
        s\left(\hat{y}_{i,j}-L_{i,j}^{\rm raw}\right),
        \qquad
        U_{i,j}^{\rm cal}
        =
        \hat{y}_{i,j}
        +
        s\left(U_{i,j}^{\rm raw}-\hat{y}_{i,j}\right).
    \end{equation}

    The minimum scale factor required for the calibrated interval to contain the known simulated value is
    \begin{equation}
        r_{i,j}
        =
        \begin{cases}
        \dfrac{\hat{y}_{i,j}-y_{i,j}}
              {\hat{y}_{i,j}-L_{i,j}^{\rm raw}},
        & y_{i,j}<\hat{y}_{i,j},
        \\[10pt]
        \dfrac{y_{i,j}-\hat{y}_{i,j}}
              {U_{i,j}^{\rm raw}-\hat{y}_{i,j}},
        & y_{i,j}\geq\hat{y}_{i,j}.
        \end{cases}
    \end{equation}
    For a given scale factor \(s\), the calibrated interval contains the known value \(y_{i,j}\) if and only if \(r_{i,j}\leq s\). Therefore, choosing the 68th percentile of the \(r_{i,j}\) values sets the empirical coverage of the calibration sample to 68\%:
    \begin{equation}
        s
        =
        Q_{0.68}
        \left(
        \left\{
        r_{i,j}
        \right\}
        \right),
    \end{equation}
    where \(Q_{0.68}\) denotes the empirical 68th percentile over the calibration simulations and wavelength bins. %

    We evaluate the calibration using five-fold cross-validation within the calibration sample. Independent base simulations are divided among the folds, with all realisations originating from the same base simulation kept together. For each fold, \(s\) is estimated using the other four folds and the resulting intervals are evaluated on the held-out simulations. This provides an out-of-fold estimate of the empirical coverage obtained by the calibration procedure. After this validation, the final factor is estimated from the complete calibration sample and applied to the PI obtained for the observed monitoring data. %

\section{A model for GJ\,1214}\label{sec:a_model_for_gj_1214}

   \subsection{The GJ\,1214 system}

    GJ\,1214 is a nearby M4 dwarf star hosting a sub-Neptune planet, GJ\,1214\,b.
    The planet was discovered by the MEarth transit survey \citep{Charbonneau_2009}.
    Early RV and transit follow-up established GJ\,1214\,b as a low-density sub-Neptune and refined its ephemeris \citep[e.g.][]{AngladaEscude_2013, Harpsoe_2013, Gillon_2014}. GJ\,1214 is active and photometrically variable due to rotating active regions \citep{Berta_2011,Narita_2013,Nascimbeni_2015}. \citet{mallonn_paper} measured a ${\sim}\,$125-day rotation period and found variability dominated by unocculted dark spots. Table~\ref{tab:system_params} summarises the properties of the system. %
    
    With a high transmission spectroscopy metric \citep{Kempton_2018}, GJ\,1214\,b is a benchmark for atmospheric studies. Early ground- and space-based transmission spectra were essentially featureless across the optical and near-infrared, consistent with either a high mean-molecular-weight atmosphere or a hydrogen-dominated atmosphere muted by high-altitude clouds/hazes \citep{Kreidberg_2014, Caceres_2014, Rackham_2017}. HST/WFC3 observations by \citet{Kreidberg_2014} showed a flat near-IR spectrum with high precision, supporting high-altitude aerosols. %

   \defcitealias{mallonn_paper}{Ma18}
   \defcitealias{Mahajan_2024}{Ma24}
   \defcitealias{Newton_2014}{Ne14}
   \defcitealias{Cloutier_2021}{Cl21}
   \begin{table*}%
        \centering
        \caption{System parameters for GJ\,1214 (top) and GJ\,1214\,b (bottom).}\label{tab:system_params}
        \renewcommand{\arraystretch}{1.3}  %
        \setlength{\dashlinedash}{1pt} %
        \setlength{\dashlinegap}{1pt}  %
        \begin{tabular}{LCCCC}
            \hline \hline
            \text{Parameter} & \text{Symbol} & \text{Value} & \text{Unit} & \text{Reference} \\\hline\rule{0pt}{10pt}  %
            \text{Mass} & M_\star & 0.1820^{+0.0042}_{-0.0041} & \mathrm{M}_\odot & \text{\citetalias{Mahajan_2024}} \\
            \text{Effective temperature} & T_\star & 3101\pm43 & \mathrm{K} & \text{\citetalias{Mahajan_2024}} \\
            \text{Spectral Type} & \mathrm{Sp.\,T} & \mathrm{M4V} & - & \text{\citetalias{Newton_2014}} \\
            \text{Surface gravity}^* & \log g & 5.0286^{+0.0078}_{-0.0088} & \mathrm{dex} & \text{\citetalias{Mahajan_2024}}  \\ %
            \text{Radius} & R_\star & 0.2162^{+0.0025}_{-0.0024} & \mathrm{R}_\odot & \text{\citetalias{Mahajan_2024}} \\
            \text{Spot temperature contrast}^{**} & \DeltaTsp & 339^{+155}_{-144} & \mathrm{K} & \text{This work} \\
             &  & 372^{+138}_{-182} &  & \text{\citetalias{mallonn_paper}} \\
            \text{Rotation period}^{**} & \Prot & 126.9^{+4.0}_{-2.7} & \mathrm{days} & \text{This work}  \\
             &  & 125\pm 5 &  & \text{\citetalias{mallonn_paper}}  \\
            \noalign{\vskip 2pt} %
            \hdashline
            \noalign{\vskip 2pt} %
            \text{Orbital period} & P_\mathrm{p} & 1.580404531^{+0.000000018}_{-0.000000017} & \mathrm{days} & \text{\citetalias{Mahajan_2024}} \\
            \text{Planet-to-star radius ratio} & p\coloneqq R_\mathrm{p}/R_\star & 0.11589\pm 0.00016 & - & \text{\citetalias{Mahajan_2024}} \\
            \text{Semi-major axis to stellar radius ratio} & a/R_\star & 14.97^{+0.12}_{-0.14} & - & \text{\citetalias{Mahajan_2024}} \\
            \text{Impact parameter} & b & 0.264^{+0.020}_{-0.023} & - & \text{\citetalias{Mahajan_2024}} \\
            \text{Eccentricity} & e & 0.0062^{+0.0079}_{-0.0044} & - & \text{\citetalias{Mahajan_2024}} \\ %
            \text{RV semi-amplitude} & K & 14.38^{+0.57}_{-0.56} & \mathrm{m\,s}^{-1} & \text{\citetalias{Mahajan_2024}} \\
            \text{Orbit inclination} & i_\mathrm{p} & 88.980^{+0.094}_{-0.085} & \mathrm{deg} & \text{\citetalias{Mahajan_2024}} \\
            \text{Transit Midpoint} & T_\mathrm{c} & 2455701.413328^{+0.000066}_{-0.000059} & \mathrm{days}\;(\mathrm{BJD}_\mathrm{TDB}) & \text{\citetalias{Cloutier_2021}} \\ \hline %
        \end{tabular}%
        \tablefoot{
            \tablefoottext{*}{For simplicity, we write $\log$ for $\log_{10}$ and $\ln$ for $\log_\mathrm{e}$.} \tablefoottext{**}{The literature values from \citet{mallonn_paper} were used to define the priors adopted in the inversion and in the simulations used for \UNSPOTTER{}. The rows labelled ``this work'' report posterior summaries from the inversion.} %
        }
        \tablebib{
            \citetalias{Newton_2014}: \citet{Newton_2014}; \citetalias{mallonn_paper}: \citet{mallonn_paper}; \citetalias{Cloutier_2021}: \citet{Cloutier_2021}; \citetalias{Mahajan_2024}: \citet{Mahajan_2024}.
        }
    \end{table*}

    Recent JWST observations have further constrained the atmosphere of GJ\,1214\,b. MIRI-LRS thermal phase-curve data provided spectrally resolved emission and evidence for a metal-rich atmosphere with a thick aerosol layer \citep{Kempton_2023}. JWST/NIRSpec G395H transmission spectroscopy reported tentative CO$_2$ and CH$_4$ features, favouring very high atmospheric metallicity beneath optically thick aerosols \citep{Schlawin_2024}. \citet{Ohno_2025} combined the HST/WFC3, JWST/NIRSpec G395H, and JWST/MIRI-LRS transmission spectrum to infer a metal-dominated atmosphere below the aerosol deck.
    Stellar activity has been reported to have a significant impact on the transit depths of this system. \citet{Rackham_2017} found optical transit depths shallower than those in the near-infrared and interpreted this as evidence for unocculted faculae. However, subsequent long-term monitoring by \citet{mallonn_paper} found the variability of GJ\,1214 to be more consistent with a spot-dominated photosphere. %

   \subsection{Observations and data reduction}

    We conducted ground-based monitoring around the JWST observations presented in \citet{Schlawin_2024} using a variety of telescopes and instruments. Table \ref{tab:observations} summarises the key aspects of the datasets used in our analysis, including monitoring periods, time spans, and number of retained nightly-binned points.%
    
    \subsubsection{Photometry}

    The Joan Oró Telescope \citep[TJO, ][]{10.1117/12.857672}, located at Montsec Observatory, is a robotic telescope featuring a $\SI{0.8}{\metre}$ primary mirror with an F/9.6 optical system in Ritchey-Chrétien configuration. It employs the Large Area Imager for Astronomy (LAIA), a high-performance CCD camera (Andor iKon XL 230-84) installed at its Cassegrain focus, which provides a non-vignetted field of view (FOV) of $\SI{30}{\arcmin}$ diameter ($0.44\,\mathrm{arcsec/pixel}$). We used the Johnson-Cousins \textit{I} filter. Raw frames were corrected for dark current and bias, and were flat-fielded using the ICAT\footnote{IEEC Calibration and Analysis Tool.} pipeline \citep{2006IAUSS...6E..11C}. Aperture photometry was performed with AstroImageJ \citep{2017AJ....153...77C}, selecting an optimal aperture size that minimised the root mean square error (RMS) of the resulting relative fluxes.

    MuSCAT2 \citep{2019JATIS...5a5001N} is a four-colour simultaneous imaging instrument mounted on the $\SI{1.52}{\metre}$ Telescopio Carlos Sánchez at the Teide Observatory, Spain. It employs four separate CCDs to simultaneously capture images in four photometric bands: \textit{g}, \textit{r}, \textit{i}, and $\textit{z}_\mathrm{s}$. MuSCAT2 uses four Princeton Instruments PIXIS 1024B-series CCD cameras (e2v CCD47-10 sensors), one per band, providing a $\SI{7.4}{\arcmin}\times\SI{7.4}{\arcmin}$ FOV ($0.44\,\mathrm{arcsec/pixel}$). The raw data were reduced using the MuSCAT2 pipeline \citep{Parviainen2019}, which performs dark and flat field calibrations, and computes aperture photometry for the target and several stars in the field. %

    The T90 telescope at Observatorio de Sierra Nevada (OSN) is a $\SI{90}{\cm}$ Ritchey-Chrétien telescope equipped with a CCD camera Andor Ikon-L DZ936N-BEX2-DD 2k$\times$2k with an FOV of $\SI{13.2}{\arcmin}\times\SI{13.2}{\arcmin}$ ($0.387\,\mathrm{arcsec/pixel}$). Typically, 20 exposures were obtained in each of the V and R filters per night, with exposure times of $\SI{150}{s}$ and $\SI{120}{s}$, respectively. Each CCD frame was corrected in a standard way for bias and flat-fielding. We performed aperture photometry testing different aperture sizes to select the optimal ones.

    In all cases, systematic trends were corrected using differential photometry with the brightest reference stars in the field that showed no variability. Then, we removed outliers and measurements affected by poor observing conditions or low signal-to-noise ratio using 3-sigma clipping. We binned intra-night points into nightly means using inverse-variance weighting, and propagated uncertainties accordingly. %
    After these quality cuts, we retained all 22 TJO \textit{I}-band nights, 29 out of 33 OSN nights and all 37 MuSCAT2 nights. We also obtained TJO \textit{B}-band data (22 nights), but excluded them from the analysis due to insufficient precision.

\begin{table}
    \centering
    \caption{Summary of observational datasets used in this study.}\label{tab:observations}
    \renewcommand{\arraystretch}{1.15}
    \begin{tabular}{lccc}
        \hline \hline
        Instrument & Period & Span (days) & $N_\mathrm{used}$ \\
        \hline
        TJO       & May--Oct 2023 & 136 & 22 \\
        OSN       & Jun--Sep 2023 &  88 & 29 \\
        MuSCAT2   & May--Oct 2023 & 139 & 37 \\
        CARMENES  & Jun--Sep 2023 &  94 & 26 \\
        \hline
    \end{tabular}
    \tablefoot{
    The period indicates the approximate monitoring window, and the span gives the elapsed time between the first and last observations used in the analysis. $N_\mathrm{used}$ is the number of nightly binned points retained after quality cuts, reported per band or time series, i.e. effectively the number of nights retained. %
    }
\end{table}

    \subsubsection{Spectroscopy}
    
    We also monitored GJ\,1214 with the CARMENES instrument, %
    a high-resolution, dual-channel spectrograph operating from the $\SI{3.5}{m}$ telescope of the Calar Alto Observatory \citep{quirrenbach_2018_carmenes}. It simultaneously captures stellar spectra in two wavelength ranges: visible-red (520--960 nm) and near-infrared (960--1710 nm), with resolving powers of $94\,600$ and $80\,400$, respectively.%

    The CARMENES spectra were processed with the CARMENES \texttt{caracal}\footnote{CARMENES Reduction And CALibration.} pipeline \citep{10.1117/12.2233574}. We derived RVs, FWHM, and BIS from the CARMENES data using the \texttt{raccoon}\footnote{Radial velocities and Activity indicators from Cross-COrrelatiON with masks. \url{https://github.com/mlafarga/raccoon}} pipeline \citep{raccoon_paper}. The RVs were corrected for nightly zero-points, instrumental drift, and secular acceleration \citep{2023A&A...670A.139R}. Outlier removal was conducted using 3-sigma clipping. After quality cuts, we retained 26 out of 30 CARMENES epochs. To isolate stellar activity signals, we subtracted the known planet-induced signal from the RV time series using the orbital parameters listed in Table~\ref{tab:system_params}. We refer to the resulting time series as the RV residuals.

    \subsubsection{JWST/NIRSpec transmission spectroscopy}

    We use the JWST/NIRSpec G395H transmission spectrum of GJ\,1214\,b from \citet{Schlawin_2024}, based on two transits observed on UT 2023-07-18 and UT 2023-07-19 (MANATEE GTO-1185) covering $2.8$--$5.1\,\mu$m. These two observed transits are hereafter referred to as $t_{143}$ and $t_{145}$, respectively, where the subscript denotes the integer part of the mid-transit time in $\mathrm{BJD}_\mathrm{TDB}-2460000\,\mathrm{days}$; e.g. $t_{143}$ has $T_c \simeq 143.93\,\mathrm{days}$. \citet{Schlawin_2024} extracted the spectra with two independent pipelines (\texttt{tshirt} and \texttt{Eureka!}), hence we adopt the published wavelength bins and transit-depth measurements from both reductions for comparison with our predicted stellar-contamination corrections. Visual inspection and the analysis in \citet{Schlawin_2024} show no evidence for spot-crossing anomalies at the achieved precision; we therefore treat stellar heterogeneities as unocculted and exclude simulated configurations that would produce spot crossings along the transit chord.

    \section{Results}\label{sec:results}

    Here, we present the application of our two stellar activity correction frameworks—inverse modelling and deep learning—to the ground-based monitoring and JWST/NIRSpec observations of GJ\,1214\,b. We analysed two distinct epoch pairs: the observed JWST/NIRSpec transits ($t_{143}$ and $t_{145}$), hereafter ``observed epochs'', and, following the same notation, the pair of transits closest to the photometric minimum ($t_{191}$ and $t_{192}$), hereafter ``activity maximum epochs'', since under our adopted spot-only modelling they correspond to a local activity maximum, i.e. a larger relative contribution of the cooler component to the disc-integrated flux.%

   In order to cover the full JWST/NIRSpec G395H wavelength range and to examine the wavelength dependence of the correction beyond it, we adapted for use in \texttt{StarSim} the
   $\mu$-resolved specific-intensity spectra $I_\lambda(\mu)$ from the state-of-the-art NewEra PHOENIX models \citep{newera_phoenix}. We used spectra computed at 127 $\mu$ values for $\log g = 4.5,5.0,5.5$ and $T_{\rm eff}=2300$--$3500$\,K, spanning the stellar parameter space required for GJ\,1214. In the present analysis, we compute corrections over $0.4$--$\SI{12}{\micro\metre}$, while comparisons with the published JWST/NIRSpec transmission spectrum are restricted to the observed G395H range. %

   \subsection{Inverse modelling correction framework} %

   We used the inverse modelling approach detailed in Section \ref{sec:inverse_approach} to fit the monitoring multi-band photometry, RVs and spectroscopic activity indicators with \texttt{StarSim}'s inverse mode. We adopted the conservative priors informed by previous studies and initial tests found in Table~\ref{tab:inversion_priors}. The remaining stellar parameters were fixed to the values found in Table~\ref{tab:system_params}. In addition, we fixed parameters that are not constrained by the available monitoring: the convective shift to $\mathrm{CS}=0$, motivated by the results of \citet{2021A&A...654A.168L}, who found no net convective blueshift or redshift for stars below $\sim 4000\,\mathrm{K}$; we assumed no differential rotation ($d\Omega=0$), and adopted a constant spot-evolution law, i.e. each active region was assumed to keep a fixed size throughout its lifetime. Each inversion evaluated $N_\mathrm{steps}=1000$ random draws of the stellar parameters and ran $N_\mathrm{iters,SA}=7000$ SA iterations. We adopt $N_\mathrm{sp}=4$ for our analysis, as favoured by the AIC (Table~\ref{tab:spot_number_daic}). Full diagnostics for $\hat{\ell}_N$, $\Delta\hat{\ell}$, AIC and BIC are reported in Appendix~\ref{app:spot_number_ic}.

   \begin{table}%
        \centering
        \caption{Inversion priors.}\label{tab:inversion_priors}
        \renewcommand{\arraystretch}{1.15}  %
        \begin{tabular}{LCC}
            \hline \hline
            \text{Parameter} & \text{Prior Distribution} & \text{Units} \\\hline\rule{0pt}{10pt}
            \phi_\mathrm{sp} & \mathcal{U}(0, 360) & \mathrm{deg} \\
            \theta_\mathrm{sp} & \mathcal{U}(0, 180) & \mathrm{deg} \\
            r_\mathrm{sp} & \mathcal{U}(0, 20) & \mathrm{deg} \\
            \Delta T_\mathrm{sp}{}^* & \mathcal{N}_\mathrm{Tr}(372, 182, 0, 900) & \mathrm{K} \\
            P_\mathrm{rot}{}^* & \mathcal{N}_\mathrm{Tr}(125, 5, 100, 150) & \mathrm{days} \\ \hline
        \end{tabular}
        \tablefoot{
            \tablefoottext{*}{$\mathcal{N}_\mathrm{Tr}(\mu,\sigma,a,b)$ denotes the truncated normal distribution with mean $\mu$, standard deviation $\sigma$, lower bound $a$ and upper bound $b$. For stellar parameters, truncated normal priors are based on values found in Table \ref{tab:system_params}, truncated within physical and/or model limits.}
            }
    \end{table}

   \begin{table}
    \centering
    \caption{Spot-number selection using $\Delta\mathrm{AIC}$ relative to the minimum AIC value.}
    \label{tab:spot_number_daic}
    \renewcommand{\arraystretch}{1.15}
    \begin{tabular}{cc}
    \hline\hline
    $N_\mathrm{sp}$ & $\Delta\mathrm{AIC}$ \\
    \hline
    1 & 26.683 \\
    2 & 2.302 \\
    3 & 2.943 \\
    4 & 0.000 \\
    5 & 4.605 \\
    \hline
    \end{tabular}
    \tablefoot{%
    $\Delta\mathrm{AIC} \coloneqq \mathrm{AIC}_N-\min_N \mathrm{AIC}_N$ for the suite of fixed-$N_\mathrm{sp}$ inversions.%
    }
    \end{table}

\begin{figure*}
      \noindent\hbox to \textwidth{%
        \begin{subfigure}[t]{0.32\textwidth}
          \vspace{0pt}\raggedright
          \includegraphics[trim={1.75ex 1ex 19.5ex 0.0ex},clip,height=0.95\linewidth]%
          {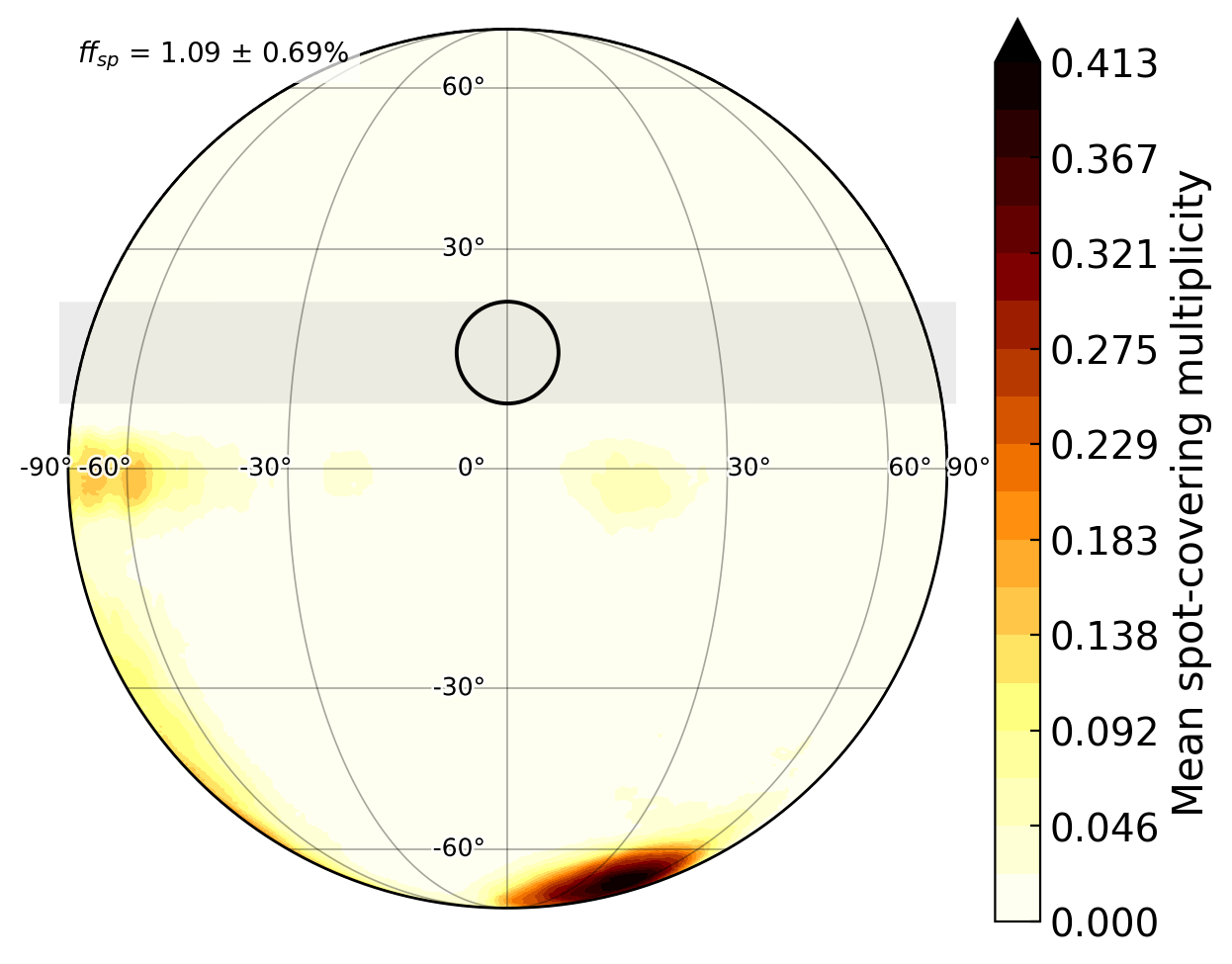}
          \label{fig:inv-1}
        \end{subfigure}%
        \hskip 0pt plus -1fil\relax%
        \begin{subfigure}[t]{0.32\textwidth}
          \vspace{0pt}\centering
          \includegraphics[trim={1.75ex 1ex 1ex 0ex},clip,height=0.95\linewidth]%
          {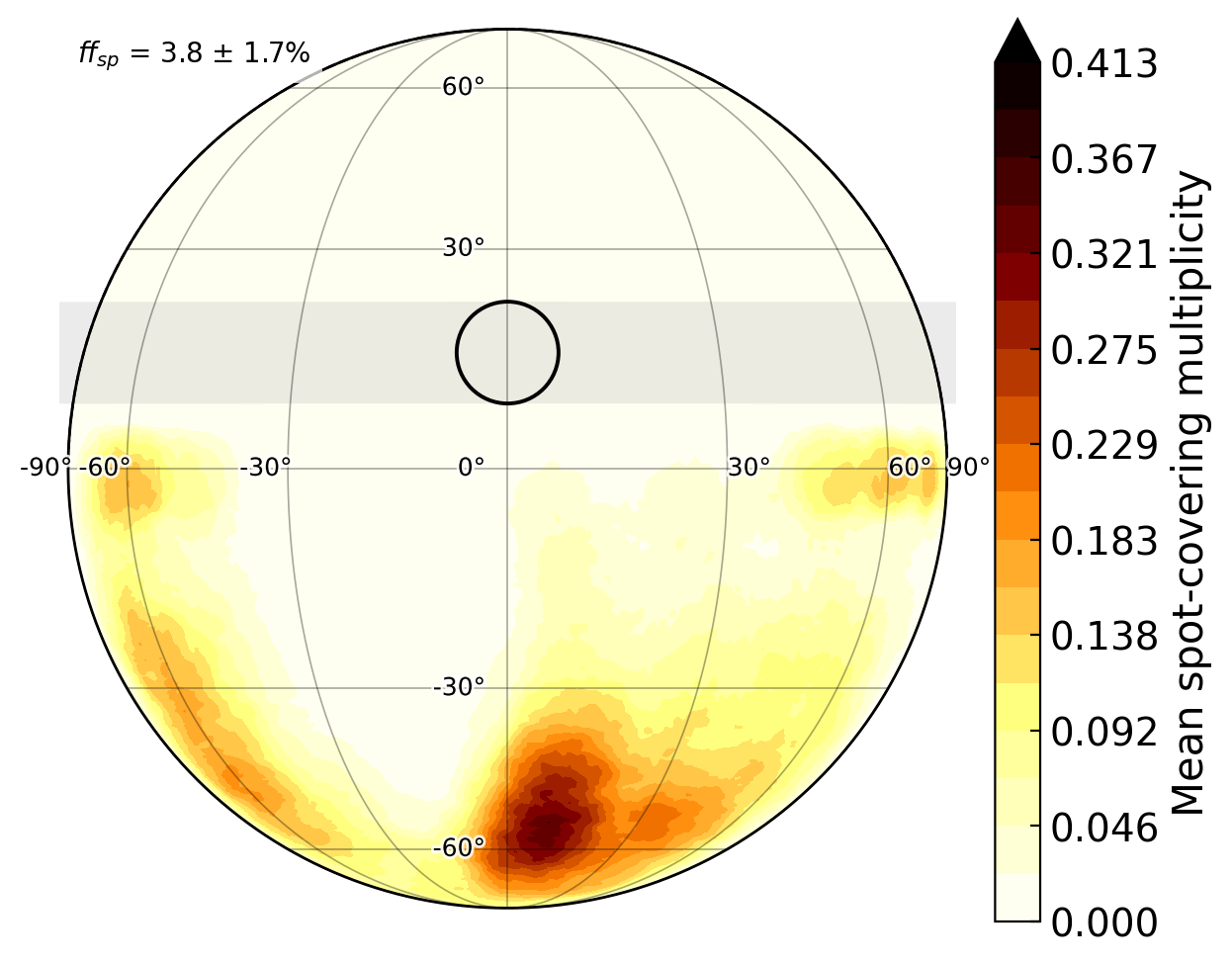}
          \label{fig:inv-2}
        \end{subfigure}%
        \hskip 0pt plus 5fil\relax%
        \begin{subfigure}[t]{0.32\textwidth}
          \vspace{0pt}\raggedleft
          \includegraphics[trim={1ex 1ex 2ex 0ex},clip,height=0.95\linewidth]%
          {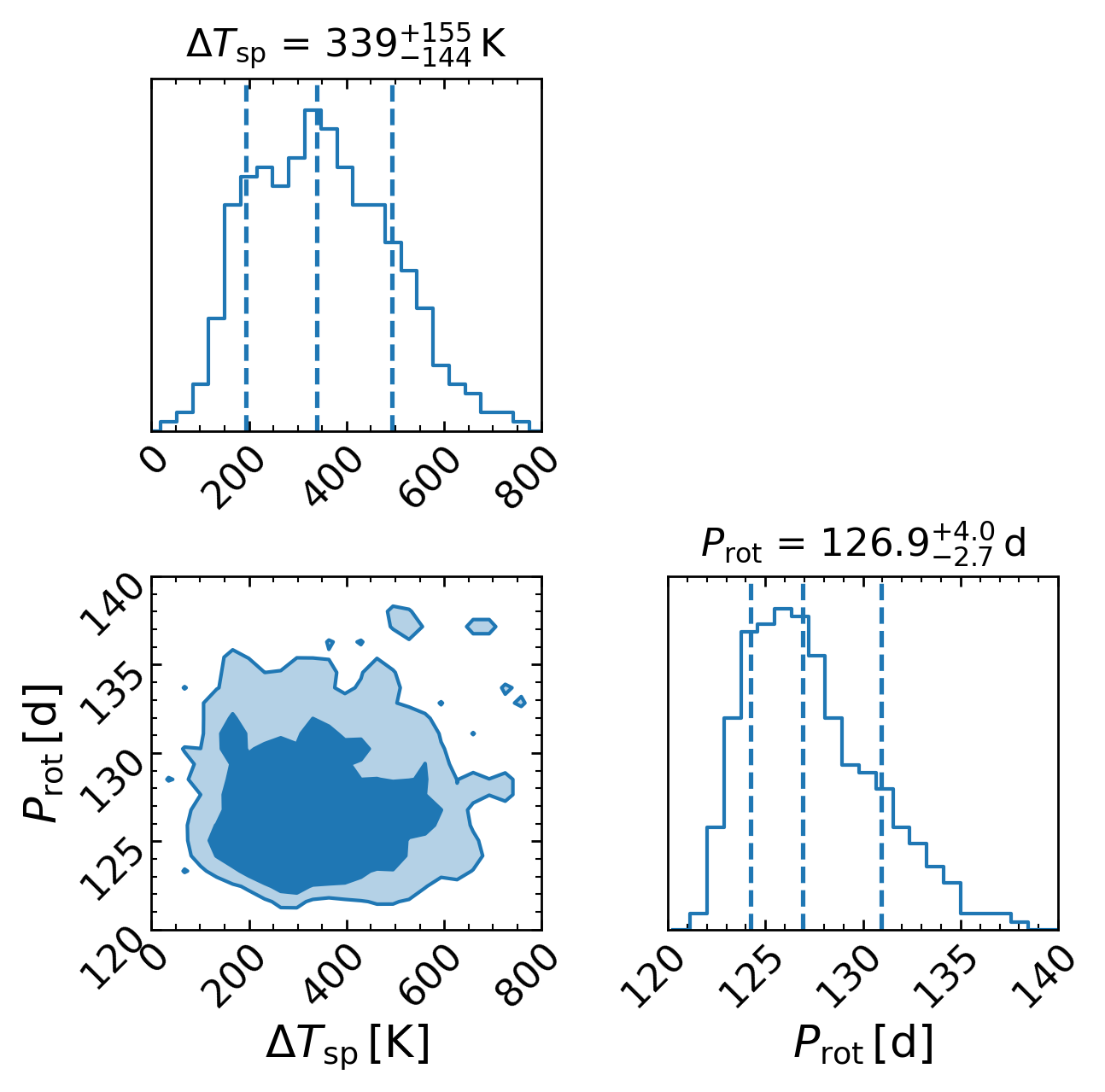}
          \label{fig:inv-3}
        \end{subfigure}%
      }%
    
      \caption{Spot coverage diagnostics at $t_{145}$ (left) and $t_{191}$ (middle) from the inversion. The colour scale shows the mean spot-covering multiplicity (expected number of spots covering each surface element, overlaps add) across retained inversion solutions, after folding to the hemisphere opposite to the transit chord for visualisation. The transit chord band is overplotted in grey. The annotated $\mathit{ff}_\mathrm{sp}$ is the disc-integrated mean spot filling factor evaluated at that epoch. Right: posterior distributions of the global parameters inferred by the inversion.}%
      \label{fig:inversion_results}
    \end{figure*}

   Figure~\ref{fig:inversion_results} illustrates the mean spot-covering multiplicity for the host star, folded to the hemisphere opposite to the transit chord (overlaps add) to account for the degeneracy between hemispheres in systems with a spin angle close to $90\,\mathrm{deg}$, and the posterior distribution of the fitted global parameters. Here, the multiplicity at a given surface element is defined as the average number of spots covering that element across the retained inversion solutions, with overlaps adding. It is therefore not a covering fraction and can exceed unity.
   The corresponding surface configurations at the JWST epochs and near activity maximum shown in the left and middle panels, respectively, indicate that the retained solutions preferentially place spot coverage at high latitudes. %
   The posterior distributions of $\DeltaTsp$ and $\Prot$ inferred by the inversion are consistent with the literature estimates from \citet{mallonn_paper} adopted in Table~\ref{tab:system_params} (right panel of Fig.~\ref{fig:inversion_results}).
   
\begin{figure*}[t]
\centering
\includegraphics[width=\linewidth]{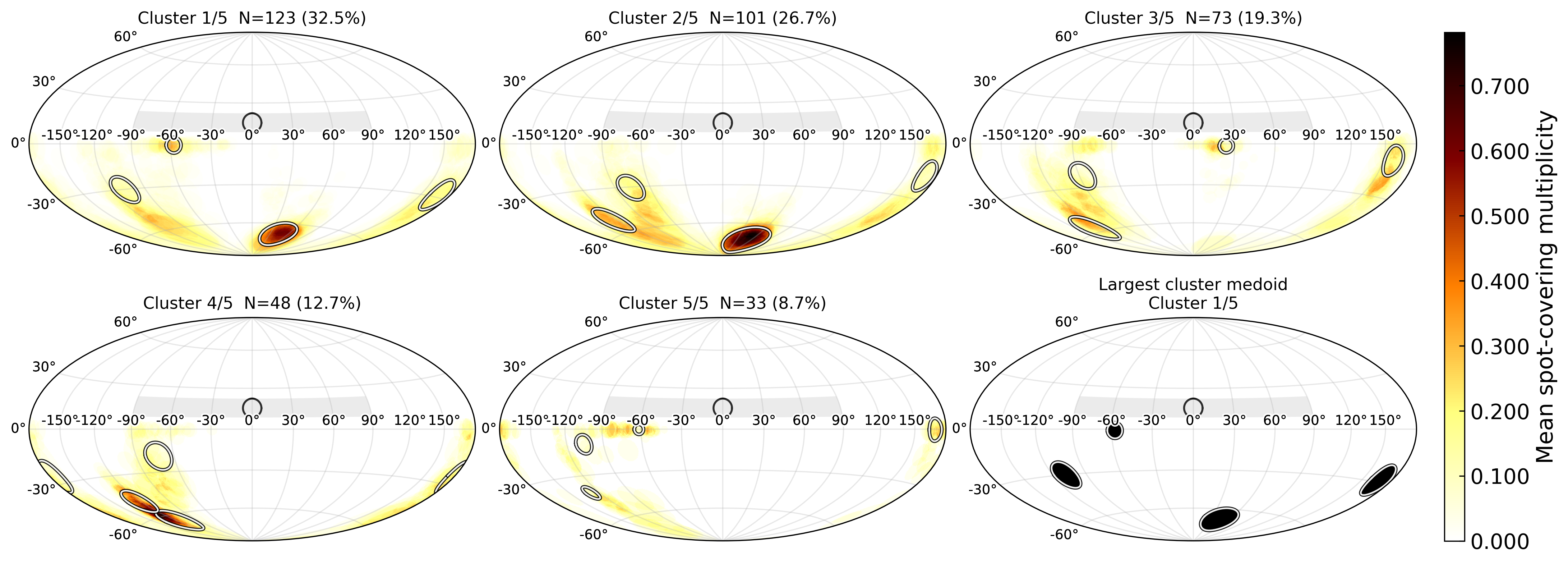}
\caption{Cluster-conditional spot-covering multiplicity maps for the retained inversion solutions at $t_{145}$, obtained from the $K$-medoids clustering analysis. The colour at a given surface element indicates the mean spot-covering multiplicity, i.e. the
average number of spots covering that element across the solutions
in the corresponding cluster, after folding to the hemisphere opposite to the transit chord (overlaps add). White curves show the spot outlines of the corresponding cluster medoids, and the transit chord band is shown in grey. The bottom right panel shows the medoid of the largest cluster at $t_{145}$. This is a representative discrete spot configuration from the dominant surface family, included to contrast with the ensemble coverage maps shown elsewhere.}
\label{fig:inv_clustering_main}
\end{figure*}

   The disc-map panels shown in this figure should be interpreted as an ensemble diagnostic, not as uniquely retrieved surface images. Darker colours indicate larger mean spot-covering multiplicity, i.e. surface elements that are covered more often by spots across the retained inversion solutions at that epoch; they do not encode spot temperature, intensity, or contrast. High values in adjacent surface elements can arise from different spot placements across solutions and should not be interpreted as a single large spot. Because several distinct spot configurations reproduce the monitoring data comparably well, we further summarised the retained inversion ensemble with a $K$-medoids clustering analysis of the full spot configurations. This favours $K=5$ recurring surface families. Figure~\ref{fig:inv_clustering_main} shows the corresponding cluster-conditional spot-covering maps for the representative epoch $t_{145}$, illustrating that the inversion does not return a unique surface map, but rather a small number of recurrent high-latitude spot configurations. The clustering metric, $K$-selection diagnostic, and additional representative medoid plots are given in Appendix~\ref{app:inv_clustering}.
   
\begin{figure*}
\centering
\includegraphics[trim={0ex 0ex 0ex 0ex},clip,width=\linewidth]{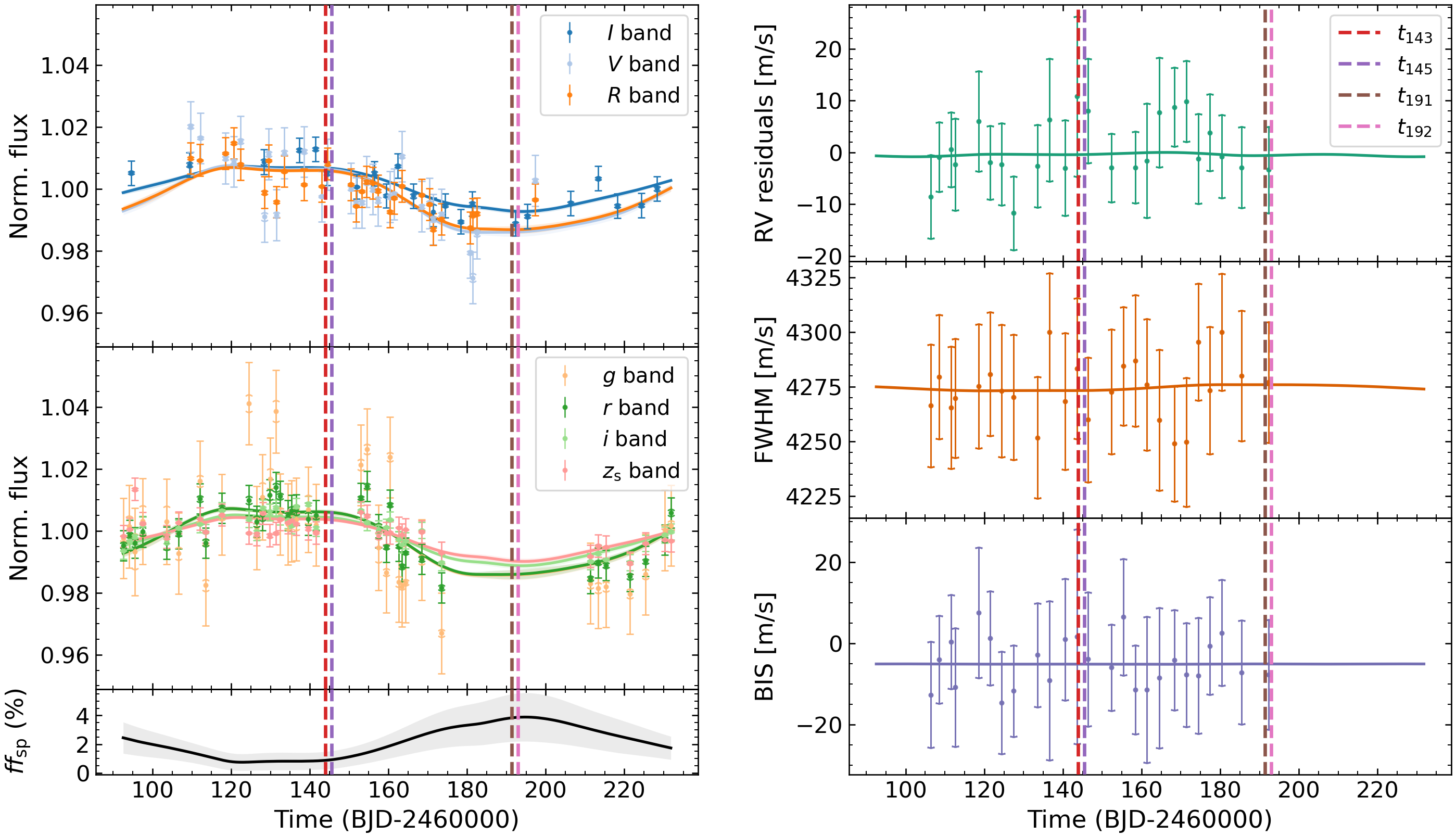}
\caption{Inversion model fits to the monitoring time series. Left: photometric light curves from TJO and OSN (top) and MuSCAT2 (middle), with the corresponding \texttt{StarSim} spot filling factor (bottom). Right: RV residuals and activity indices from CARMENES.}%
\label{fig:inversion_monitoring}
\end{figure*}

   The multi-band light curves exhibit a coherent low-amplitude rotational modulation that is reproduced by the inversion models across filters, as shown in the left panels of Fig.~\ref{fig:inversion_monitoring}, %
   indicating a stable large-scale asymmetry over the monitoring baseline. The lower left panel of Fig.~\ref{fig:inversion_monitoring} also shows the spot filling factor $\mathit{ff}_\mathrm{sp}$ predicted by the retained inversion solutions, defined as the ratio of the projected area of visible active regions to the projected stellar disc area. This is therefore a model-dependent geometric quantity, computed from the explicit surface maps at each epoch, and depends on the spot distribution, visibility, and projection effects. It should not be directly identified with filling factors inferred from other prescriptions, such as disc-integrated spectroscopic multi-component fits, which instead infer flux- or spectrum-weighted component fractions from the unresolved stellar spectrum and typically do not explicitly model the surface geometry or its associated centre-to-limb and chord-versus-disc effects. 
   
   A direct comparison can be made with \citet{mallonn_paper}, who also inferred projected spot filling factors for GJ\,1214 using \texttt{StarSim}. Their values of approximately $5$--$10\%$ during the more active seasons of their monitoring are somewhat larger than the ${\sim}0.5$--$4\%$ inferred here, although our value near activity maximum ($3.8\pm1.7\%$ at $t_{191}$) approaches the lower end of their range. Given the different observing epochs and the long-term evolution of GJ\,1214 reported by \citet{mallonn_paper}, our results are therefore qualitatively consistent with previous evidence for time-variable spot coverage at the percent level. 
   
   In contrast, the RV residuals, BIS and FWHM show comparatively weak variability and are also reproduced by the joint fit (Fig.~\ref{fig:inversion_monitoring}, right). Fitting the photometry and spectroscopic indicators jointly therefore disfavours solutions that match the photometric modulation by invoking large spot-induced Doppler signals, helping restrict the allowed spot latitudes, coverage and temperature contrast. %

   \subsection{Deep learning correction framework}\label{subsec:results_unspotter}

   We applied the \UNSPOTTER{} correction framework described in Section \ref{sec:deep_learning_correction_framework} to GJ\,1214, performing $N_\mathrm{sims}=100\,000$ simulations, and drawing stellar and planet parameters from the priors in Table~\ref{tab:NN_stellar_and_planet_params}. The noise augmentation (Section \ref{subsubsec:synthetic_dataset_generation}) was instantiated using the actual GJ\,1214 data for $N_\mathrm{noisy}=6$ realisations. We use timestamps in $\mathrm{BJD}_\mathrm{TDB}$ (days). For the temporal normalisation (Section~\ref{sec:preproc}), we set $t_{\rm ref}=\SI{2460000}{d}$ and $t_{\rm scale}=\SI{1}{d}$. %

    \begin{table}
        \caption{\label{tab:NN_hyperparameters}NN hyperparameters selected using \texttt{Optuna}.}
        \centering
        \renewcommand{\arraystretch}{1.15}
        \begin{tabular}{LCC}
        \hline\hline
        \text{Parameter} & \text{Search space} & \text{Adopted value}\\
        \hline
        d_\mathrm{temp}   & \{4,5,\dots,16\}                 & 14\\
        d_\mathrm{emb}    & \{2,3,\dots,8\}                  & 4\\
        d_\mathrm{model}  & \{16,32,64,96,128\}               & 32\\
        N_\mathrm{blocks} & \{1,2,3,4,5\}                  & 4\\
        N_\mathrm{heads}  & \{2,3,\dots,8\}                  & 4\\
        d_\mathrm{FFN}    & \{64,128,256\}                 & 128\\
        \mathrm{dropout}_\mathrm{tr} & [0,\,0.5]          & 8.47\times10^{-3}\\
        N_\mathrm{dense}  & \{1,2,3\}                  & 2\\
        d_\mathrm{dense}  & \{64,128,256,512\}                & 128\\
        \mathrm{dropout}_\mathrm{dense} & [0,\,0.5]        & 7.51\times10^{-2}\\
        \eta_0 & \mathrm{LogUniform}(10^{-4},\,10^{-3}) & 4.61\times10^{-4}\\
        \text{Batch size} & \{32,64,128,256\}                 & 64\\
        \hline
        \end{tabular}
        \tablefoot{Trials with $d_\mathrm{model}\bmod N_\mathrm{heads}\neq 0$ were pruned to enforce an integer attention head dimension. $\mathrm{LogUniform}(a,b)$ denotes a distribution for which $\log x$ is sampled uniformly between $\log a$ and $\log b$.}
    \end{table}

   We optimised the NN hyperparameters for $t_{145}$ using \texttt{Optuna} and adopted the best-performing configuration, listed in Table~\ref{tab:NN_hyperparameters}, for all four epoch-specific ensembles of $M=20$ independently initialised networks for each transit, with $D=50$ Monte Carlo dropout passes for each of $R=1000$ realisations of the input sequence. The models share the same input representation and output wavelength grid, motivating a common architecture and training configuration to keep the comparison between epochs controlled, while each ensemble is trained independently. %

   Generating the synthetic dataset on a machine equipped with two Intel Xeon E5-2690 v4 CPUs (28 physical cores in total) required approximately $348\,\mathrm{h}$ (14.5 days) for the four epochs considered here. Training \UNSPOTTER{} on a single NVIDIA Tesla V100 GPU required approximately $127\,\mathrm{h}$ (5.3 days) for all four epochs.

\begin{figure}[t]
    \centering
    \includegraphics[width=\linewidth]{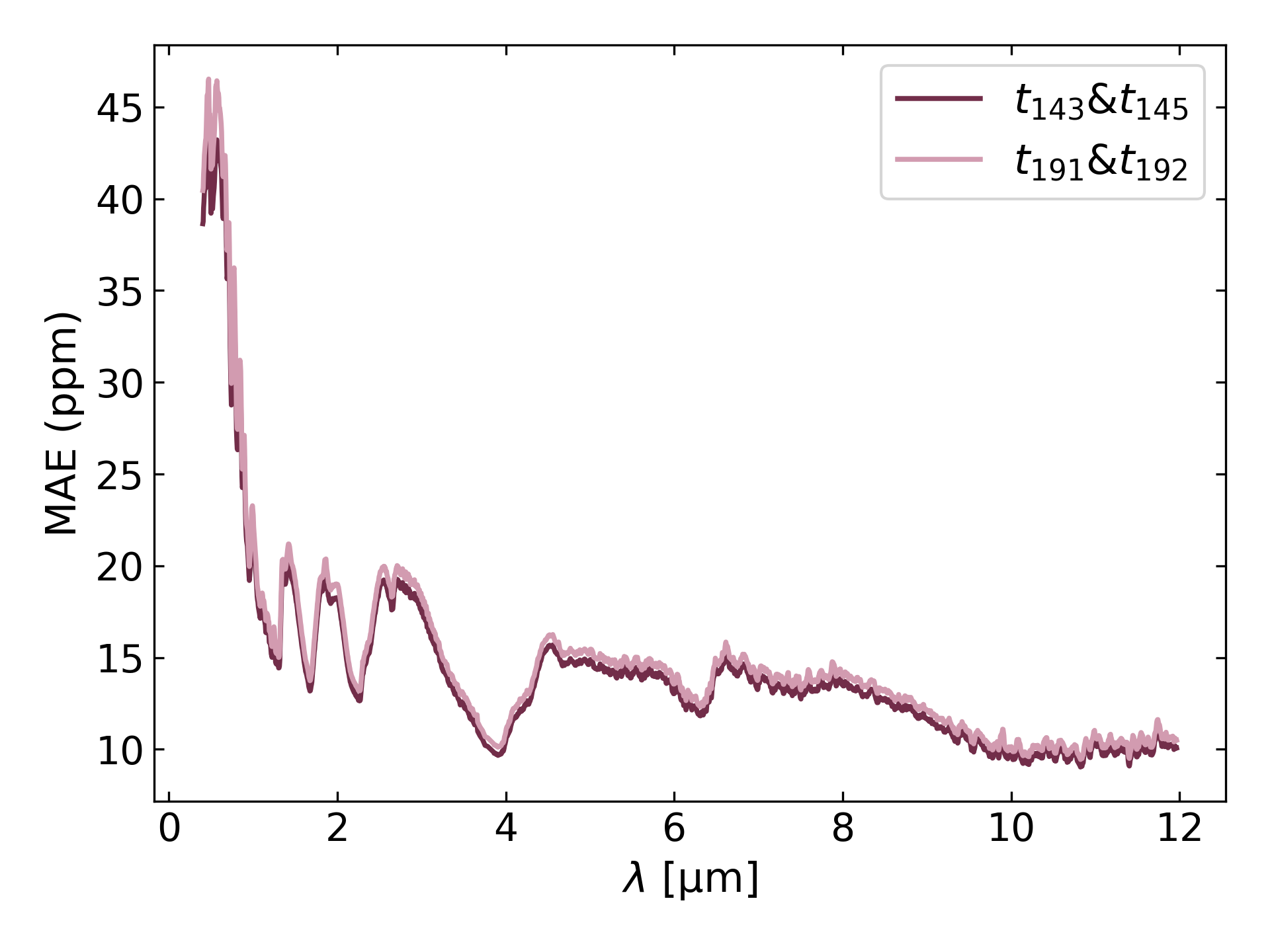}
    \caption{\UNSPOTTER{} validation on test set simulations. Mean absolute error of the predicted stellar-contamination spectrum as a function of wavelength.}
    \label{fig:unspotter_validation}
    \end{figure}

    We evaluate \UNSPOTTER{} on the independent synthetic test sets before applying it to the observations. Figure~\ref{fig:unspotter_validation} shows the wavelength-dependent mean absolute error (MAE) of the predicted stellar-contamination spectrum. Across the four epoch-specific NN ensembles, the test-set MAE over the \(0.4\)--\(\SI{12}{\micro\metre}\) range is \(14.29\)--\(14.93\,\mathrm{ppm}\), with corresponding root mean squared errors (RMSEs) of \(26.68\)--\(27.22\,\mathrm{ppm}\). The comparable test errors support the suitability of this configuration across the four epochs. The larger errors towards the optical reflect the stronger wavelength dependence of the stellar-contamination signal in this regime. %

    The uncertainty calibration described in Sect.~\ref{subsubsec:nn_posteriors} yields multiplicative scale factors $s$ of about \(1.9\) for the observed epochs (\(t_{143}\) and \(t_{145}\)) and \(2.4\) for the activity-maximum epochs (\(t_{191}\) and \(t_{192}\)). These factors adjust the absolute scale of the uncertainty intervals to achieve 68\% coverage while preserving their wavelength-dependent structure and case-to-case variation. The calibrated intervals remain close to the nominal 68\% coverage over most of the wavelength range, becoming mildly conservative towards optical wavelengths at the activity-maximum epochs, by about $4$ percentage points.

    We also tested the dependence of the predictive accuracy on the number of independent training simulations and found that the adopted training-set size lies in a regime of diminishing returns: increasing the training sample reduces the mean test MAE by an amount comparable to the run-to-run dispersion between independently trained networks (Appendix~\ref{app:unspotter_learning_curve}).

   \begin{table*}%
        \centering
        \caption{Simulation intervals for the stellar (top) and planet (bottom) parameters for the deep learning correction framework.}\label{tab:NN_stellar_and_planet_params}
        \renewcommand{\arraystretch}{1.15}  %
        \setlength{\dashlinedash}{1pt} %
        \setlength{\dashlinegap}{1pt}  %
        \begin{tabular}{LCCC}
            \hline \hline
            \text{Stellar parameter} & \text{Symbol} & \text{Unit} & \texttt{StarSim}\text{ simulation interval}\\\hline\rule{0pt}{10pt}
            \text{Effective temperature}\tablefootmark{a} & T_\star & \mathrm{K} & \mathcal{N}_\mathrm{Tr} (3101,\,43,\,2300,\,3500) \\
            \text{Spot temperature contrast}\tablefootmark{a} & \DeltaTsp & \mathrm{K} & \mathcal{N}_\mathrm{Tr} (372,\,182,\,0,\,T_\star-2300) \\ %
            \text{Mass} & M_\star & \mathrm{M}_\odot & \mathcal{N}_\mathrm{Tr} (0.1820,\,0.0042,\,0,\,1) \\
            \text{Radius} & R_\star & \mathrm{R}_\odot & \mathcal{N}_\mathrm{Tr} (0.2162,\,0.0025,\,0,\,1) \\
            \text{Surface gravity}\tablefootmark{a} & \log g & \mathrm{dex} & \mathcal{N}_\mathrm{Tr} (5.0286,\,0.0088,\,4.5,\,5.5) \\
            \text{Convective shift}\tablefootmark{b} & CS & \mathrm{CS}_\odot & \mathcal{U}(0,\,0.5) \\ %
            \text{Differential rotation}\tablefootmark{c} & d\Omega & \mathrm{d\Omega}_\odot & 0 \\ 
            \text{Inclination}\tablefootmark{d} & i_\star & \mathrm{deg} & 90 \\ 
            \text{Rotation period} & \Prot & \mathrm{days} & \mathcal{N}_\mathrm{Tr}(125,\,5,\,0,\,300) \\
            \text{Number of spots}\tablefootmark{e} & N_\mathrm{sp} & - & \mathcal{U}\{2,\,10\} \\
            \text{Spot radius\,}\tablefootmark{f} & r_\mathrm{sp} & \mathrm{deg} & \mathrm{Exp}_\mathrm{Tr}(N_\mathrm{sp}^2/200,\,1,\,30) \\
            \text{Spot longitude} & \phi_\mathrm{sp} & \mathrm{deg} & \mathcal{U}(0,\,360) \\
            \text{Spot colatitude}\tablefootmark{g} & \theta_\mathrm{sp} & \mathrm{deg} & \Big(90^\circ+\tfrac{180^\circ}{\pi}[\arccos U_1-\arccos U_2]\Big)\big|_{[0,\theta_\mathrm{-}]\cup[\theta_\mathrm{+},180^\circ]} \\ %
            \noalign{\vskip 2pt} %
            \hdashline
            \noalign{\vskip 2pt} %
            \text{Orbital period} & P_\mathrm{p} & \mathrm{days} &
            \mathcal{N}_\mathrm{Tr}\left(1.580404531,\,0.000000018,\,1.5804,\,1.5805\right) \\
            \text{Planet-to-star radius ratio} & p & - &
            \mathcal{N}_\mathrm{Tr}\left(0.11589,\,0.00016,\,0.1,\,0.2\right) \\
            \text{Impact parameter} & b & - &
            \mathcal{N}_\mathrm{Tr}\left(0.264,\,0.023,\,0,\,1\right) \\
            \text{RV semi-amplitude} & K & \mathrm{m\,s}^{-1} &
            \mathcal{N}_\mathrm{Tr}\left(14.38,\,0.57,\,0,\,100\right) \\
            \text{Transit midpoint}\tablefootmark{h} & T_\mathrm{c} & \mathrm{days}\; 
            
            (\mathrm{BJD}-2460000) & \mathcal{N}_\mathrm{Tr}\left(-4298.586672,\,0.000066,\,-4298.7,\,-4298.5\right) \\
            \hline
        \end{tabular}
        \tablefoot{
            \tablefoottext{a}{The lower/upper limits are chosen since the stellar atmosphere models have $T_\star$ ranging from $2300\,\mathrm{K}$ to $3500\,\mathrm{K}$ and $\log g$ from $4.5\,\mathrm{dex}$ to $5.5\,\mathrm{dex}$.}
            \tablefoottext{b}{In units of the solar convective shift $\mathrm{CS}_\odot$.}%
            \tablefoottext{c}{In units of the solar differential rotation $\mathrm{d\Omega}_\odot$.}%
            \tablefoottext{d}{Defined from $0^\circ$ (pole-on) to $180^\circ$ (opposite pole-on), with $i_\star=90^\circ$ corresponding to equator-on.}%
            \tablefoottext{e}{$\mathcal{U}\{a,b\}$ denotes a discrete uniform distribution over the integers $a, \dots, b$, $b\geq a$. This is distinct from $\mathcal{U}(a,b)$, a continuous uniform distribution over the real interval $(a,b)$.}
            \tablefoottext{f}{$\mathrm{Exp}_\mathrm{Tr}(\lambda,a,b)$ denotes the truncated exponential distribution with rate parameter $\lambda$, lower bound $a$, and upper bound $b$, i.e. with PDF $f(x)=\frac{\lambda e^{-\lambda(x-a)}}{1-e^{-\lambda(b-a)}}\text{, } a\leq x\leq b$. }
            \tablefoottext{g}{$U_{1,2}\sim\mathcal{U}(0,1)$, and the subscript denotes truncation to avoid spot crossings, with  $\theta_{\pm}=\frac{180^\circ}{\pi}\arccos(b\mp p)\pm(r_{\mathrm{sp}}+\delta)$, where $\delta$ is a small angular margin (accounting for the \texttt{StarSim} grid resolution) to ensure the spot and planet are not assigned to the same surface cell.}%
            \tablefoottext{h}{Epoch based on \citet{Cloutier_2021}.}
            }
    \end{table*}

\begin{figure}[t]
\centering
\includegraphics[trim={3.5ex 2.0ex 1.5ex 1.5ex},clip,width=\linewidth]{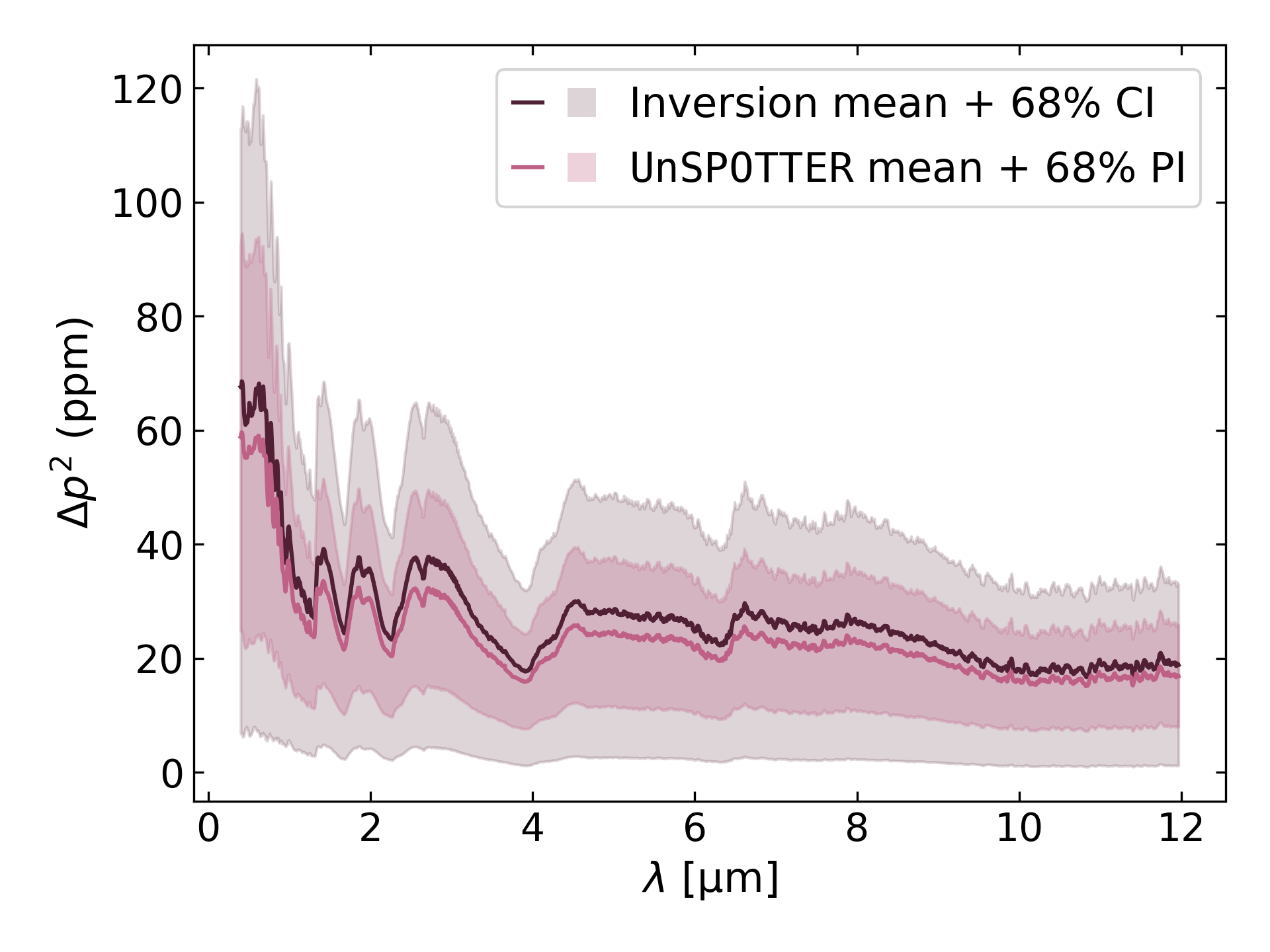}
\caption{Activity correction $\Delta p^2(\lambda)$ over $0.4$--$\SI{12}{\micro\metre}$ for the observed JWST epochs ($t_{143}$, $t_{145}$).}
\label{fig:corrections_full_jwst}
\end{figure}

\begin{figure}[t]
\centering
\includegraphics[trim={3.5ex 2.0ex 1.5ex 1.5ex},clip,width=\linewidth]{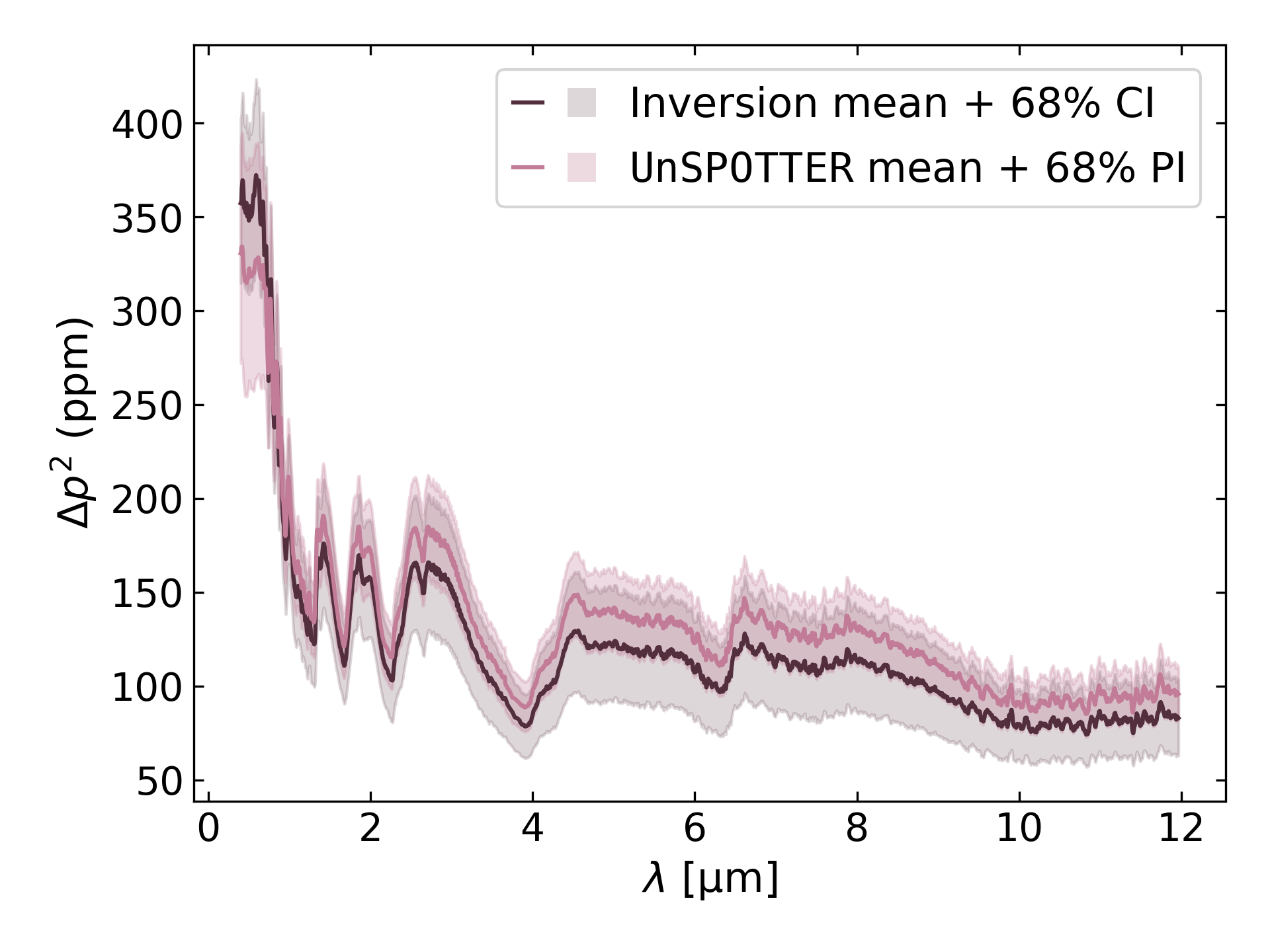}
\caption{Same as Fig.~\ref{fig:corrections_full_jwst}, but for epochs near activity maximum ($t_{191}$, $t_{192}$).}
\label{fig:corrections_full_max}
\end{figure}

\begin{figure*}[t]
\centering
\includegraphics[trim={2.75ex 0.25ex 1.5ex 1.0ex},clip,width=\linewidth]{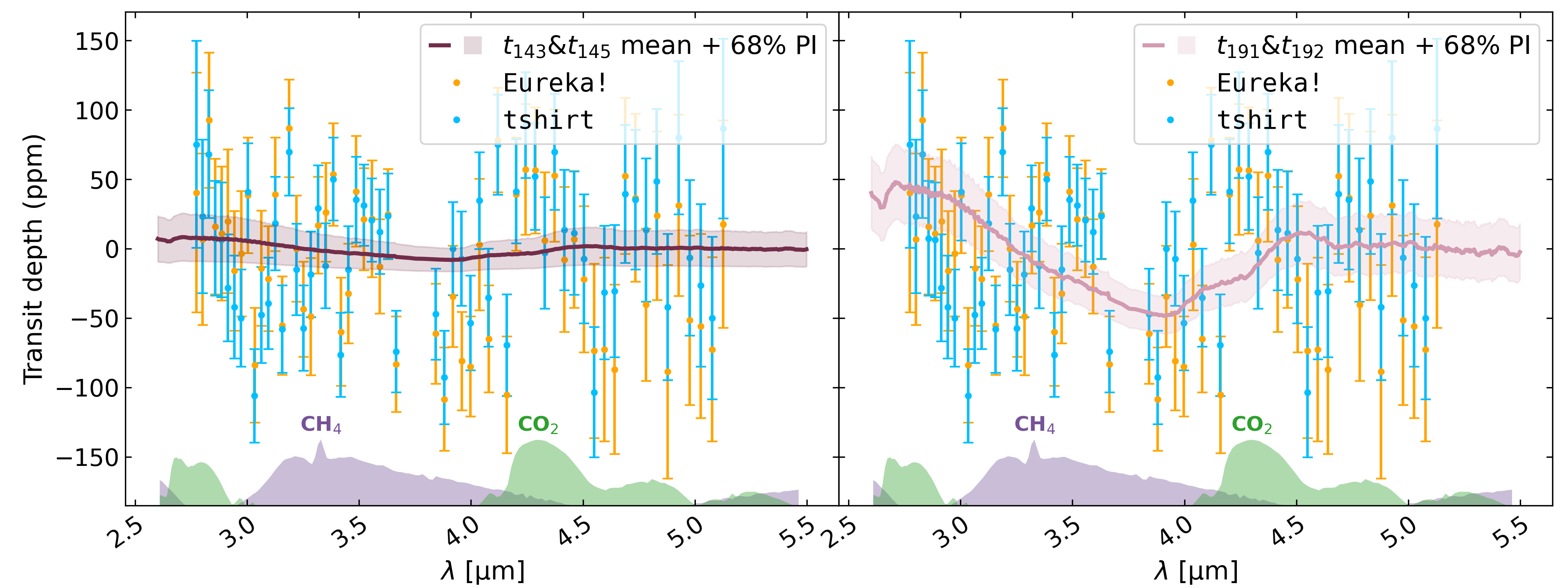}
\caption{Comparison of predicted activity corrections over the JWST/NIRSpec G395H range with the transmission spectrum from \citet{Schlawin_2024} (both \texttt{tshirt} and \texttt{Eureka!} reductions), obtained using \UNSPOTTER{}. Left: observed epochs ($t_{143}$, $t_{145}$). Right: epochs near activity maximum ($t_{191}$, $t_{192}$). The same published JWST/NIRSpec transmission spectrum is shown in both panels. Curves and data are centred around zero for visual clarity. The purple CH$_4$ and green CO$_2$ fills are illustrative molecular profiles based on \texttt{petitRADTRANS} calculations \citep{Molliere2019,Blain2024}, shown at arbitrary vertical scale.}
\label{fig:corrections_comp_jwst_unspotter}
\end{figure*}

    \subsection{Predicted stellar contamination}

    Figure~\ref{fig:corrections_full_jwst}
    shows the posterior summaries of $\Delta p^2(\lambda)$ over $0.4$--$\SI{12}{\micro\metre}$ for the inversion and \UNSPOTTER{} frameworks, 
    for the transits corresponding to the JWST observations ($t_{143}$, $t_{145}$). 
    The inferred corrections are epoch dependent. At the observed JWST epochs ($t_{143}$, $t_{145}$), both frameworks predict corrections
    with maximum--minimum differences below $\sim$20\,ppm
    and weakly wavelength-dependent across the NIRSpec/G395H range.
    For comparison, Fig.~\ref{fig:corrections_full_max} shows the same as Fig.~\ref{fig:corrections_full_jwst} but for transits at the epochs near the photometric minimum or local activity maximum ($t_{191}$, $t_{192}$). Both frameworks predict larger, structured corrections, with a maximum--minimum difference of up to ${\sim}100\,\mathrm{ppm}$ over $2.6$--$\SI{5.5}{\micro\metre}$.
    We report pairwise summaries for ($t_{143}$, $t_{145}$) because the published JWST/NIRSpec G395H transmission spectrum combines the two observed transits. We use the same paired presentation for ($t_{191}$, $t_{192}$) for consistency.
    The two transits within each epoch pair yield mutually consistent correction posteriors: for a given framework, the posterior means of the two transits differ by $\lesssim 10\,\mathrm{ppm}$, well within the corresponding uncertainty intervals (see Appendix~\ref{appendix:all_times_appendix} for the individual-transit corrections).
    Figure~\ref{fig:corrections_comp_jwst_unspotter} compares the corresponding predictions over $2.6$--$\SI{5.5}{\micro\metre}$ to the \citet{Schlawin_2024} spectrum (centred at zero for visual clarity). The analogous inversion comparison is shown in Appendix~\ref{app:inv_comp_with_real} (Fig.~\ref{fig:corrections_comp_jwst_inversion}). 

    The principal structures discussed by \citet{Schlawin_2024} are the increased transit depths near $\SI{4.3}{\micro\metre}$ and $\SI{2.8}{\micro\metre}$, attributed to possible CO$_2$ absorption, together with CH$_4$ around $\SI{3.3}{\micro\metre}$. At the observed epochs, our predicted stellar-activity corrections are considerably smaller and more weakly chromatic than these features and do not reproduce their spectral structure. The predicted correction therefore leaves the candidate molecular features essentially unchanged. %

    We also examine whether the predicted corrections are compatible with the earlier HST/WFC3 transmission spectrum of \citet{Kreidberg_2014} in Appendix~\ref{app:hst_comparison} (Fig.~\ref{fig:corrections_comp_hst_unspotter}).

\begin{figure}[t]
\centering
\includegraphics[trim={3.5ex 3.7ex 3.0ex 3.5ex},clip,width=1.0\linewidth]{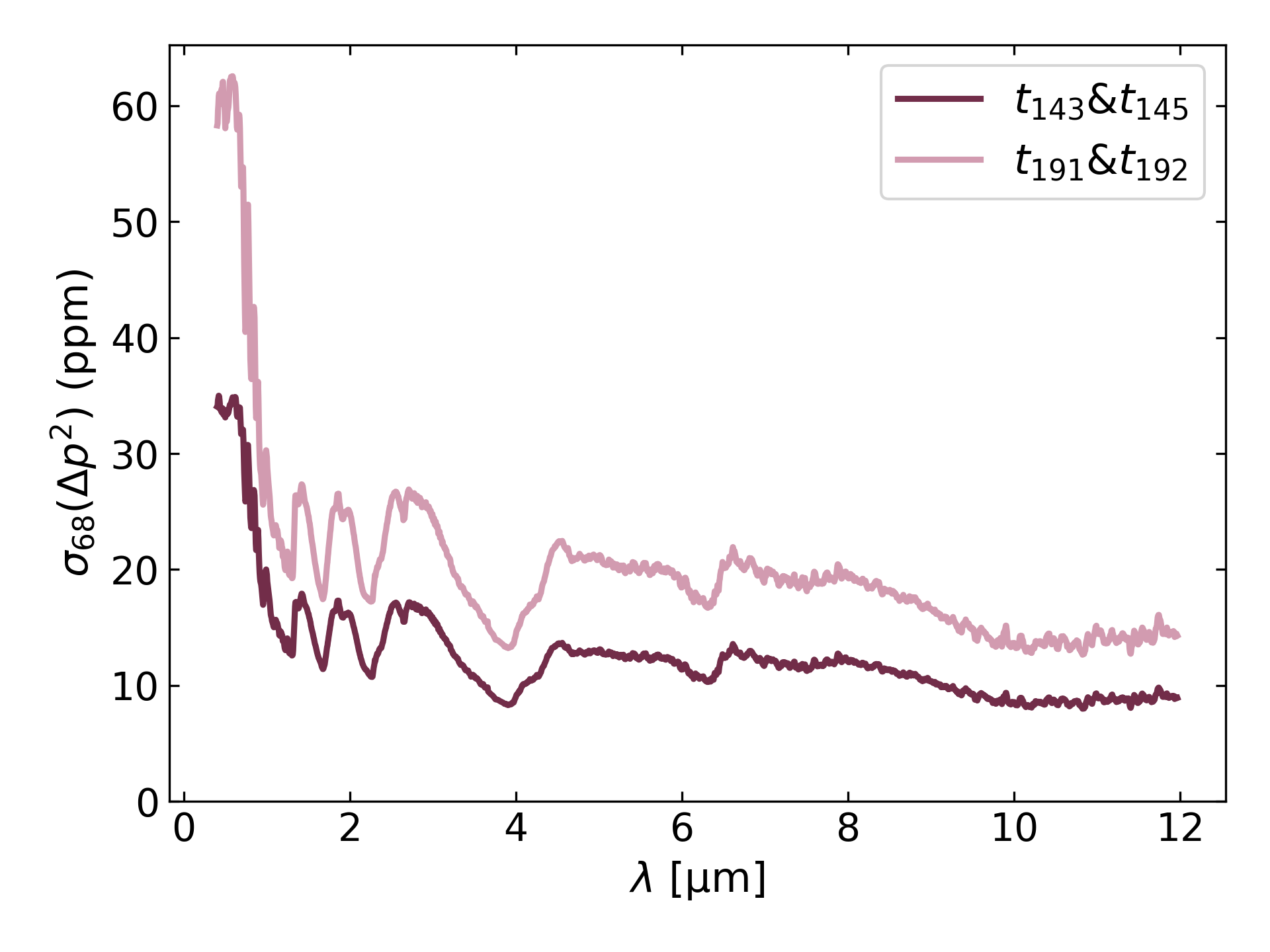}
\caption{Half-width of the calibrated 68\% predictive interval for the \UNSPOTTER{} stellar-contamination corrections.} %
\label{fig:nn_sigma68}
\end{figure}

To quantify how precisely the stellar-contamination correction can be inferred from the monitoring data, Fig.~\ref{fig:nn_sigma68} shows $\sigma_{68}$,
the half-width of the calibrated 68\% PI, for the \UNSPOTTER{} predictions at the two epoch pairs. Smaller values correspond to more tightly constrained corrections. Across the JWST/NIRSpec G395H range, %
the uncertainty of the correction
is typically of order $10$--$20$ ppm or lower for the observed epochs and remains at the $20$--$30$ ppm level or lower near activity maximum. The mean half-width across this range is $12.4\,\mathrm{ppm}$ for the observed epoch pair and $19.7\,\mathrm{ppm}$ for the activity-maximum pair. Over the broader $0.4$--$\SI{12}{\micro\metre}$ range, the uncertainty increases towards the blue edge, reaching ${\sim}35$ ppm for the observed epochs and ${\sim}60$ ppm near activity maximum close to $\lambda\sim\SI{0.4}{\micro\metre}$. %
From roughly $\lambda\gtrsim\SI{1}{\micro\metre}$ onwards, the correction precision is below $30$ ppm.

    \section{Discussion}\label{sec:discussion}

    The inversion and \UNSPOTTER{} yield consistent corrections for GJ\,1214. This agreement provides a robustness check: the two approaches are sensitive to different modelling risks (discrete surface-family assumptions for the inversion; training domain coverage/generalisation for the NNs), so their agreement reduces the likelihood that the inferred chromatic structure is method-specific. %

    The two methods should therefore be interpreted as complementary tools with different roles, rather than as interchangeable estimators. The inversion provides interpretability (maps and physical parameters) and diagnostics when fits or assumptions fail. \UNSPOTTER{} naturally marginalises over a broader activity parameter set defined by the simulation priors, and is designed to provide the sharper predictive correction once the target is well represented by the training domain. Using both therefore separates \emph{explanation} (what surface maps best explain the data, and what $\Delta p^2$ follows) from \emph{prediction} (averaging over many surface maps consistent with stars and monitorings like these, what $\Delta p^2$ should we expect), and flags regimes where either the monitoring is insufficient or the NNs are out-of-distribution.

    The uncertainty bands of the two frameworks also differ in a meaningful way. In the inversion, the quoted interval is obtained from the spread of $\Delta p^2(\lambda)$ across all retained surface-map solutions that reproduce the monitoring comparably well. Because the inversion is formulated at the level of explicit spot configurations, this spread reflects not only the uncertainty in the contamination spectrum itself, but also residual degeneracies among the reconstructed surface maps within the adopted four-spot model, as well as possible mismatch between that simplified model and the true stellar photosphere. Different four-spot configurations can fit the monitoring similarly well while still producing somewhat different corrections, and this effect can be amplified if the true surface is more complex than the adopted model, so that several approximate four-spot reconstructions reproduce the monitoring comparably well. We therefore interpret the inversion interval as reflecting both map-level degeneracy and model uncertainty within the adopted inversion framework. These map-level degeneracies thus broaden the inversion interval. The choice of $N_\mathrm{sp}=4$ is deliberate: it provides a practical balance between fit quality and model complexity, while avoiding the stronger degeneracies and convergence difficulties that arise for a higher number of spots. \UNSPOTTER{}, by contrast, is trained to predict $\Delta p^2(\lambda)$ directly from the monitoring observables, and can therefore marginalise more efficiently over surface-configuration differences that are weakly constrained by the data but have limited relevance for the final correction. In our application this leads to narrower uncertainty intervals, suggesting that the NN provides a more precise predictive estimate of the stellar-contamination correction while remaining consistent with the inversion at the level of the posterior means.

    The results also show a clear epoch dependence: the corrections are smallest at the observed epochs ($t_{143}$, $t_{145}$), near a photometric maximum, and largest near the photometric minimum ($t_{191}$, $t_{192}$). This is expected as, under our adopted two-component model, a photometric maximum corresponds to a smaller contribution of the cooler component to the disc-integrated flux, reducing the mismatch between the transit chord and the disc average and hence yielding smaller $\Delta p^2$. %
    
    At the observed JWST epochs, the predicted corrections across the NIRSpec/G395H range remain well below the scale of the tentative spectral features reported by \citet{Schlawin_2024}, suggesting that stellar spots are unlikely to be the dominant origin of those features in the published transmission spectrum. In contrast, if the observations had been scheduled at activity maximum, the predicted peak-to-peak amplitudes are comparable to, and in places exceed, the reported spectral structure. Crucially, this highlights that visit timing relative to the activity cycle can determine whether stellar contamination is negligible or retrieval-limiting without an explicit correction.

    The largest differences among correction frameworks occur towards the blue edge of our plotted range ($\lambda \sim \SI{0.4}{\micro\metre}$), particularly near activity maximum. Although this regime lies closer in wavelength to the monitoring photometry than the mid-infrared does, the bluest monitoring bands are also the noisiest and we excluded the $B$-band data from the analysis. In addition, spot contrast becomes more pronounced towards shorter wavelengths, so a given uncertainty or bias in the parameters that control it produces a larger effect on the inferred corrections. This regime is therefore plausibly more affected by modelling systematics and simulation--reality mismatch. This is consistent with recent JWST observations of TOI-3884, where empirical M-dwarf spot contrasts are reproduced by 1D atmosphere models at $\lambda \gtrsim \SI{1}{\micro\metre}$ but are underestimated at shorter wavelengths \citep{2026ApJ..1003L..45M}. %

    The quality of the inferred stellar-contamination corrections can be improved mainly through better physical inputs and better monitoring. On the modelling side, the physical accuracy of the corrections will improve as more realistic specific-intensity libraries and active-region spectra become available, in particular improved centre-to-limb behaviour and magnetic-region contrasts from modern 3D MHD calculations. On the observational side, denser and more contemporaneous monitoring would improve both approaches. In particular, better rotational-phase coverage and denser sampling close to the transmission-spectroscopy epochs would constrain the activity state more tightly, while improved precision and wavelength coverage in the bluest photometric bands would help at short wavelengths, where the corrections are most sensitive to spot contrast. Combining multi-band photometry with high-quality spectroscopic activity indicators remains especially valuable for reducing the main activity degeneracies, such as between spot surface coverage and temperature contrast. Extensions that include occulted heterogeneities, when warranted by the data quality, would also broaden the applicability of the framework. In this sense, the inversion is useful for diagnosing which aspects of the activity model and monitoring are weakly constrained, while \UNSPOTTER{} is particularly well-suited to delivering precise corrections once trained within an appropriate simulation domain.

    Application of \UNSPOTTER{} to other active stars would require target-specific simulations using suitable stellar intensity spectra and stellar and activity priors, with synthetic simulations matched to the observed timestamps and uncertainties. The network would then be retrained, validated and calibrated for the new application. The required sampling and phase coverage depend on the complexity and temporal evolution of the activity pattern, as well as the precision and type of monitoring available. For sufficiently stable patterns, phase coverage can be accumulated across several rotations. 
    
    Contemporaneous observations are key to constraining the rotational phase and activity state at the transmission spectroscopy epoch, allowing the correction to incorporate information about the star during that particular visit rather than relying only on its long-term activity level.

    \section{Conclusions}\label{sec:conclusions}

      In this work,
      we present two frameworks, inversion and \UNSPOTTER{}, that use ground-based monitoring around transmission spectroscopy observations to produce epoch-specific chromatic corrections to stellar contamination, with uncertainties that can be propagated to atmospheric retrievals. %
    
      We applied these frameworks to the planetary system hosted by
      GJ\,1214.
      We recover
      consistent corrections at the JWST epochs using both methodologies, which
      clearly suggest that observations occurred at a favourable activity phase with low chromatic contamination.
    For \UNSPOTTER{}, the inferred stellar-contamination corrections can be constrained to a precision of $\sim$15 ppm across the JWST/NIRSpec G395H range for the observed epochs. The uncertainty increases towards the blue edge of the broader $0.4$--$\SI{12}{\micro\metre}$ range, reaching about 35 ppm for the observed epochs and about 60 ppm near activity maximum around $\lambda\sim\SI{0.4}{\micro\metre}$, while from roughly $\lambda\gtrsim\SI{1}{\micro\metre}$ onwards it remains below the 30-ppm level.

      A counterfactual evaluation suggests that the same system would imprint strongly chromatic contamination that could plausibly mimic atmospheric features near the local activity maximum (photometric minimum), highlighting the key value of monitoring-informed scheduling and activity correction to accurately interpret transmission spectra.

      The inversion yields a set of interpretable surface maps consistent with the monitoring data, rather than a single unique reconstruction. In particular, the clustering analysis shows that the retained inversion solutions organise into a small number of recurrent surface families, providing a useful diagnostic of the remaining map degeneracy and of which aspects of the inferred stellar-surface structure are robust. For GJ\,1214, these recurrent families are high-latitude, indicating that while the exact spot map remains non-unique, the broader stellar-surface structure is more robustly constrained. \UNSPOTTER{} marginalises over a broader ensemble of spot configurations learned from simulations, which makes the predicted correction less sensitive to nuisance surface-map degeneracies that are poorly constrained by the monitoring data but have limited impact on the inferred correction. The two methods are complementary: the inversion provides physical interpretability; the NNs provide sharper constraints on the stellar contamination corrections. The inversion can also inform the simulation priors and noise model used to train the NNs. In our target- and epoch-specific implementation, the NNs carry an upfront training cost, but then enable rapid inference. This makes \UNSPOTTER{} particularly well-suited to scalable uses (e.g. across many epochs/targets within a trained domain). %

      This methodology can potentially be extended to other active stars with rotational modulation. Such applications would require target-specific simulations, retraining, and validation tailored to the available monitoring data. With these conditions, given photometric and/or spectroscopic monitoring around transmission spectroscopy observations, the framework may deliver activity-level forecasts for scheduling and consistent corrections for atmospheric analyses, supporting high-precision exoplanet atmosphere characterisation for JWST-class transmission spectroscopy and future missions (e.g. \textit{Ariel}).

\FloatBarrier

\begin{acknowledgements}
    We thank the two anonymous referees and the editor for their constructive comments and suggestions, which helped improve the manuscript. We thank Mark R. Swain for a helpful discussion. This publication has been made possible by funding from the European Research Council (ERC) under the European Union’s Horizon Europe programme (ERC Advanced Grant SPOTLESS; no. 101140786). Views and opinions expressed are however those of the author(s) only and do not necessarily reflect those of the European Union or the European Research Council. Neither the European Union nor the granting authority can be held responsible for them. Financial support has also been provided by Spanish grants PID2020-120375GB-I00, PID2021-125627OB-C31, PID2022-137241NB-C43, PID2023-150491NB-I00, and PID2024-158486OB-C31 funded by MCIU/AEI/10.13039/501100011033 and by “ERDF A way of making Europe”, by MCIU/AEI grant CNS2022-136050, and by the Generalitat de Catalunya via SGR 01526/2021 and the CERCA programme. This work was also partly supported by the Spanish program Unidad de Excelencia María de Maeztu CEX2020-001058-M awarded to ICE and Severo Ochoa CEX2021-001131-S awarded to IAA, financed by MCIN/AEI/10.13039/501100011033, and by the MaX-CSIC Excellence Award MaX4-SOMMA-ICE, and by the Marie Sk\l{}odowska-Curie Actions grant agreement No 101149286 (INCITE). %
      We thank Calar Alto Observatory for allocation of director's guaranteed time to this programme (DGT.23A.313). %
      Data were partly collected with the 90-cm telescope at Sierra Nevada Observatory (OSN), operated by the IAA-CSIC. %
      The Joan Oró Telescope (TJO) at the Montsec Observatory (OdM) is owned by the Catalan Government and operated by the Institute of Space Studies of Catalonia (IEEC).
      This article is based on observations made with the MuSCAT2 instrument, developed by ABC, at Telescopio Carlos Sánchez operated on the island of Tenerife by the IAC in the Spanish Observatorio del Teide. This work is partly supported by JSPS KAKENHI Grant Numbers JP24H00017, JP24K00689, JP24K17083, JP25K24620, JP26H01402, JP26K00755, and JSPS Grant-in-Aid for JSPS Fellows Grant Number JP24KJ0241. We acknowledge financial support from the Agencia Estatal de Investigaci\'on of the Ministerio de Ciencia e Innovaci\'on MCIN/AEI/10.13039/501100011033 and the ERDF “A way of making Europe” through projects PID2021-125627OB-C32 and PID2024-158486OB-C32. This work is supported by the European Union (ERC AdvG SPEAR, GA 101200674). Views and opinions expressed are however those of the authors only and do not necessarily reflect those of the European Union or the European Research Council. Neither the European Union nor the granting authority can be held responsible for them. %
      This work is part of the first author's doctoral thesis, within the framework of the Doctoral Program in Physics at the Universitat Autònoma de Barcelona.
      The data production, processing and analysis tools for this paper have been developed, implemented and operated in collaboration with the Port d’Informació Científica (PIC) data center. PIC is maintained through a collaboration agreement between the Institut de Física d’Altes Energies (IFAE) and the Centro de Investigaciones Energéticas, Medioambientales y Tecnológicas (CIEMAT).
\end{acknowledgements}

\bibliographystyle{aa_url} %
\bibliography{bibliography} %

\appendix

\section{Preparation of synthetic spectra for \texttt{StarSim}}\label{app:preparation_synthetic_spectra}

The main steps of the procedure that we implemented to adapt the synthetic spectra to the instrument’s characteristics and the desired spectral channels are:

\subsection{Spectral degradation to instrumental resolution}

We degrade each specific intensity spectrum $I_\lambda(\lambda,\mu)$ to the instrument’s wavelength-dependent resolving power $R(\lambda)$ using a Gaussian line-spread function (LSF). With
\begin{equation}
R(\lambda)=\frac{\lambda}{\mathrm{FWHM}_{\rm LSF}(\lambda)}, 
\qquad
\sigma_\lambda(\lambda)=\frac{\mathrm{FWHM}_{\rm LSF}(\lambda)}{2\sqrt{2\ln 2}},%
\end{equation}
the \emph{continuous} convolution targeted at grid point $\lambda_j$ is
\begin{equation}
\widetilde{I}_\lambda(\lambda_j,\mu)=
\frac{\displaystyle \int I_\lambda(\lambda',\mu)\,
\exp\!\left[-\frac{(\lambda'-\lambda_j)^2}{2\sigma_j^2}\right]\;d\lambda'}
{\displaystyle \int \exp\!\left[-\frac{(\lambda'-\lambda_j)^2}{2\sigma_j^2}\right]\;d\lambda'},
\end{equation}
where $\sigma_j\coloneqq\sigma_\lambda(\lambda_j)$.

On the \emph{discrete, non-uniform} grid $\{\lambda_k\}$, we approximate the integrals by Riemann sums over a finite window $\mathcal{W}_j=\{k:\;|\lambda_k-\lambda_j|\le 3\sigma_j\}$:
\begin{equation}
\widetilde{I}_\lambda(\lambda_j,\mu)=
\frac{\displaystyle \sum_{k\in\mathcal{W}_j}
I_\lambda(\lambda_k,\mu)\,
\exp\!\left[-\frac{(\lambda_k-\lambda_j)^2}{2\sigma_j^2}\right]\,
\Delta\lambda_k}
{\displaystyle \sum_{k\in\mathcal{W}_j}
\exp\!\left[-\frac{(\lambda_k-\lambda_j)^2}{2\sigma_j^2}\right]\,
\Delta\lambda_k},
\end{equation}
with $\Delta\lambda_k$ the local pixel width (e.g. $\Delta\lambda_k\approx\tfrac{1}{2}(\lambda_{k+1}-\lambda_{k-1})$ or, if available, the tabulated bin width $w_k=\lambda_k^+-\lambda_k^-$).
To avoid applying an undersampled kernel, if the local width in pixels %
\begin{equation}
\sigma_{\rm pix,j}\;\coloneqq\;\frac{\sigma_j}{\Delta\lambda_j},
\qquad
\Delta\lambda_j\approx\tfrac{1}{2}(\lambda_{j+1}-\lambda_{j-1}),
\end{equation}
falls below $0.5$, we leave the point unchanged. Outside the tabulated $R(\lambda)$ range we fix $\mathrm{FWHM}_{\rm LSF}$ to the edge value (constant beyond the blue/red edges).

This procedure is applied independently for each $\mu$ value. %

\subsection{Flux-conserving resampling to instrumental sampling}

After degradation, we resample $\widetilde{I}_\lambda(\lambda,\mu)$ from the model grid to the instrumental sampling using the \texttt{FluxConservingResampler} class of \texttt{specutils} \citep{specutils_citation,2022ApJ...935..167A}, which implements the SpectRes flux-conserving algorithm \citep{carnall2017spectresfastspectralresampling}. In the sense of their Eq.~(1), each input value $\widetilde{I}_\lambda(\lambda_j,\mu)$ is the bin-averaged specific intensity across the input pixel with edges
\begin{equation}
[\lambda^-_j,\lambda^+_j],\qquad w_j\coloneqq \lambda^+_j-\lambda^-_j.
\end{equation}
For an output pixel $i$ with edges $[\lambda^-_i,\lambda^+_i]$ define the geometric overlap
\begin{equation}
\Delta\lambda_{ij}=\max\bigl[0,\;\min(\lambda^+_j,\lambda^+_i)-\max(\lambda^-_j,\lambda^-_i)\bigr],
\end{equation}
and the covered width $W_i \coloneqq \sum_j \Delta\lambda_{ij}$. The resampled intensity is the overlap-weighted average of the input bin-averaged intensities,
\begin{equation}
I^{\rm samp}_\lambda(\lambda_i,\mu)=
\frac{\sum_{j}\widetilde{I}_\lambda(\lambda_j,\mu)\,\Delta\lambda_{ij}}
{\sum_{j}\Delta\lambda_{ij}},
\end{equation}
which corresponds to Eq.~(3) of \citet{carnall2017spectresfastspectralresampling} upon identifying $\Delta\lambda_{ij}=P_{ij}\,w_i$ and swapping $i\leftrightarrow j$.

\subsection{Binning to published transmission-spectrum channels}

Given channel centres $\lambda^{\rm mid}_m$ and widths $\Delta\lambda^{\rm bin}_m$, we form edges %
\begin{equation}
\lambda^{\rm min}_m = \lambda^{\rm mid}_m - \tfrac{1}{2}\Delta\lambda^{\rm bin}_m,
\qquad
\lambda^{\rm max}_m = \lambda^{\rm mid}_m + \tfrac{1}{2}\Delta\lambda^{\rm bin}_m,
\end{equation}
and compute the channel-averaged specific intensity from the instrument-sampled spectrum:
\begin{equation}
\langle I_\lambda\rangle_m(\mu)=
\frac{1}{\lambda^{\rm max}_m-\lambda^{\rm min}_m}
\int_{\lambda^{\rm min}_m}^{\lambda^{\rm max}_m}
I^{\rm samp}_\lambda(\lambda,\mu)\,d\lambda,
\end{equation}
evaluated numerically by trapezoidal integration over the instrument grid. Bins whose edges are not fully contained in the model coverage are discarded.

\subsection{Densified theoretical sampling}

To plot smooth theoretical curves while remaining consistent with the observed channels, we insert $N$ uniformly spaced intermediate bins between consecutive published bins $m$ and $m{+}1$ by linearly interpolating both centres and widths:
\begin{align}
\lambda^{\rm mid}_{m,j} &= (1-\alpha_j)\,\lambda^{\rm mid}_m+\alpha_j\,\lambda^{\rm mid}_{m+1},\\
\Delta\lambda^{\rm bin}_{m,j} &= (1-\alpha_j)\,\Delta\lambda^{\rm bin}_m+\alpha_j\,\Delta\lambda^{\rm bin}_{m+1},\\
\alpha_j&=\frac{j}{N+1},\qquad j=1,\dots,N.
\end{align}

Outside the published range we extend with equispaced centres (edge spacing) and fixed widths (edge width). For each bin, we compute the channel-averaged intensity via the same procedure defined above. %

\section{On the choice of limb darkening laws}\label{appendix:limb_darkening_laws}

\begin{figure*}
    \centering

    \begin{subfigure}{0.49\textwidth}
        \centering
        \includegraphics[trim={3.5ex 3.7ex 3.0ex 3.5ex},clip,width=\linewidth]{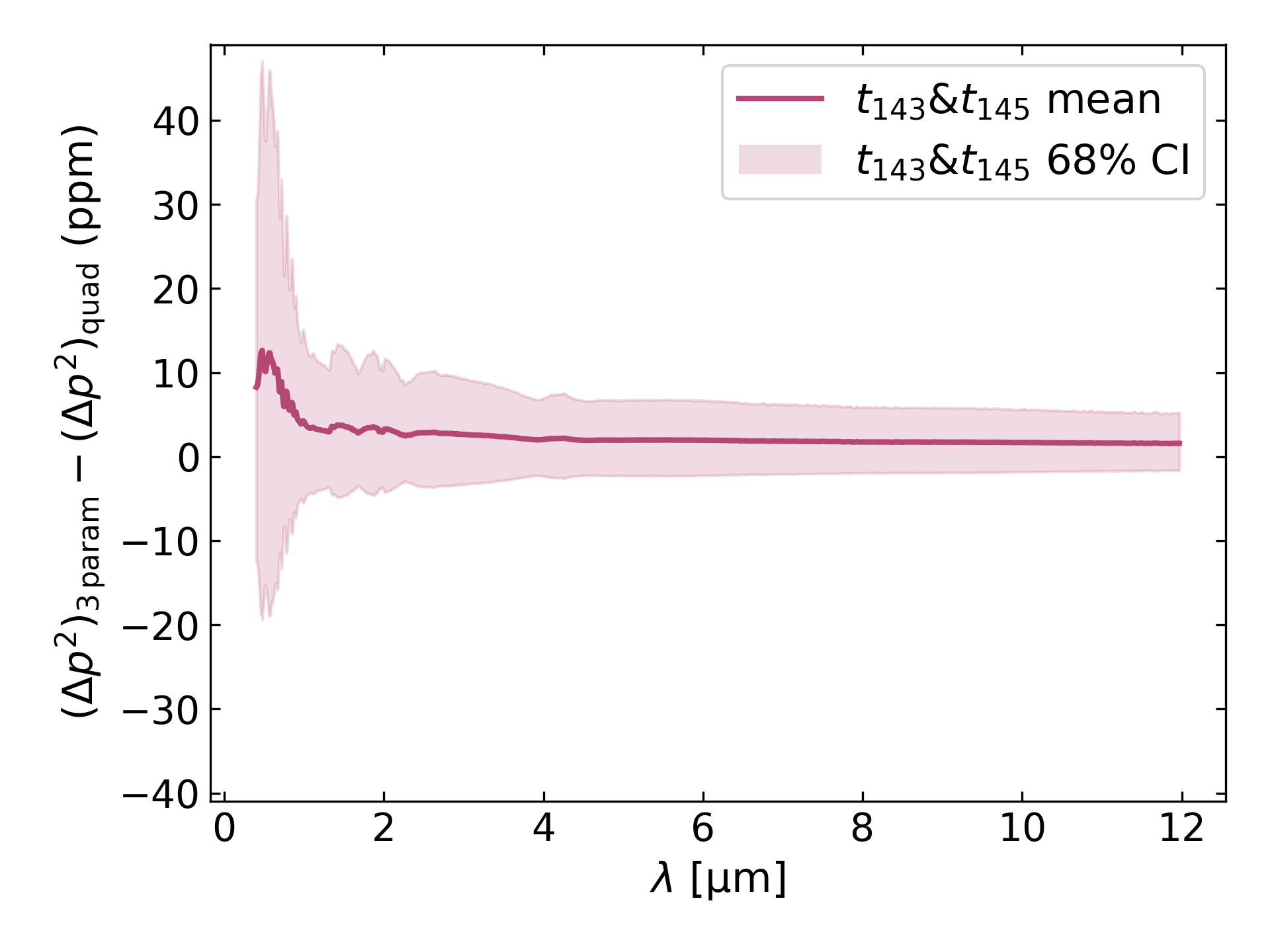}
        \caption{Observed epochs ($t_{143}$ and $t_{145}$).}
        \label{fig:iqr_1_transp}
    \end{subfigure}
    \hfill
    \begin{subfigure}{0.49\textwidth}
        \centering
        \includegraphics[trim={3.5ex 3.7ex 3.0ex 3.5ex},clip,width=\linewidth]{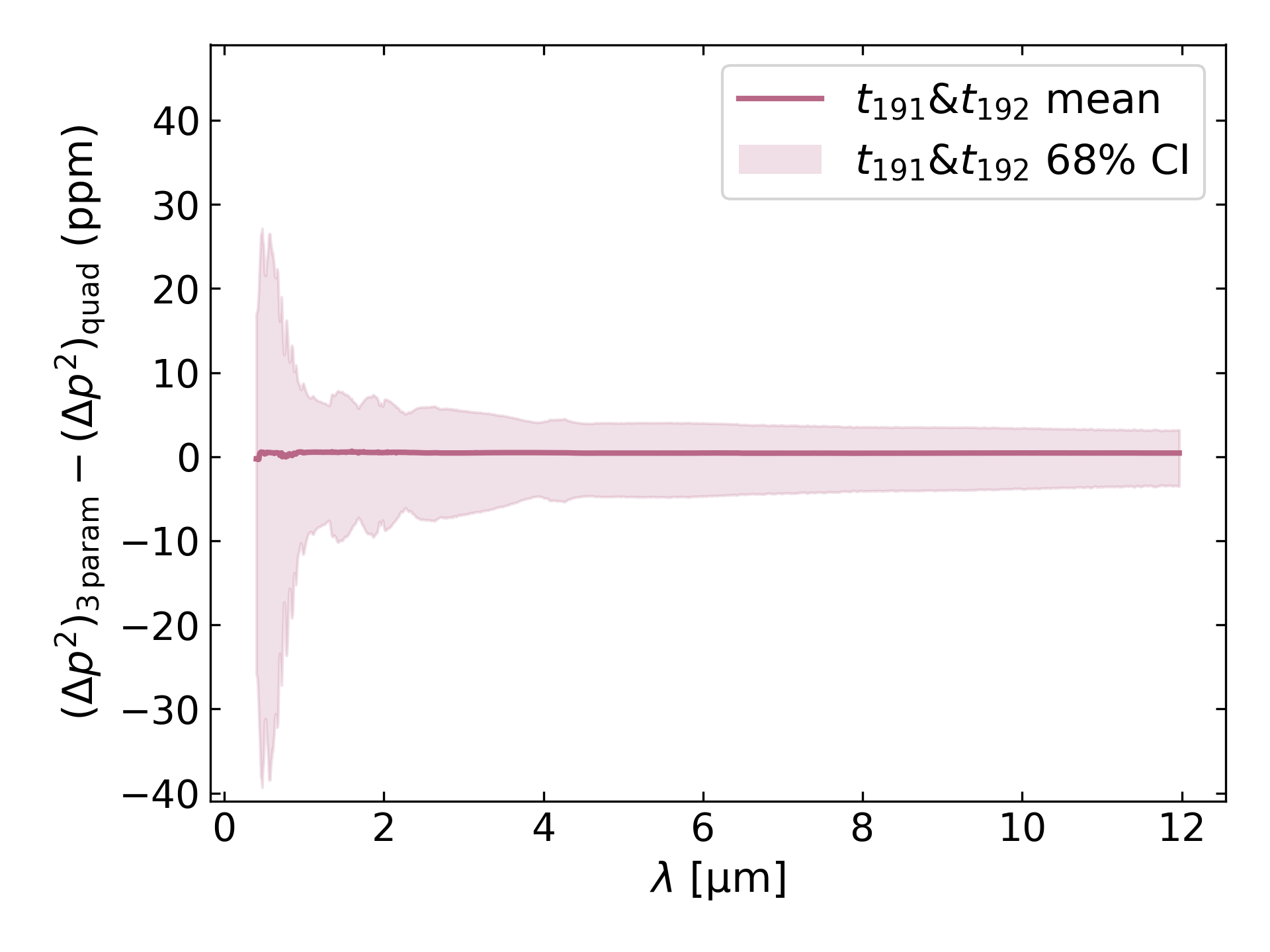}
        \caption{Activity maximum epochs ($t_{191}$ and $t_{192}$).}
        \label{fig:iqr_3_transp}
    \end{subfigure}

    \caption{Difference between LD prescriptions in activity corrections derived using the inversion framework.}
    \label{fig:inversions_quad_vs_3param}
\end{figure*}

\begin{figure*}
    \centering

    \begin{subfigure}{0.49\textwidth}
        \centering
        \includegraphics[trim={3.5ex 3.7ex 3.0ex 3.5ex},clip,width=\linewidth]{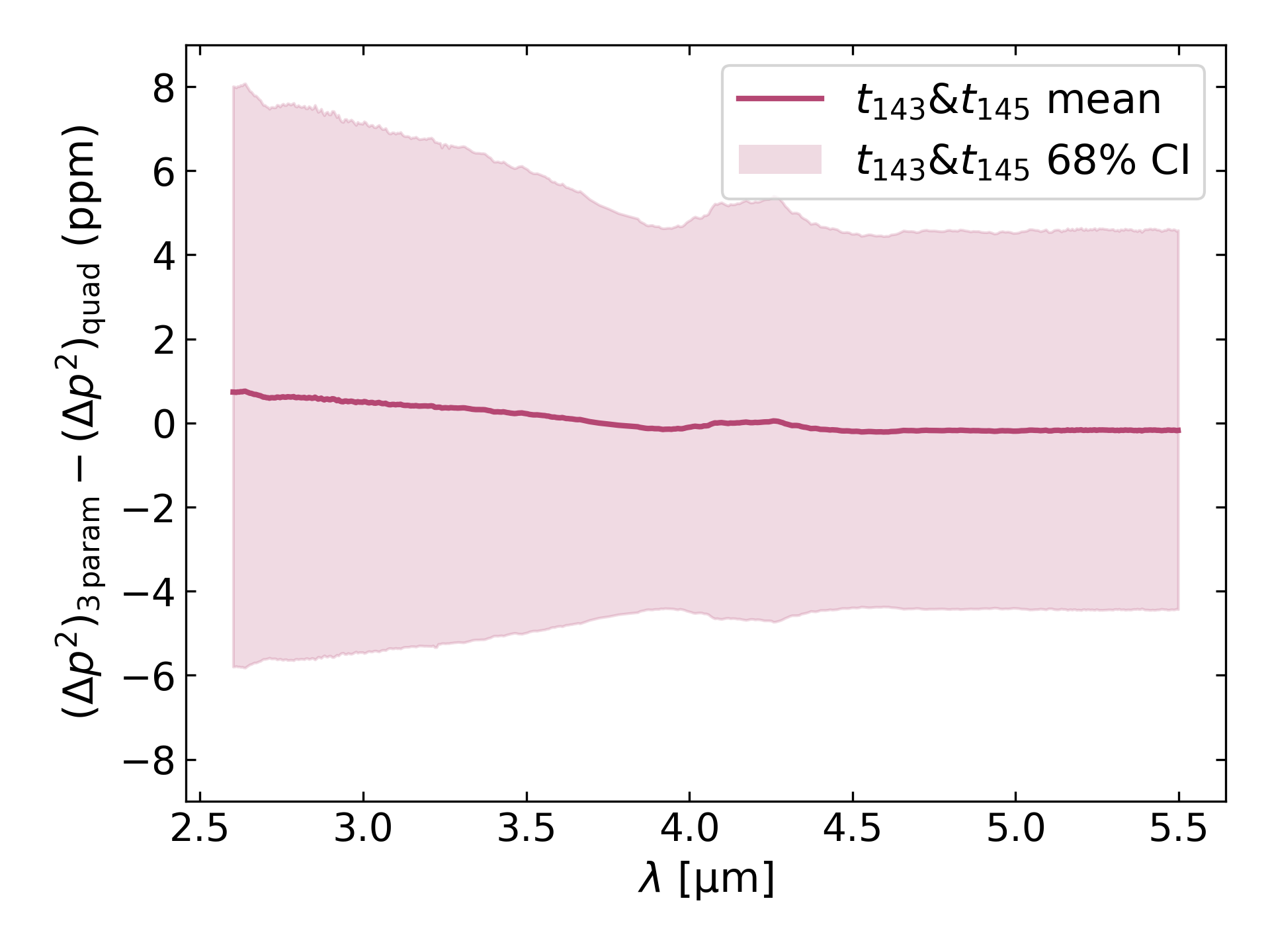}
        \caption{Observed epochs ($t_{143}$ and $t_{145}$).}
        \label{fig:iqr_1_transp_comp}
    \end{subfigure}
    \hfill
    \begin{subfigure}{0.49\textwidth}
        \centering
        \includegraphics[trim={3.5ex 3.7ex 3.0ex 3.5ex},clip,width=\linewidth]{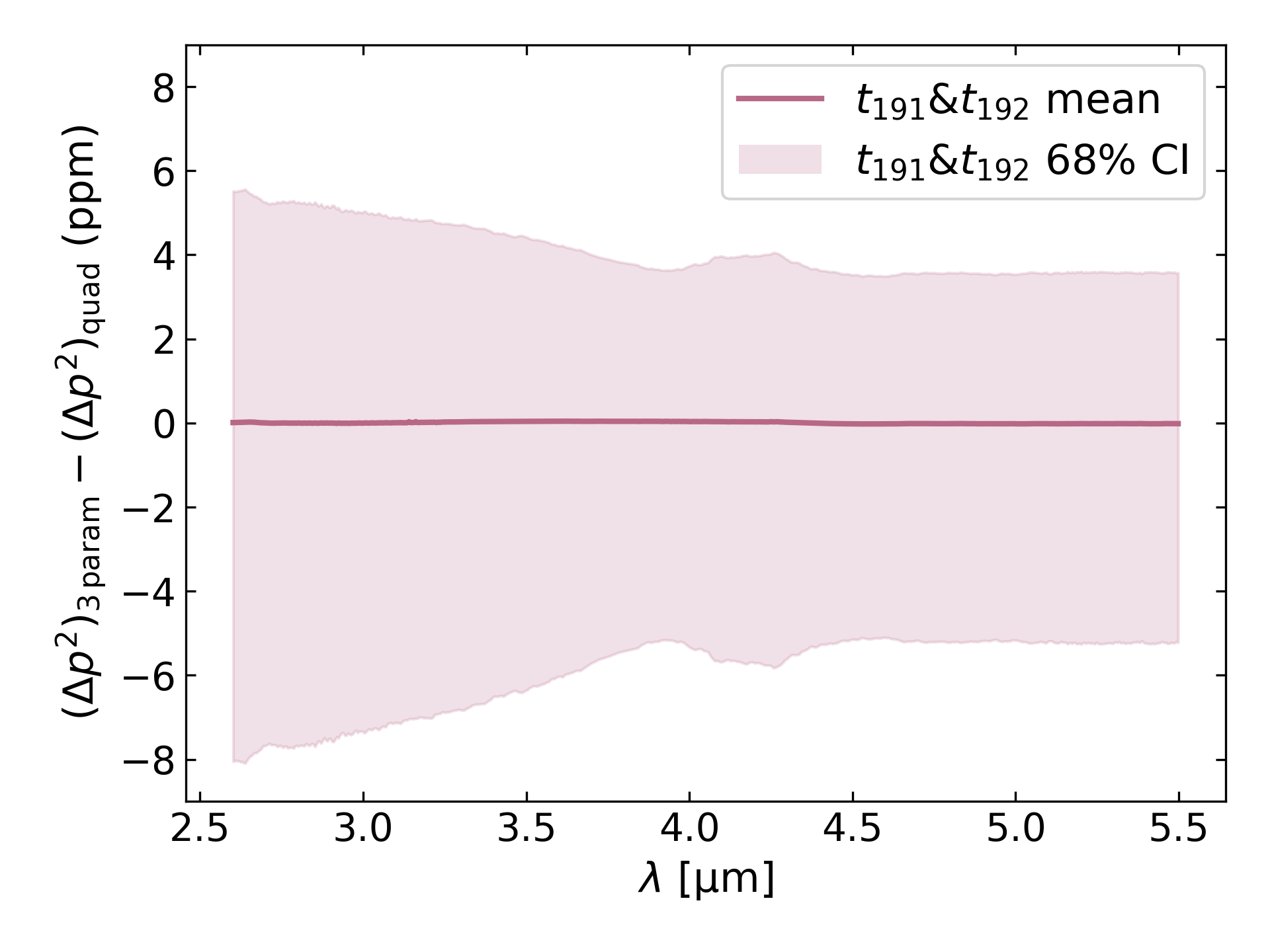}
        \caption{Activity maximum epochs ($t_{191}$ and $t_{192}$).}
        \label{fig:iqr_3_transp_comp}
    \end{subfigure}

    \caption{Same quantity as in Fig.~\ref{fig:inversions_quad_vs_3param}, but restricted to the wavelength range used for comparison with the \citet{Schlawin_2024} transmission spectrum. Curves have been centred around $0$ for clarity.}%
    \label{fig:inversions_comparison_quad_vs_3param}
\end{figure*}

In Figures~\ref{fig:inversions_quad_vs_3param} and \ref{fig:inversions_comparison_quad_vs_3param} we show the difference between the activity corrections derived with the three-parameter and quadratic LD laws for the inversion framework. %

For each visit, we compare the two LD prescriptions on a solution-by-solution basis. That is, for each retained inversion solution, corresponding to the same underlying stellar-surface configuration, we recompute the correction spectrum $\Delta p^2(\lambda)$ with both LD laws and compute the difference. %

Over the full analysed $0.4$--$\SI{12}{\micro\metre}$ range, the combined mean difference remains small and $0$ lies within the 68\% interval at every wavelength bin for both epoch pairs. However, the 68\% interval broadens towards the blue edge of the range, reaching several tens of ppm near $\lambda\sim\SI{0.4}{\micro\metre}$. In contrast, over the JWST/NIRSpec G395H range ($2.8$--$5.1\,\mu$m), both the mean difference and the 68\% interval remain small: the combined mean difference stays below $0.8$ ppm for the observed epochs and below $0.1$ ppm near activity maximum, while the corresponding 68\% intervals remain at the level of only a few ppm half-width. %

These values are negligible compared to the tens-of-ppm scale of the reported spectral structure in the JWST transmission spectrum (e.g. an in--out CO$_2$ band depth difference of $58\pm15\,\mathrm{ppm}$; \citealt{Schlawin_2024}). We therefore conclude that, for GJ\,1214\,b and our observing geometry ($b\simeq0.26$), the choice between these two LD prescriptions has a negligible impact on the inferred stellar-contamination corrections in the NIRSpec G395H bandpass for this particular target, although larger effects are expected towards $\lambda\sim\SI{0.4}{\micro\metre}$ and when using the quadratic law for high-precision transmission spectroscopy, particularly for planets with moderate to high impact parameters ($b \gtrsim 0.5$) \citep{2024ApJ...977L...7K}. %

\section{Spot-number selection diagnostics}
\label{app:spot_number_ic}

In regular nested models, the likelihood-ratio statistic $2\Delta\hat{\ell}$ is asymptotically $\chi^2$-distributed with degrees of freedom equal to the number of additional free parameters. Spot models are non-regular (e.g. boundary solutions such as $r_\mathrm{sp}\rightarrow 0$ and multi-modal likelihood surfaces), and our optimisation is stochastic; therefore we use $2\Delta\hat{\ell}$ as a heuristic diagnostic of fit improvement rather than as a formal hypothesis test.

AIC is prediction-oriented, whereas BIC applies a stronger complexity penalty scaling with $\ln n_f$, where $n_f$ is the sample size. We compute $n_f$ as the number of fitted data points entering the likelihood. Because our monitoring data are time series with correlated residuals and shared systematics across indicators, $n_f$ should be interpreted as an effective sample size; hence, in this regime, BIC tends to be conservative. We therefore selected $N_\mathrm{sp}$ based on AIC and used BIC and $2\Delta\hat{\ell}$ as diagnostic cross-checks.

The AIC is defined as
\begin{equation}
\mathrm{AIC}_N = 2k_N - 2\hat{\ell}_N,
\end{equation}
where $\hat{\ell}_N$ is the maximised log-likelihood of the $N$-spot model and $k_N$ is its number of free parameters, and %
the Bayesian information criterion (BIC) as
\begin{equation}
\mathrm{BIC}_N = k_N \ln n_f - 2\hat{\ell}_N.%
\end{equation}

In our \texttt{StarSim} setup all non-spot parameters are common to all spot-count families. Substituting $k_N = k_0 + 3N$ into AIC and BIC gives
\begin{align}
\mathrm{AIC}_N &= 2k_0 + \underbrace{\left(6N - 2\hat{\ell}_N\right)}_{\mathrm{AIC}^\star_N},\\
\mathrm{BIC}_N &= k_0\ln n_f + \underbrace{\left(3N\ln n_f - 2\hat{\ell}_N\right)}_{\mathrm{BIC}^\star_N}.
\end{align}
The additive constants $2k_0$ and $k_0\ln n_f$ are identical for all spot-count families and therefore cancel in model comparisons. We thus report $\mathrm{AIC}^\star_N$ and $\mathrm{BIC}^\star_N$, and their differences relative to the minimum across $N_\mathrm{sp}$.

For GJ\,1214, we use $n_f=306$ (the total number of fitted data points across all time series). The resulting diagnostics $(\hat{\ell}_N,\Delta\hat{\ell},2\Delta\hat{\ell},\mathrm{AIC}^\star,\mathrm{BIC}^\star)$ are listed in Table~\ref{tab:spot_number_ic}.

The diagnostics show the expected trade-off between fit quality and complexity: $\mathrm{AIC}^\star$ is minimised at $N_\mathrm{sp}=4$, whereas $\mathrm{BIC}^\star$ favours $N_\mathrm{sp}=2$ because of its stronger complexity penalty. Since our goal is predictive performance and BIC is likely overly conservative in this non-regular, time-correlated setting, we adopt AIC as the primary criterion and therefore use $N_\mathrm{sp}=4$ in the main analysis.

\begin{table*}
\centering
    \caption{Spot-number selection diagnostics for the \texttt{StarSim} inversions of GJ\,1214.}
    \label{tab:spot_number_ic}
    \begin{tabular}{cccccccc}
    \hline\hline
    $N_\mathrm{sp}$ & $\hat{\ell}_N$ & $\Delta\hat{\ell}$ & $2\Delta\hat{\ell}$ & $\mathrm{AIC}^\star$ & $\Delta\mathrm{AIC}^\star$ & $\mathrm{BIC}^\star$ & $\Delta\mathrm{BIC}^\star$ \\
    \hline
    1 & 572.987 & --    & --    & -1139.974 & 26.683 & -1128.803 & 13.211 \\
    2 & 588.178 & 15.191& 30.381& -1164.356 & 2.302  & -1142.014 & 0.000  \\
    3 & 590.857 & 2.680 & 5.359 & -1163.715 & 2.943  & -1130.203 & 11.811 \\
    4 & 595.329 & 4.471 & 8.943 & -1166.658 & 0.000  & -1121.975 & 20.039 \\
    5 & 596.026 & 0.697 & 1.395 & -1162.052 & 4.605  & -1106.199 & 35.815 \\
    \hline
    \end{tabular}
    \tablefoot
    {
     We report $\hat{\ell}_N$, adjacent-model gains $\Delta\hat{\ell}_{N-1\rightarrow N}=\hat{\ell}_{N}-\hat{\ell}_{N-1}$, and the ranking-equivalent information criteria $\mathrm{AIC}^\star_N$ and $\mathrm{BIC}^\star_N$ (see text), along with differences relative to the minimum (lower is better). Derived quantities are computed from full-precision values and rounded to three decimals. %
    }%
\end{table*}

\section{Clustering analysis of inversion surface-map solutions}
\label{app:inv_clustering}

The inversion produces an ensemble of $N_{\rm sol}$ retained spot configurations that fit the monitoring data comparably well. To summarise the structure of this ensemble in surface-configuration space, we cluster the full $N_{\rm sp}$-spot solutions at a given epoch using a $K$-medoids algorithm, where each cluster representative (the medoid) is an \emph{actual} inversion solution \citep{KaufmanRousseeuw1990}. This analysis is used only to aid interpretation of dominant families of surface solutions; our contamination corrections $\Delta p^2(\lambda)$ are still computed from the full retained ensemble as described in Section~\ref{subsubsec:inv_ensemble_inference}.

Because each inversion solution is a full surface configuration, clustering requires a notion of dissimilarity between two $N_{\rm sp}$-spot solutions. We represent the $i$-th retained solution as an unordered set of spots,
\begin{equation}
\mathcal{S}_i \;=\; \{\, s_{i,1},\ldots,s_{i,N_{\rm sp}} \,\},
\qquad
s_{i,n} \;=\; (\theta_{i,n},\,\phi_{i,n},\,r_{i,n}),
\end{equation}
where $(\theta,\phi)$ is the spot-centre position (colatitude and longitude) and $r$ is the angular radius.

We define a spot dissimilarity metric that combines an angular separation term and a radius difference penalty. For two spots $s=(\theta,\phi,r)$ and $s'=(\theta',\phi',r')$,
\begin{equation}
\delta(s,s') \;=\; w_{\rm ang}\,\Delta\psi\big((\theta,\phi),(\theta',\phi')\big)\;+\;w_r\,|r-r'|,
\end{equation}
where $\Delta\psi$ is the great-circle distance, and $(w_{\rm ang},w_r)$ are fixed weights (we adopt $w_{\rm ang}=1$ and $w_r=0.3$). %

The distance between two $N_{\rm sp}$-spot solutions is then defined by optimally matching their spots (minimum-cost assignment) under $\delta$:
\begin{equation}
d(\mathcal{S}_i,\mathcal{S}_j)
\;=\;
\min_{\pi\in \Pi}\,\sum_{n=1}^{N_{\rm sp}}
\delta\!\left(s_{i,n},\,s_{j,\pi(n)}\right),
\end{equation}
where $\Pi$ is the set of permutations of $\{1,\ldots,N_{\rm sp}\}$.

For a given $K$, $K$-medoids selects a set of medoid indices
$M_K=\{m_1,\ldots,m_K\}\subset\{1,\ldots,N_{\rm sol}\}$, where each $m_k$ corresponds to an actual inversion solution (medoid) $\mathcal{S}_{m_k}$.
Given a candidate medoid set $M_K$, each solution $\mathcal{S}_i$ is assigned to its nearest medoid under the configuration distance $d$, and the total within-cluster dissimilarity is the sum of these nearest-medoid distances. The objective for a fixed $K$ is therefore
\begin{equation}
J(K)\;\coloneqq\;\min_{\substack{M_K\subset\{1,\ldots,N_{\rm sol}\}}}
\;\sum_{i=1}^{N_{\rm sol}} \min_{m_k\in M_K} d\!\left(\mathcal{S}_i,\,\mathcal{S}_{m_k}\right).
\end{equation}
For a fixed $K$, we minimise $J(K)$ with the standard Partitioning Around Medoids (PAM) algorithm \citep{KaufmanRousseeuw1990}, which iteratively updates the medoid set until no further improvement in $J(K)$ is found. %

\begin{figure*}[t]
\centering
\begin{subfigure}[t]{1.0\linewidth}
  \centering
  \includegraphics[width=0.65\linewidth]{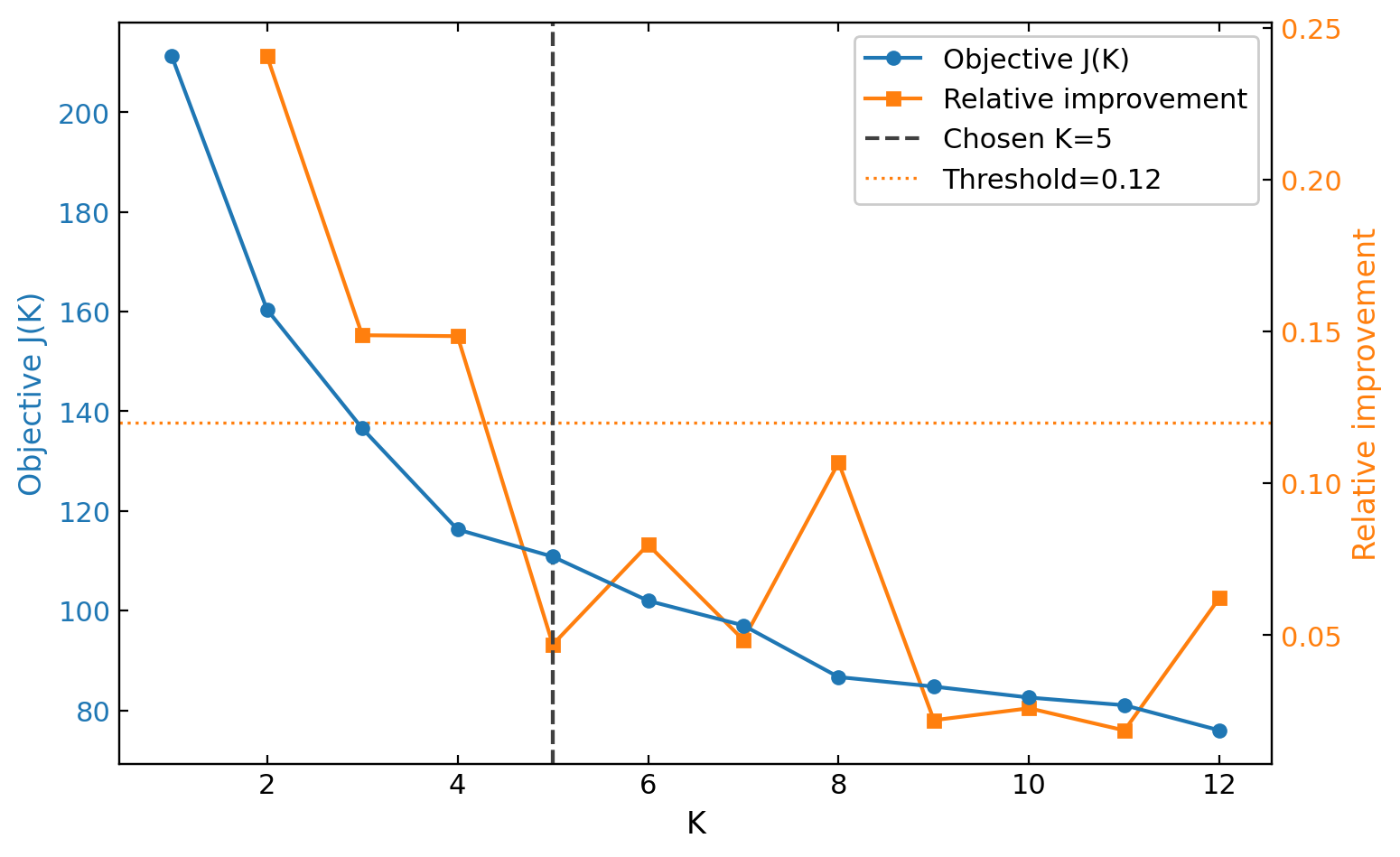}
\end{subfigure}

\begin{subfigure}[t]{1.0\linewidth}
  \centering
  \includegraphics[width=0.65\linewidth]{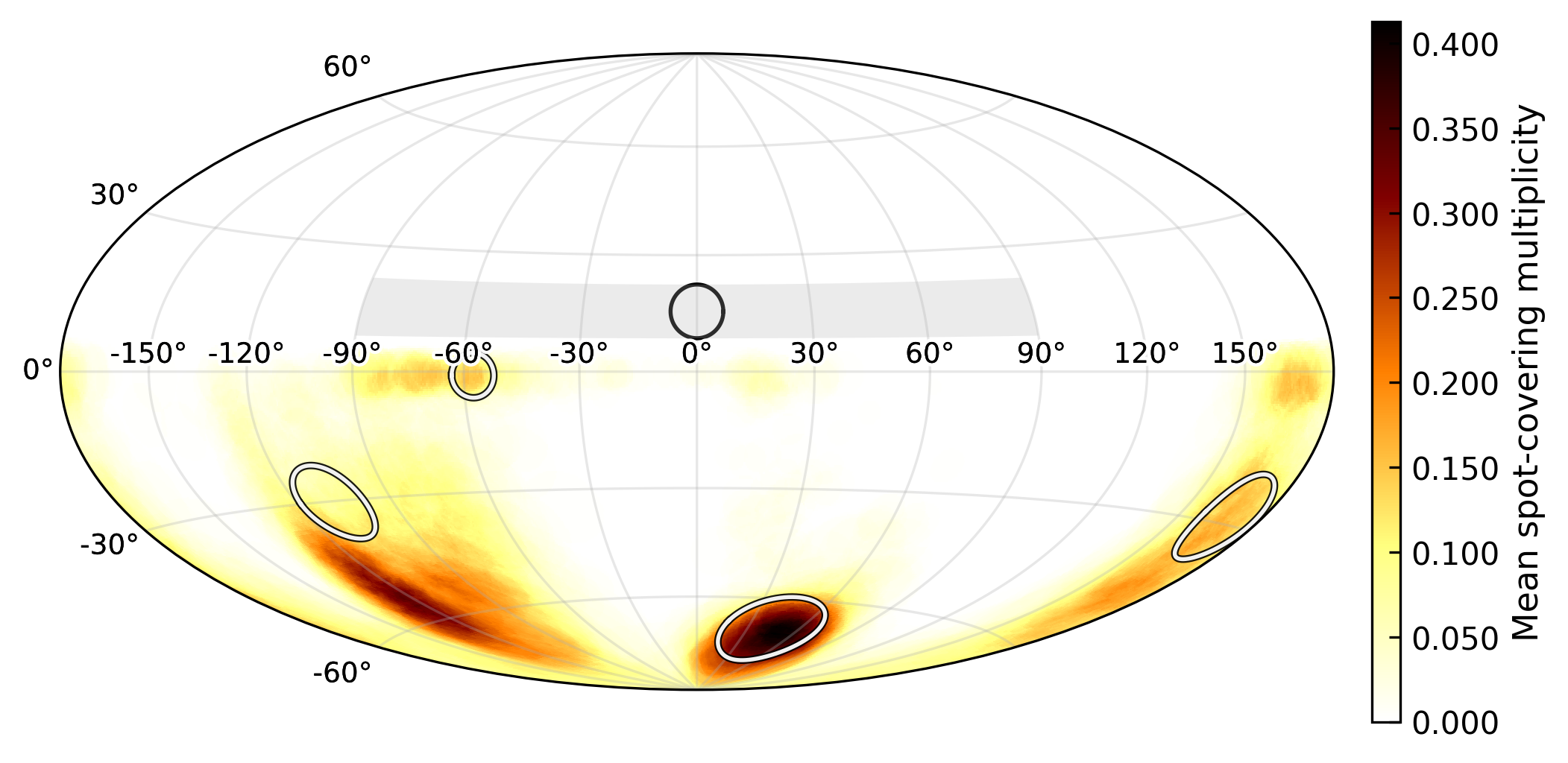}
\end{subfigure}

\caption{Additional clustering diagnostics for the inversion surface-map ensemble at the $t_{145}$ epoch. Top: Elbow diagnostic for selecting $K$, showing the $K$-medoids objective $J(K)$ and relative improvement $\mathrm{RI}(K)$. The horizontal line marks the adopted threshold $\tau=0.12$; the vertical line marks $K=5$, the smallest value with $\mathrm{RI}(K)<\tau$. Bottom: All-solution mean spot-covering multiplicity map in a Hammer projection, with the spot outlines of the largest cluster's medoid overplotted in white.} %
\label{fig:inv_clustering}
\end{figure*}

We select $K$ with an elbow criterion based on the relative improvement
\begin{equation}
\mathrm{RI}(K) \;\coloneqq\; \frac{J(K-1)-J(K)}{J(K-1)} \qquad (K\ge 2),
\end{equation}
adopting a threshold $\tau=0.12$ and choosing the smallest $K$ such that $\mathrm{RI}(K)<\tau$. For $t_{145}$ this yields $K=5$ (Fig.~\ref{fig:inv_clustering}, top).

For each cluster we compute a cluster-conditional mean spot-covering multiplicity map on a latitude--longitude grid folded to the anti-planet hemisphere (overlaps add), and overplot the medoid spot outlines along with the transit chord band for context; these cluster-conditional maps are shown in the main text (Fig.~\ref{fig:inv_clustering_main}). In this appendix, we additionally show the corresponding all-solution mean spot-covering multiplicity map with the medoid of the largest cluster overplotted (Fig.~\ref{fig:inv_clustering}, bottom).%

\section{\UNSPOTTER{} learning curve}
\label{app:unspotter_learning_curve}

To assess whether the size of the synthetic training set limits the predictive performance of \UNSPOTTER{}, we performed a controlled learning-curve experiment. We used the network configuration adopted in the main analysis and progressively reduced the number of independent base simulations used for training to 12.5\%, 25\%, 50\%, 75\%, and 100\% of the adopted training sample, while keeping the validation and test sets fixed. The experiment was performed for the $t_{145}$ network. As in the production training, each independent base simulation contributes six noisy realisations. Three networks with different random initialisations were trained at each training-set size.

\begin{figure*}[t]
\centering
\includegraphics[width=0.6\linewidth]{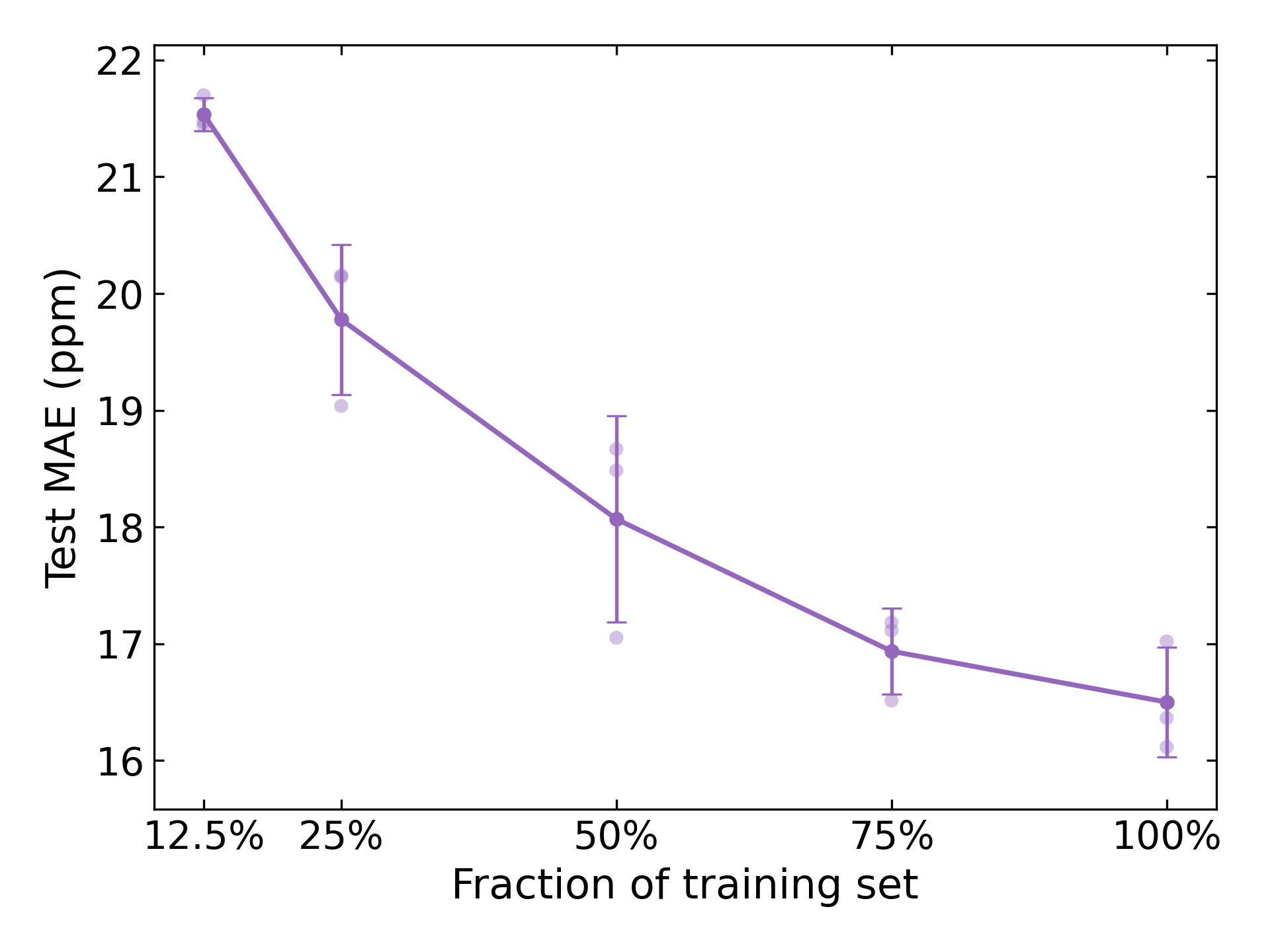}
\caption{
Learning curve for the \UNSPOTTER{} network at the $t_{145}$ epoch.
The horizontal axis gives the fraction of the complete training set used for training.
Small points show the test-set MAE for three independently initialised trainings at each sample size, while filled markers and error bars show their mean and standard deviation.
The validation and test sets are held fixed throughout the experiment.
}
\label{fig:unspotter_learning_curve}
\end{figure*}

Figure~\ref{fig:unspotter_learning_curve} shows the resulting test-set MAE. The error decreases systematically as the training sample is increased, from $21.54\pm0.14$ ppm at 12.5\% of the training set to $16.50\pm0.47$ ppm for the full training set. The improvement becomes progressively smaller towards the largest training-set sizes: increasing the training sample from 75\% to 100\% decreases the mean MAE from $16.94$ to $16.50$ ppm, an improvement of approximately $0.44$ ppm ($2.6\%$), comparable to the run-to-run dispersion between independently initialised networks at these sample sizes.

The learning curve therefore places the adopted training-set size in a regime of diminishing returns, as it indicates that the predictive performance is no longer strongly limited by the number of synthetic training simulations, and that substantially enlarging the simulation set would be expected to yield only modest further gains relative to the additional computational cost.

\section{Comparison with the JWST/NIRSpec transmission spectrum for the inversion framework}
\label{app:inv_comp_with_real}

\begin{figure*}[t]
\centering
\includegraphics[trim={2.75ex 0.25ex 1.5ex 1.0ex},clip,width=\linewidth]{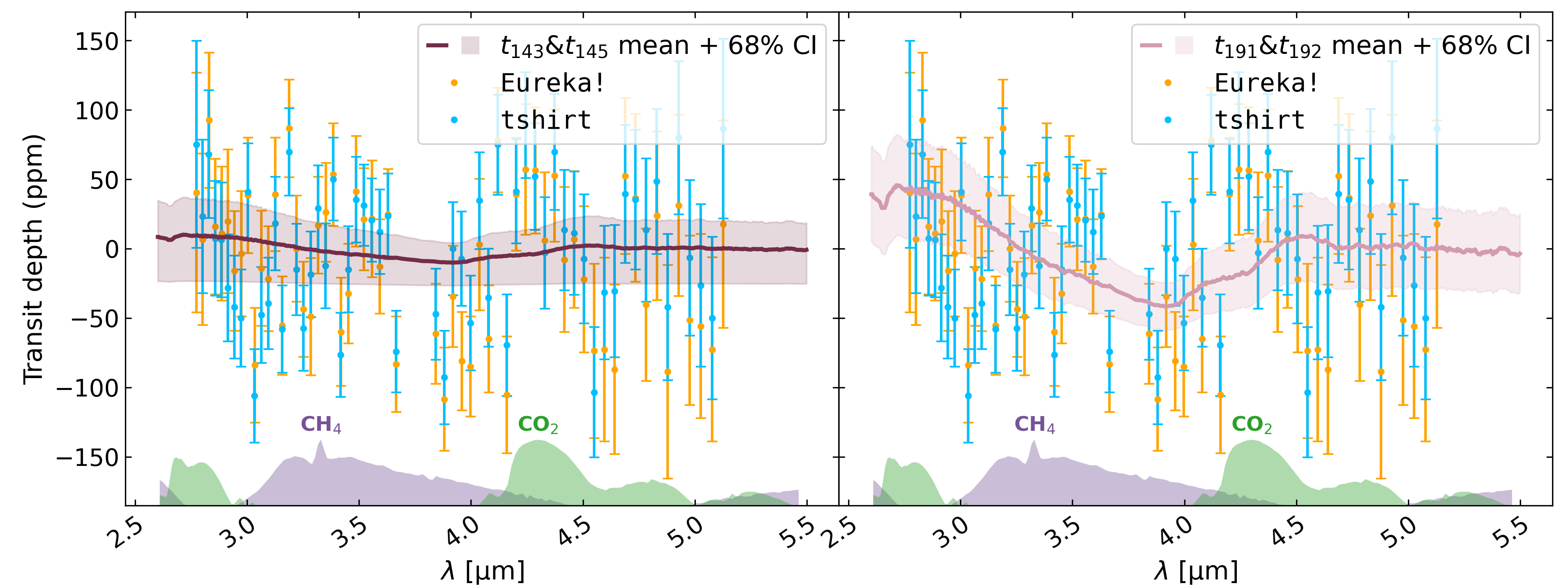}
\caption{Same as Fig.~\ref{fig:corrections_comp_jwst_unspotter}, but for the inversion.}
\label{fig:corrections_comp_jwst_inversion}
\end{figure*}

Figure~\ref{fig:corrections_comp_jwst_inversion} shows the same comparison with the \citet{Schlawin_2024} transmission spectrum as in Figure~\ref{fig:corrections_comp_jwst_unspotter}, but using the inversion-based corrections. %

\section{Corrections for individual transits}\label{appendix:all_times_appendix}

\begin{figure*}
    \centering

    \begin{subfigure}{0.49\textwidth}
        \centering
        \includegraphics[trim={3.5ex 3.7ex 3.0ex 3.5ex},clip,width=\linewidth]{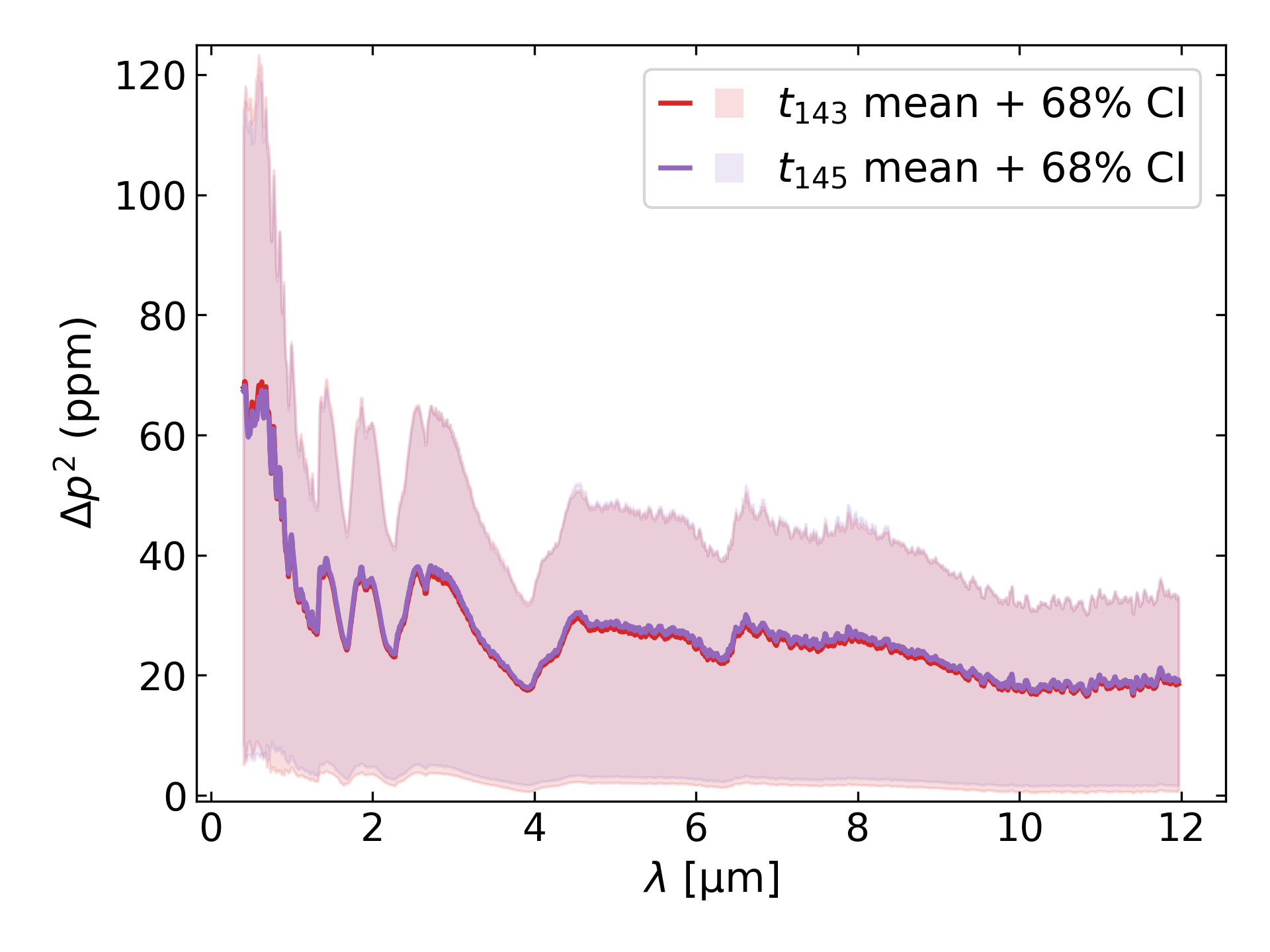}
        \caption{$t_{143}$ and $t_{145}$}
        \label{fig:iqr_1_transp_all_times_appendix}
    \end{subfigure}
    \hfill
    \begin{subfigure}{0.49\textwidth}
        \centering
        \includegraphics[trim={3.5ex 3.7ex 3.0ex 3.5ex},clip,width=\linewidth]{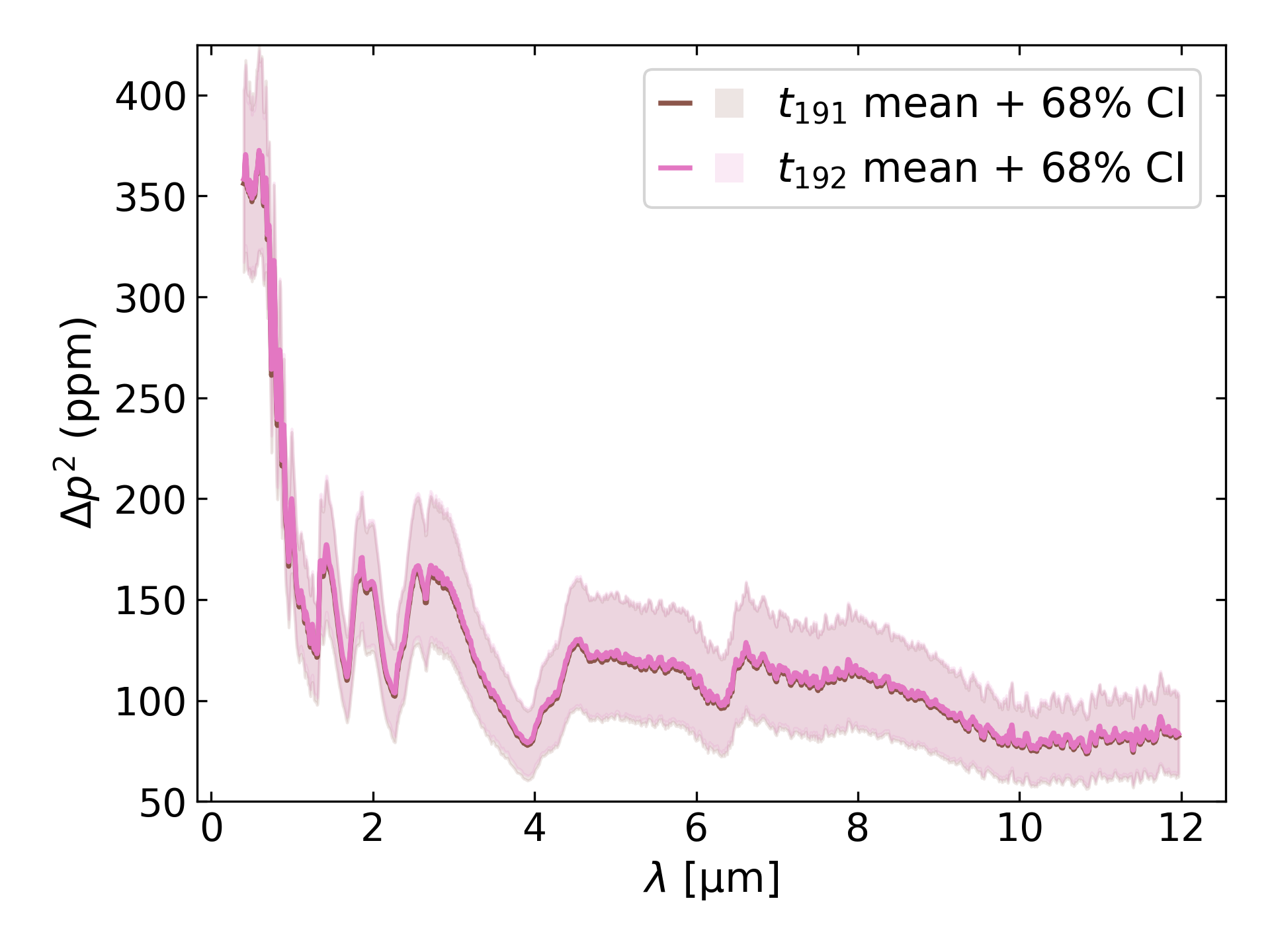}
        \caption{$t_{191}$ and $t_{192}$}
        \label{fig:iqr_3_transp_all_times_appendix}
    \end{subfigure}

    \caption{Comparison of transit depth corrections for inversions.}
    \label{fig:inversions_all_times_appendix}
\end{figure*}

\begin{figure*}
    \centering

    \begin{subfigure}{0.49\textwidth}
        \centering
        \includegraphics[trim={3.5ex 3.7ex 3.0ex 3.5ex},clip,width=\linewidth]{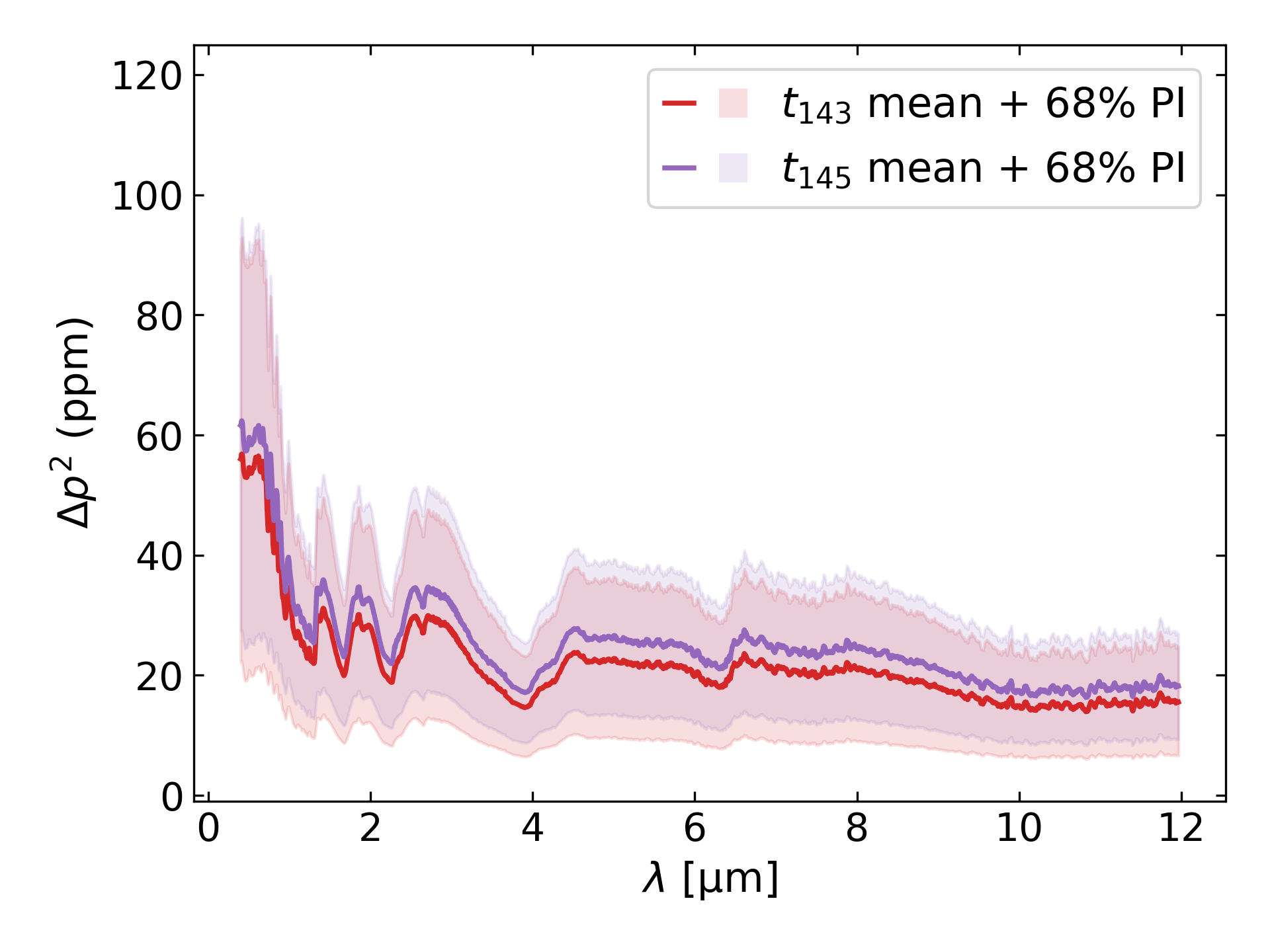}
        \caption{$t_{143}$ and $t_{145}$}
        \label{fig:NN_t143_all_times_appendix}
    \end{subfigure}
    \hfill
    \begin{subfigure}{0.49\textwidth}
        \centering
        \includegraphics[trim={3.5ex 3.7ex 3.0ex 3.5ex},clip,width=\linewidth]{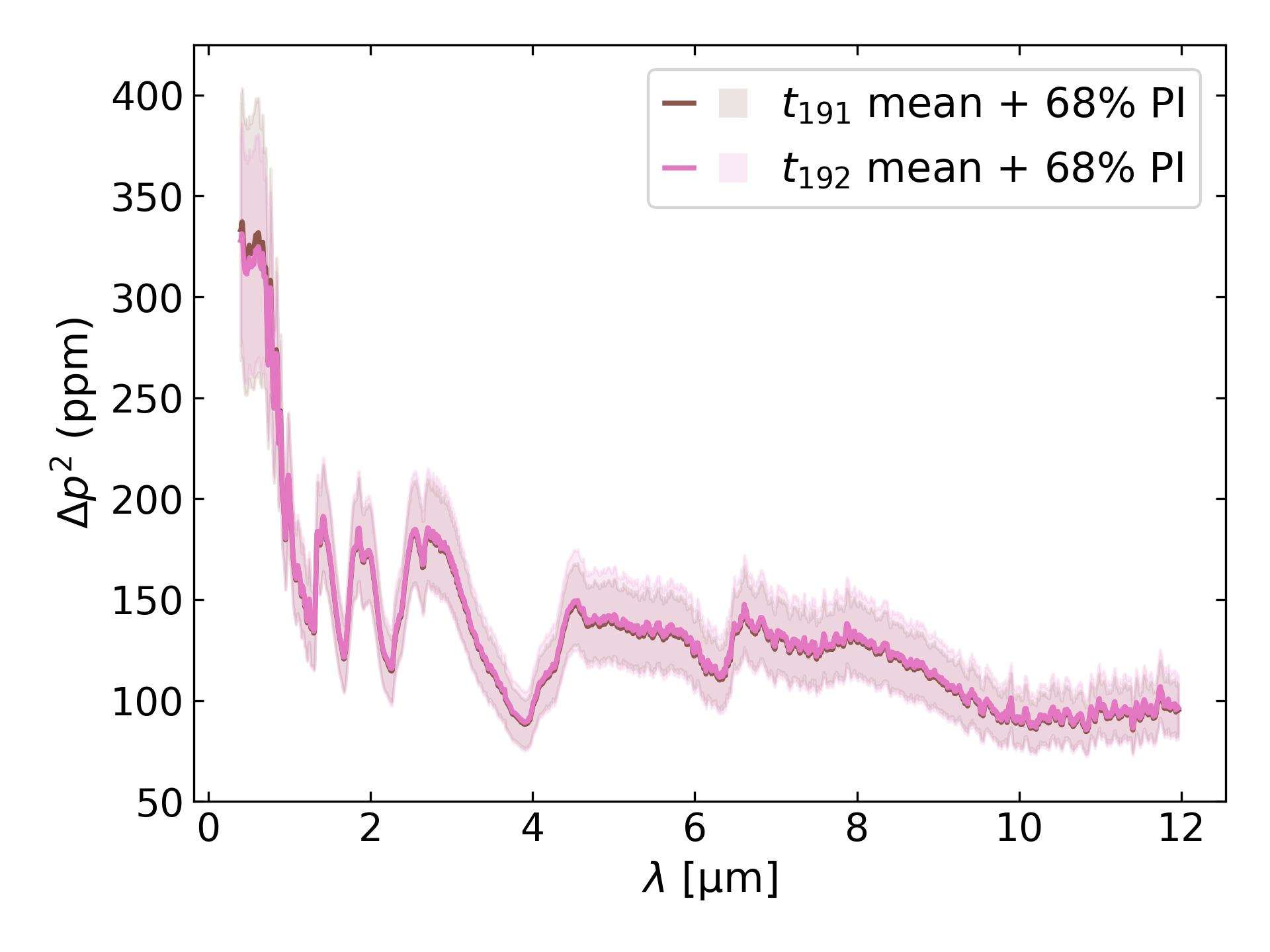}
        \caption{$t_{191}$ and $t_{192}$}
        \label{fig:NN_t145_all_times_appendix}
    \end{subfigure}

    \caption{Comparison of transit depth corrections for the \UNSPOTTER{} framework.}
    \label{fig:NN_all_times_appendix}
\end{figure*}

Figures \ref{fig:inversions_all_times_appendix} and \ref{fig:NN_all_times_appendix} show the inversion and \UNSPOTTER{} corrections derived for each individual epoch ($t_{143}$, $t_{145}$, $t_{191}$ and $t_{192}$).

\section{Comparison with the HST/WFC3 transmission spectrum}
\label{app:hst_comparison}

\begin{figure*}[t]
\centering
\includegraphics[width=\linewidth]{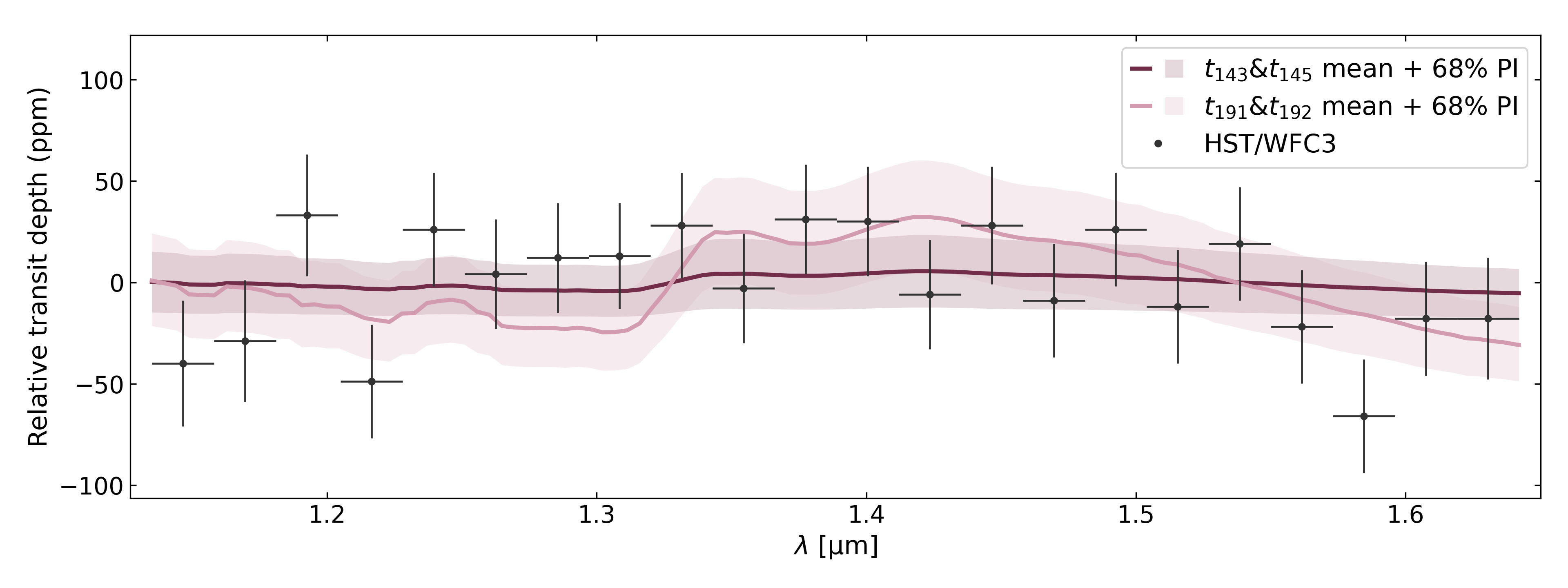}
\caption{Illustrative comparison of \UNSPOTTER{} activity corrections with the HST/WFC3 transmission spectrum from \citet{Kreidberg_2014}. Curves and data are centred around zero for visual clarity.}
\label{fig:corrections_comp_hst_unspotter}
\end{figure*}

Figure~\ref{fig:corrections_comp_hst_unspotter} compares the \UNSPOTTER{} corrections with the HST/WFC3 transmission spectrum of \citet{Kreidberg_2014}, to assess whether the corrections inferred from our monitoring are compatible with the previously measured, featureless spectrum. The HST/WFC3 spectrum combines twelve transits observed in 2012--2013, during a different observing period from our monitoring and potentially sampling different stellar activity states. We therefore make an illustrative comparison with the corrections inferred for both the observed epochs and the epochs near activity maximum in our monitoring. %

Assuming a flat intrinsic planetary transmission spectrum, we compare the HST measurements with each fixed mean correction profile averaged over the HST wavelength bins, fitting only a constant depth offset. Using the quoted HST uncertainties as independent Gaussian errors gives $\chi^2=20.0$ and $21.0$ for 21 degrees of freedom for the observed and near-maximum epoch pairs, respectively. The HST measurements are therefore statistically consistent with the chromatic corrections predicted for both activity levels examined here. %

\end{document}